\documentclass[journal]{IEEEtran}

\usepackage{cite}
\usepackage{graphicx}
\usepackage{amsfonts}
\usepackage{caption}
\usepackage{subcaption}
\usepackage{diagbox}
\usepackage{upgreek}
\usepackage{comment}
\usepackage{multirow}
\usepackage{stfloats}
\usepackage{hhline}
\usepackage{xcolor}

\usepackage{amsmath}
\usepackage{algorithmic}
\usepackage{array}
\newcolumntype{C}[1]{>{\centering\let\newline\\\arraybackslash\hspace{0pt}}m{#1}}
\usepackage{url}
\usepackage{color,soul}

\usepackage{stfloats}
\usepackage{url}
\usepackage{footnote}

\begin{document}
\title{Survey on Multi-Access Edge Computing for \\Internet of Things Realization}
%
%
%

\author{Pawani~Porambage,~\IEEEmembership{Student Member,~IEEE,}
        Jude~Okwuibe,~\IEEEmembership{Student Member,~IEEE,}
        Madhusanka~Liyanage,~\IEEEmembership{Member,~IEEE,}  
        ~Mika~Ylianttila,~\IEEEmembership{Senior Member,~IEEE,}
        and~Tarik~Taleb~\IEEEmembership{Senior Member,~IEEE}
\thanks{Pawani Porambage, Jude Okwuibe, Madhusanka Liyanage, and Mika Ylianttila are with the Center for Wireless Communications, University of Oulu, Finland. e-mail:\{firstname.lastname\}@oulu.fi}
\thanks{Tarik Taleb is with Department of Communications and Networking, Aalto University, Finland, and Sejong University, Korea. e-mail: tarik.taleb@aalto.ﬁ
}
}




\maketitle

\begin{abstract}
The Internet of Things~(IoT) has recently advanced from an experimental technology to what will become the backbone of future customer value for both product and service sector businesses. This underscores the cardinal role of IoT on the journey towards the fifth generation~(5G) of wireless communication systems. IoT technologies augmented with intelligent and big data analytics are expected to rapidly change the landscape of myriads of application domains ranging from health care to smart cities and industrial automations. The emergence of Multi-Access Edge Computing~(MEC) technology aims at extending  cloud computing capabilities to the edge of the radio access network, hence providing real-time, high-bandwidth, low-latency access to radio network resources. IoT is identified as a key use case of MEC, given MEC's ability to provide cloud platform and gateway services at the network edge. MEC will inspire the development of myriads of applications and services with demand for ultra low latency and high Quality of Service~(QoS) due to its dense geographical distribution and wide support for mobility. MEC is therefore an important enabler of IoT applications and services which require real-time operations. In this survey, we provide a holistic overview on the exploitation of MEC technology for the realization of IoT applications and their synergies. We further discuss the technical aspects of enabling MEC in IoT and provide some insight into various other integration technologies therein.

\end{abstract}


\IEEEpeerreviewmaketitle


\section{Introduction}
\label{sec:Introduction}
\begin{figure}[ht]
  \centering
  \includegraphics[width=0.48\textwidth]{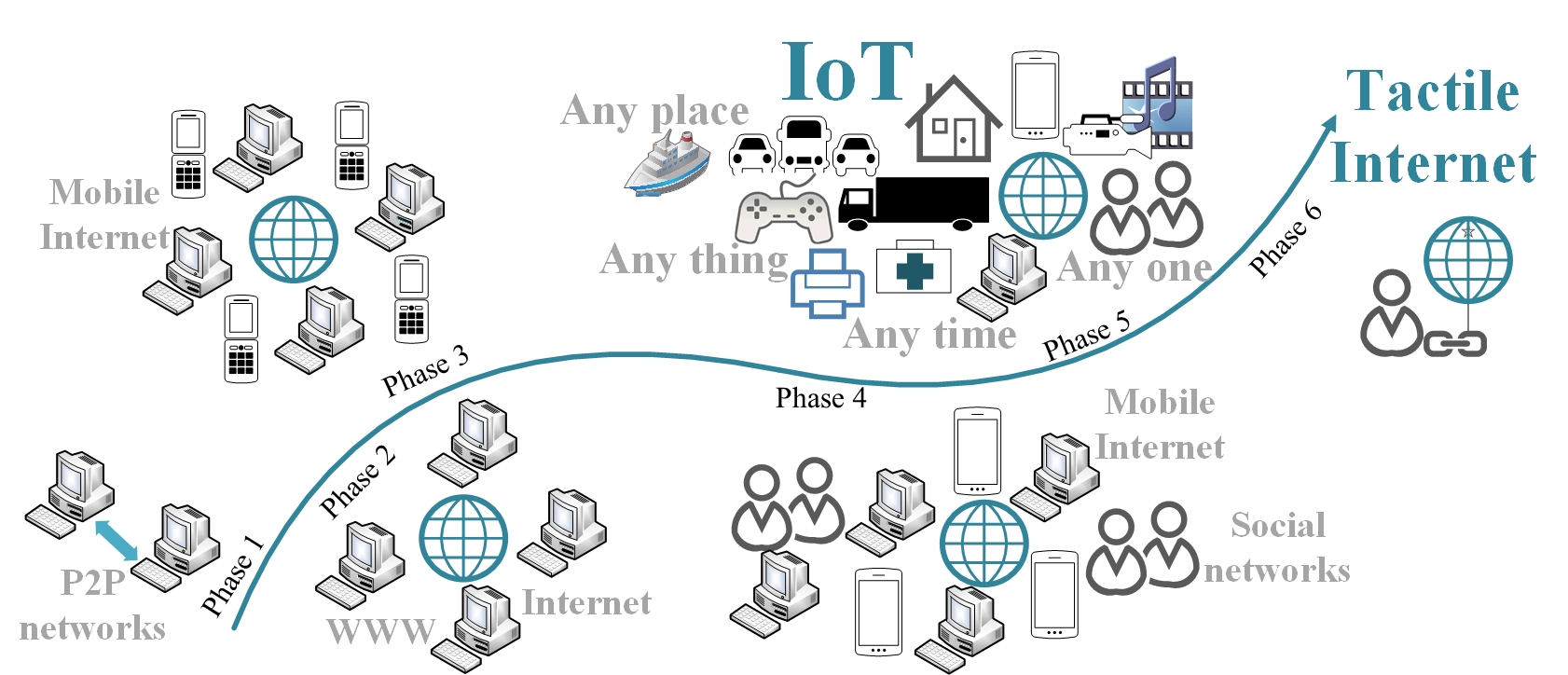}
  \caption{Evolution of the Internet.}
  \label{fig_IoTevolution}
\end{figure}

\IEEEPARstart{O}{ver} the last four decades, the Internet has evolved from peer-to-peer networking to world-wide-web, and mobile-Internet to the Internet of Things~(IoT) (Figure~\ref{fig_IoTevolution}). IoT emerged as a huge paradigm shift by connecting a versatile and massive collection of smart objects to the Internet. With IoT, people and things are able to connect at any time to any place with anything and anyone, ideally using any path or network and any available services~\cite{IoTdefinition}. From the user and application points of view, fifth generation~(5G) wireless networks will be highly capable mobile networks with high bandwidth (e.g., 10 Gbps), very low latency (e.g., 1~ms), and low operational cost which will lead to highly improved quality of service and quality of experience. Another significant advancement of the Internet will be the Tactile Internet; which is a highly advanced use case of human-to-machine and machine-to-machine interaction characterized by ultra low latency with extremely high availability, reliability and security. 

IoT system is poised to induce a significant surge in demand for data, computing resources, as well as networking infrastructures in order to accommodate the anticipated myriads of interconnected devices. Meeting these extreme demands will necessitate a modification to existing network infrastructures as well as cloud computing technologies. 

Mobile Edge Computing was introduced by the European Telecommunications Standards Institute~(ETSI) Industry Specification Group~(ISG) as a means of extending intelligence to the edge of the network along with higher processing and storage capabilities~\cite{hu2015mobile}. From 2017, the ETSI industry group renamed it to “Multi-Access Edge Computing”~(MEC), since the benefits of MEC technology reached beyond mobile and into Wi-Fi and fixed access technologies. Nevertheless, the name change conveniently allows ETSI to retain the MEC acronym, which has become widely recognized among stakeholders in the industry.

The underlying principle of MEC is to extend cloud computing capabilities to the edge of cellular networks. This will minimize network congestion and improve resource optimization, user experience and the overall performance of the network. By leveraging on the Radio Access Networks~(RANs), MEC will improve heavily on latency and bandwidth utilization, making it easier for both application developers and content providers to access network services. Several technologies are identified as enabling technologies for MEC realization, these include Software Defined Networking (SDN), Network Function Virtualization (NFV), Information Centric Networking (ICN) and  Network Slicing.

\subsection{Role of MEC for IoT}
Generally, cloud computing enables the outsourcing of storage and processing functionalities of IoT data to a third party in order to ease the hazel involved in self-management and data protection. However, the centralized nature of conventional cloud servers may face several challenges such as the single point of failure, lack of location awareness, reachability, and latencies associated with typical Wide Area Networks (WANs). On the other hand, many IoT applications need to be served with decentralized systems which need mobility management, geo-distribution, location awareness, scalability, and ultra-low latency. 
Mission critical communication IoT use cases need latency as low as 1~ms and reliability as high as 99.99~$\%$. For instance factory automation applications may typically require a reliability of $10^{-9}$ packet loss rate and a latency range of 250~$\mu$s to 10~ms~\cite{schulz2017latency}. Therefore, the conjugation of IoT applications and centralized cloud servers may introduce several limitations and vulnerabilities. In addition, the rapid growth of IoT devices and big data sets may also create cumbersome traffic on telecommunications networks. 

Edge computing was conceived in a bid to fill the gap between the centralized cloud and IoT devices. Apart from MEC, there are other edge computing paradigms such as Mobile Cloud Computing~(MCC), fog computing, and cloudlets. They tend to coexist with MEC in many technical contexts, hence the tendency for a misappropriation of these technologies given that they all have similar origin. However, these technologies are intrinsically different and each of them comes with its unique value proposition to both existing and future mobile networks as summarized in Table~\ref{tab:edgecomparison}.

\begin{figure}[ht]
  \centering
  \includegraphics[width=0.48\textwidth]{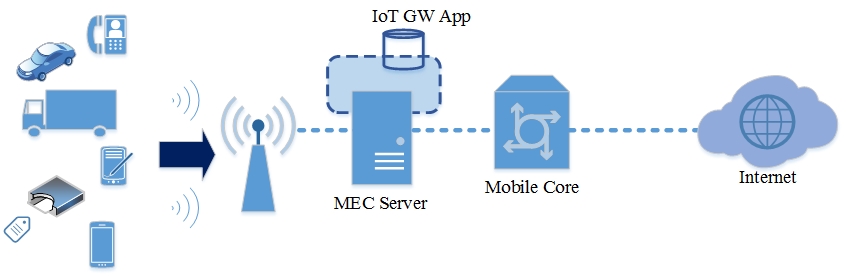}
  \caption{IoT gateway service scenario~\cite{hu2015mobile}.}
  \label{fig_IoTGatewayMEC}
\end{figure}

\begin{table*}[hb]
  \centering
        \caption{High level comparison of edge computing paradigms.}
        \label{tab:edgecomparison}
  \begin{tabular}{|p{2.4cm}|p{3.35cm}|p{3.35cm}|p{3.35cm}|p{3.25cm}|}
  \hline
         & \textbf{MEC} 
         & \textbf{Fog computing} 
         & \textbf{Cloudlet} 
         & \textbf{MCC}\\
  \hline
 		Initial promotion 
		& ETSI (2014) 
		& Cisco (2011) 
		& Carnegie Mellon Uni. (2013)
		&Aepona (2010)\\
 \hline
    	Objective
       & \multicolumn{4}{c|}{Bring cloud computing capabilities closer to User Equipment~(UE)}\\
 \hline
 		Infrastructure owners 
        & Telecom operator 
        & \multicolumn{3}{c|}{Private entities / individuals}\\
 \hline
        Node location
        & Radio network controller or macro base station
        & \multicolumn{3}{c|}{Any strategic location between end user device and cloud}\\
  \hline
 		SW architecture
        & Mobile orchestrator based
        & Fog abstraction layer based
        & Cloudlet agent based
        & Service oriented\\
   \hline
 		Service accessibility
        & \multicolumn{3}{c|}{Direct access from the closest UE} 
        & Via Internet connection\\     
 \hline
 Latency and jitter
        & \multicolumn{3}{c|}{Low}
        & High\\     
        \hline
 Context awareness
        & High
        & Medium
        & Low 
        & High\\
        \hline
        Storage capacity and computation power
        &\multicolumn{3}{c|}{Limited}
        &High\\\hline
 Relevance to IoT
       &\multicolumn{3}{c|}{High}
        & Low\\
        
 \hline
  \end{tabular}
  \end{table*}
ETSI has identified IoT as one of the key use cases of MEC~\cite{hu2015mobile}. MEC has opened many new frontiers for network operators, service and content providers to deploy versatile and uninterrupted services on IoT applications. MEC and IoT facilitate each other with mutual advantages. MEC empowers tiny IoT devices with significant additional computational capabilities through computation offloading. Similarly, IoT expands MEC services to all types of smart objects ranging from sensors and actuators to smart vehicles. As shown in Figure~\ref{fig_IoTGatewayMEC}, MEC servers can perform as gateway nodes which can aggregate and process the small data packets generated by IoT services before they reach the core network. As summarized in~\cite{taleb2017multi}, the three key benefits of the collaboration between IoT and MEC are: 1) lowering the amount of traffic passing through the infrastructure; 2) reducing the latency for applications and services; and 3) scaling network services diversely. Among these, the most significant is the low latency introduced by MEC due the reduced physical and virtual communication distance.

\subsection{Paper motivation}
\label{subsec:paper motivation}

At present, IoT has become a fairly mature technology. As a result, the recent decade has seen a plethora of surveys published in multiple research areas on IoT including enabling concepts~\cite{atzori2010internet}, visions and challenges~\cite{gubbi2013internet}, technologies~\cite{al2015internet}, standardization~\cite{gazis2017survey}, architecture~\cite{weyrich2016reference}, security~\cite{sicari2015security,granjal2015security}, privacy~\cite{porambage2016quest}, trust~\cite{yan2014survey}, Social Internet of Things~(SIoT)~\cite{atzori2012social}, communication~\cite{raza2017low}, context awareness~\cite{perera2014context}, and future directions~\cite{gubbi2013internet,stankovic2014research}. Few other papers are focused on the combined aspects of IoT research and their potential application scenarios~\cite{lin2017survey,al2015internet,miorandi2012internet,atzori2017understanding}. Some of these surveys were published during the time when IoT was more of a visionary paradigm than a real world platform. Many future research possibilities discussed in those papers have already been achieved and commercialized with high market values.
However, there is yet to be a sufficient number of publications on MEC technology, given that is relatively a novel technology which lies at the intersection of mobile cloud computing and wireless communication. In Table~\ref{tab:surveys}, we summarize the recently published surveys on MEC. These articles are focused on MEC taxonomy, future research directions, and more specific MEC attributes such as communication, computation offloading, security, and virtualization. These studies are quite shallow in addressing the MEC integration with IoT, they are mostly focusing on the requirements and usability of MEC in IoT applications. In this short magazine article~\cite{sabella2016mobile}, the authors discuss the examples of MEC deployment, with special reference to IoT use cases. 
\begin{table*}[hb]
  \centering
        \caption{Summary of important surveys on MEC.}
        \label{tab:surveys}
  \begin{tabular}{| p{2.3cm}|p{0.6cm} |p{6.62cm}|p{6.2cm}|}
  \hline
    	\textbf{Aspect} 
         & \textbf{Ref.} 
         & \textbf{Main contribution}
         &\textbf{Relevance to IoT} \\
   \hline
  \multirow{5}{*}{ Research directions}
   &\cite{garcia2015edge}
   &An elaboration of edge-centric vision and its future research challenges.
   &No explicit focus on IoT. \\
   \hhline{~---}
   &\cite{shahzadi2017multi}
   & A comprehensive overview on sate-of-the-art and future research directions for MEC.
&Concisely describes how MEC can improve latency and support big data handling in different IoT deployments. \\
\hhline{~---}
&\cite{ahmed2017mobile}
&A presentation of MEC related definitions, applications, opportunities, and research challenges. 
&Provides no detailed description on IoT. Identifies IoT data handling as a key use case of MEC.\\
\hhline{~---}
&\cite{ai2017edge}
&A concise tutorial of three edge computing technologies, including MEC, cloudlets, and fog computing.
&Describes the exploitation of edge computing technologies for IoT with respect to standardization efforts, principles, architectures, and applications. \\
\hhline{~---}
&\cite{abbas2017mobile}
&A comprehensive survey of relevant research and technological developments in the area of MEC.
&Identifies MEC services for IoT big-data analytics.\\
   \hline
   \multirow{2}{*}{Taxonomy}
   &\cite{ahmed2016survey}
   &A taxonomy of MEC based on different aspects including its characteristics, access technologies, applications, and objectives.
&Classifies MEC applications as computational offloading, collaborative computing, memory replication in IoT and content delivery. 
\\
   \hhline{~---}
   &\cite{beck2014mobile}
   &A classification of applications deployed in MEC systems.
&No explicit focus on IoT. \\
   \hline
 	    Architecture and Computation Offloading
        & \cite{mach2017mobile}
        & A detailed study on decision on computation offloading, allocation of computing resources, and mobility management along with a summary of MEC use cases and standardization efforts.
        &Describes MEC acting as an IoT gateway.\\
		\hline
		\multirow{3}{*}{Virtualization}
        &\cite{taleb2017multi}
        & A survey of 5G network edge cloud architecture and orchestration with a summary of MEC virtualization technologies including Virtual Machines~(VMs), SDN, NFV and network slicing.
        &Explains how MEC platform can encompass a local IoT gateway functionality capable of performing data aggregation and big
data analytics for application domains.\\
        \hhline{~---}
        
        &\cite{baktir2017can}
        &An investigation on how to exploit SDN for enabling edge computing. 
        &Discuses SDN scenarios based on IoT and edge Computing, and the future research.\\
        \hhline{~---}
         &\cite{afolabi2018network}
        &An elaboration of network slicing from an E2E perspective on principles, enabling technologies and solutions. 
        &Describes the role of massive IoT as a key  use case of 5G and  network slicing.\\
        \hline
       \multirow{2}{*}{Communication}
        &\cite{mao2017survey}
        & An comprehensive survey on joint radio-and-computational resource management in MEC systems.
        &Briefly introduces the role of MEC in IoT.\\
        \hhline{~---}
        
        &\cite{wang2017survey}
        &A comprehensive survey of issues on computing,
caching and communication techniques in MEC.
&Describes specific applications and use cases of MEC in IoT including healthcare, wireless sensor systems, smart grid, smart home, and smart city.
\\
		\hline
        MEC-IoT
        &\cite{sabella2016mobile}
        &An overview about the role of MEC in IoT use cases.
        & Provides examples of MEC deployments for IoT cases: Security, safety, and data analytics; Vehicle to infrastructure communication; Computation offloading to edge cloud.
        \\
        \hline
        Security
        &\cite{roman2018mobile}
        &A discussion of the security threats and challenges in the edge paradigms, along with the promising solution for each specific challenge.
        &No explicit discussion on IoT. Briefly discusses how IoT will  benefit from edge computing and related security threats. 
        \\
 \hline
  \end{tabular}
  \end{table*}

To the best of our knowledge there is not a single survey which addresses broader range of areas about MEC and its influence on IoT realization. Since both MEC and IoT are very essential to the realization of 5G, it is vital to express their associativity in terms of application scenarios and key technical attributes. Our goal is to broaden the horizons of potential inter-dependencies of MEC and IoT technologies and their related applications in future 5G and beyond. 

Furthermore, in our previous survey~\cite{taleb2017multi}, we discuss the role of MEC in 5G network edge cloud architecture and orchestration. There we do not explicitly address the integration of MEC for the realization of IoT and related applications. In addition to MEC integration technologies like SDN, NFV, and network slicing discussed in ~\cite{taleb2017multi}, we consider ICN in this work. Therefore, this survey sets to provide a comprehensive  overview of the state-of-the-art technologies which are required for the complementary integration of MEC with IoT. In this survey, our contributions manifold into three main categories:
\begin{enumerate}
\item {Providing a comprehensive survey on the exploitation of MEC technology for the realization of different IoT applications.} 
\item {Presenting a holistic overview of related works and the future research directions in areas of scalability, communication, computation offloading, resource allocation, mobility management, security, privacy, and trust management of MEC-IoT integration.}
\item {Providing a concise summary of the state-of-the-art MEC integrating technologies for IoT and related projects.}
\end{enumerate}


\subsection{Paper organization}
\label{subsec:paper summary}

The rest of the paper is organized as follows: Section~\ref{sec:IoT and MEC applications} summarizes the well-known IoT applications that require a noteworthy assistance of MEC like edge computing technologies. Section~\ref{sec:technical aspects} is particularly focused on technological aspects of MEC enabled IoT systems in terms of scalability, communication, computation offloading, resource management, mobility management, security, privacy, and trust management. Each technical aspect is described with its requirements and related works. Section~\ref{sec:Technologies} and~\ref{sec:projects} respectively summarize the related work on different MEC integration technologies and the  proceeding research projects in the respective areas. Section~\ref{sec:lessons_learned} describes the lessons learned and the future research directions. Finally, Section~\ref{sec:Discussion} concludes the paper. We provide the definitions of frequently used acronyms in Table~\ref{tab:acronyms}. 



\begin{table*}[ht]
  \centering
        \caption{Summary of important acronyms.}
        \label{tab:acronyms}
  \begin{tabular}{|p{2.3cm} p{5.7cm}|p{2.3cm} p{5.7cm}|}
  \hline
          \textbf{Acronym} 
         & \textbf{Definition} 
         & \textbf{Acronym} 
         & \textbf{Definition}\\
  \hline
   \hline
 		
		&  
		& 
		&\\
3GPP & Third Generation Partnership Project
& 5G & Fifth Generation Wireless Network\\
AI & Artifical Intelligence
&AR & Augmented Reality\\
BLE & Bluetooth Low Energy 
&CaPC & Cloud-aware Power Control\\
CPS & Cyber Physical System 
&C-RAN & Cloud Radio Access Network\\
D2D & Device-to-device 
&DDoS & Distributed Denial of Service\\
DoS & Denial of Service 
&E2E & End-to-end\\
EC & Edge Computing
&eMBB & enhance Mobile Broadband\\
EMM & Energy-aware Mobility Management 
&eNodeB & Evolved Node B\\
ETSI & European Telecommunications Standards Institute 
&EU & European Union\\
FiWi & Fiber-enable Wireless
&F-RAN & Fog Radio Access Network\\
GDPR & General Data Protection Regulation
&ICN & Information Centric Networking\\
ICT & Information Communication Technology
&IIoT & Industrial Internet of Things\\
IoT & Internet of Things 
&ISG & Industry Specification Group\\
KDN & Knowledge-Defined Networking
&LPWAN & Low-power Wide Area Network\\
LTE & Long Term Evolution
&M2M & Machine-to-machine\\
MANO & Management and Orchestration
&MCC & Mobile Cloud Computing\\
MEC & Multi-Access Edge Computing
&MIFaaS & Mobile-IoT-Federation-as-a-Service\\
MitM & Man-in-the-Middle
&mmW & millimeter-Wave\\
MR & Mixed Reality
&NB-IoT & Narrow-band IoT\\
NFV & Network Function Virtualization
&PbD & Privacy by Design\\
QoE & Quality of Experience
&QoS & Quality of Service\\
RAN & Radio Access Networks
&RAT & Radio Access Technology\\
RFID & Radio-Frequency Identification
&RNC & Radio Network Controller\\
SCeNB & Small Cell eNodeBs
&SDLB & Software Load Balancer\\
SDN & Software Defined Networking
&SDP & Software Defined Privacy\\
SIoT & Social Internet of Things
&TDMA & Time-division Multiple Access\\
UAV & Unmanned Aerial Vehicles
&UE & User Equipment\\
V2V & Vehicle to Vehicle
&V2X & Vehicle to Everything\\
VANET & Vehicular Ad-hoc Network
&VM & Virtual Machine\\
VNF & Virtual Network Function
&VR & Virtual Reality\\
VRARA & Virtual Reality/Augmented Reality Association
&WAN & Wide Area Networking\\
WAP & Wireless Access Point
&WIoT & Wearable Internet of Things\\
WLAN & Wirless Local Area Networking
&WSN & Wireless Sensor Network\\

 \hline
  \end{tabular}
  \end{table*}

\section{IoT and MEC application scenarios}
\label{sec:IoT and MEC applications}

\begin{figure*}[htbp]
  \centering
  \includegraphics[width=0.96\textwidth]{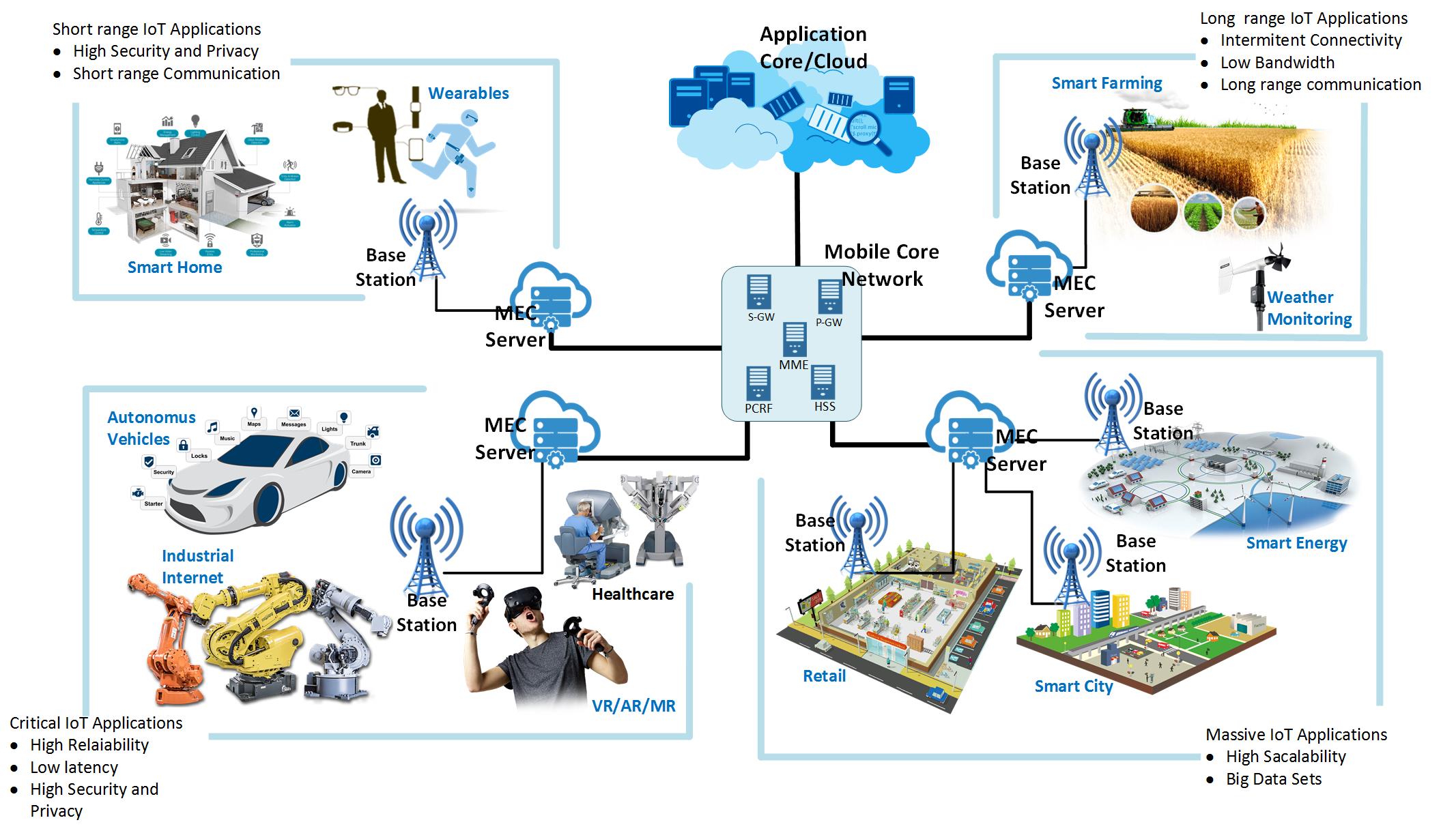}
  \caption{IoT and MEC application scenarios.}
  \label{fig_MECIOT}
\end{figure*}

This section focuses on how IoT can leverage MEC technology in various application scenarios. IoT itself is a classic application of MEC where the key value proposition of MEC is exemplified in a variety of application scenarios (Figure \ref{fig_MECIOT}). These values become evident in the utility factor measured by the end user experience while using such IoT related services. 

Table~\ref{tab:mecIoTcharacteristics} and~\ref{tab:mecIoTApp_benefits} respectively show the characteristics of different IoT applications and how each application benefits from MEC-IoT integration. In addition, Table~\ref{tab:mecIoTApp} summarizes the reviewed state-of-the-art applications in MEC-IoT domains.

\begin{table*}[htbp]
  \centering
        \caption{Characteristics of Different IoT  application.}
        \label{tab:mecIoTcharacteristics}
  \begin{tabular}{|p{2.8cm}|p{2cm}|p{3.4cm}|p{2cm}|p{2cm}|p{2.8cm}|}
  \hline
      {\textbf{\centering{IoT Application}}}	
         &{Data type}
         &{Data Capacity}
         &{Backhaul Connectivity}
         &{Expected latency}
         &{Number of IoT Devices}\\
  &        &        &  & & \\
    \hline
Smart home    &    Stream / \newline Historical data    &  $\geq$ 10~MB
of data per household per day    &     Realtime & 1~ms -1000~s &$\geq$10-100 per house \\
\hline
Smart city     &    Stream /\newline Massive data    & $\geq$10-100 million GB of data per city per day     &   Realtime  &$\leq$1ms&$\geq$1000-1million per city \\
\hline
Remote surgery  \cite{perez2016impact}  	  &   Stream data      & $\geq$1.5 million per year     &   Realtime & $\leq$200~ms &$\geq$10-100 per surgery \\
\hline
Remote consultancy  & Stream data    & $\geq$ 500 million visits per year  &   Realtime& 1~ms-100~s&1-10 per appointment \\
\hline
Autonomous vehicles	  &   Stream / \newline Massive data     &  $\geq$ 100 GB per vehicle per day  &   Realtime &$\leq$1~ms&50-200 per vehicle  \\
\hline
AR\cite{ar_vr}    &    Stream /\newline Massive data    & $\geq$1~GBps     & Realtime & $\leq$1~ms &$\geq$0.2 million globally  \\
\hline
VR \cite{ar_vr}      & Stream /\newline Massive data&$\geq$1~GBps&Realtime &   $\leq$1ms &$\geq$0.2 million globally  \\
\hline
Gaming\cite{ar_vr}	  &    Stream /\newline Massive data     &$\geq$10~Mbps    &     Realtime  & $\leq$10~ms& $\geq$1 billion globally\\
\hline
Retail~\cite{IntelWhitepaper_retail}  &   Stream / \newline Historical data &  100~Mbps - 1~Gbps &   Realtime/ Intermittent & $\leq$1~ms & $\geq$100-1000 per shop\\
\hline
WIoT  		  &  Stream data       &   $<$ 1 GB per device     &   Intermittent &  Several Hours&$\geq$1-10 per person\\
\hline
Farming     &    Historical data    &   $\geq$ 1 GB per farm    &  Intermittent & Several hours & 100-100,000 per farm \\
\hline
Smart energy	  &   Stream / \newline Massive data      &  $\geq$  100,000 GB per day  &  Realtime/ Intermittent & 1ms - 10 mins & $\geq$ 1 billion per grid \\
\hline
Industrial Internet \cite{iiot}	  &    Stream / \newline Massive data & $\geq$ 100,000 GB per day   & Realtime  & $\leq$1~ms & $\geq$ 1 million per factory \\

  \hline
  \end{tabular}
  \end{table*}

\newcommand{\spheading}[2][11.5em]{
  \rotatebox{90}{\parbox{#1}{\raggedright #2}}}
\begin{table*}[htbp]
  \centering
        \caption{MEC and IoT benefits for each application.}
        \label{tab:mecIoTApp_benefits}
  \begin{tabular}{|p{3cm}|p{5cm}|c|c|c|c|c|c|c|c|c|c|c|c|c|c|}
  \hline
\textbf{Required characteristics of MEC and IoT	}
      & \textbf{Description}
         & \spheading{Smart home}
         &\spheading{Smart city}
         &\spheading{Remote surgery}
         &\spheading{Remote health consultancy}
         &\spheading{Autonomous vehicles }
         &\spheading{Augmented Reality (AR)}
         &\spheading{Virtual Reality (VR)}
         &\spheading{Gaming}
         &\spheading{Retail}
         &\spheading{Wearable IoT}
         &\spheading{Farming}
         &\spheading{Smart energy}
         &\spheading{Industrial Internet}\\

    \hline
Low Latency   &  Optimize to process a very high volume of data messages with minimal delay  &     &\checkmark         &\checkmark  & \checkmark        &\checkmark &\checkmark         &\checkmark  &\checkmark         &         &\checkmark         &\checkmark &\checkmark         &\\
\hline
Increased Bandwidth & Ability move a large set amount of data rapidly & \checkmark         &\checkmark         &\checkmark  &         &\checkmark &\checkmark         &\checkmark  &\checkmark         &\checkmark         &\checkmark         &\checkmark &\checkmark         &\checkmark\\
\hline
Content Awareness   & Adaptation of network characteristics according the local services requirements 	&\checkmark &\checkmark	&\checkmark & & &\checkmark & \checkmark & \checkmark & \checkmark & \checkmark & \checkmark & \checkmark &\\ \hline
Low power devices & Support for low power devices which has limited transmission powers 	& &	& & & \checkmark & \checkmark & \checkmark & & & \checkmark &\checkmark &\checkmark & \checkmark\\
\hline
Fixed wireless support &  Operation of wireless systems used to connect two fixed locations with a  wireless link	& \checkmark & \checkmark 	&\checkmark & & &\checkmark & & \checkmark & \checkmark &\checkmark &\checkmark &\checkmark & \checkmark\\
\hline
Fast inter-RAT handoff   & Speed up the handover takes place between different RATs   & \checkmark         &\checkmark         &\checkmark        &         &         &\checkmark         &\checkmark         &\checkmark         &\checkmark         &\checkmark         &       &  &\\
\hline
Caching   &  Keeping frequently accessed information in a location close to the requester   & \checkmark         &\checkmark         & &         &         &\checkmark         &\checkmark         &\checkmark         &\checkmark         &\checkmark & & &\\
\hline
Edge Analytics & An automated analytical computation is performed on data at a sensor, network switch or other device instead of waiting for the data to be sent back to a centralized data store.	& \checkmark         &\checkmark         &\checkmark         &         &  &\checkmark
         &\checkmark         &\checkmark         &\checkmark         &\checkmark         &   & &\checkmark\\
\hline
Application virtualization between edge and cloud & On demand application and service migration from centralized cloud to the edge cloud   &       &\checkmark         &\checkmark          &         &\checkmark          &\checkmark 
         &\checkmark          &\checkmark          &\checkmark          &\checkmark          &\checkmark          &\checkmark          &\checkmark \\
\hline
Private or local network & Limit the communication and data exchanges to a certain network segment  &\checkmark &\checkmark	&\checkmark & &\checkmark &\checkmark &\checkmark &\checkmark &\checkmark &\checkmark &\checkmark &\checkmark &\checkmark\\
\hline
Security & Provide localized security	& 			 & \checkmark 	& &\checkmark & &\checkmark & & \checkmark & & & \checkmark & \checkmark & \checkmark \\
\hline
Privacy   & Provide localized Privacy		&	 & 	\checkmark	& \checkmark & \checkmark & \checkmark & \checkmark & & \checkmark& & \checkmark & & & \\
\hline
Fast Mobility		& 	Enable the ability to  move or be moved fast  within the network or network coverable area 	&	 & 	\checkmark	&\checkmark &  &\checkmark &\checkmark & &\checkmark & &\checkmark &\checkmark &\checkmark & \\
\hline
  \end{tabular}
  \end{table*}

\subsection{Smart home and Smart city}
One of the pioneering applications of the IoT technology has been in the areas of home automation and consumer electronics~\cite{stojkoska2017review}. Several smart home applications that are built on the basis of IoT concept are already available in most consumer markets. These range from the simple thermostat sensors to other more sophisticated automation systems like smart metering, smart heating and lighting, cleaning services, and home entertainment systems. That notwithstanding, the amount of data that would be generated on a typical IoT network like the smart home is expected to be huge. Hence transferring such data to the centralized cloud servers will be impractical with most pre-MEC techniques. As a solution, MEC leverages specialized and reliable local services for processing and storage capabilities for the large IoT traffic created within a building. The conventional gateways which allow IoT applications to run on the centralized cloud can be empowered with MEC-server functionalities~\cite{vallati2016mobile,morabito2016enabling}. This extends gateway functionalities to the edge of the network with reduced communication latency. Since such appliances are statically deployed in smart home or smart building environments, the cooperation with MEC servers will offer some other features such as easy instantiation, relocation, privacy preservation, and upgrading when necessary~\cite{sabella2016mobile,sun2016edgeiot}.

Correspondingly, IoT technology has advanced from home to community, and even city scale applications. We see numerous future promises for public safety, health care, utility, tourism, and the transport sectors. Enormous IoT data traffic produced in smart cities can be ideally processed at the edge of the network providing low latency and location awareness~\cite{nguyen2016virtual,taleb2017mobile}. 
In particular, a video cameras (i.e. deployed for surveillance) connected with a Long Term Evolution~(LTE) network can convey video streams to the MEC server for real-time processing and anomaly detection~\cite{sabella2016mobile}. Collaborative edge paradigms that connect multiple MEC servers (i.e., dedicated for different services) will advocate the applications which need to process geographically distributed data. For instance, a  connected health care application requires to collaborate with entities from multiple domains such as hospital, pharmacy, insurance, logistics, and government~\cite{shi2016edge}.



\subsection{Healthcare}

Mobile health and telemedicine are identified as important use cases of 5G. Wearable low power IoT medical sensors for monitoring health related data and tracking records are now popular in public healthcare facilities~\cite{hossain2016cloud}. Although IoT technologies are widely adopted in the health sector~\cite{islam2015internet}, their performance goals will not be achievable without edge computing solutions like MEC~\cite{IntelWhitepaper_retail,shi2016promise,tran2017collaborative}. For instance, humanoid robots sitting next to an elderly person may need tactile feedback in 1ms latency for his or her care taking services. Mission critical use cases like remote surgeries require ultra-low latency, uninterrupted communication links, and collaborations among surgeons present in different locations. Remote patient monitoring is another use case which enables consultants in major cities to interact with patients residing far away from the medical facility. The frequent updates of health records for an elderly person or someone with a chronic disease needs to proceed ubiquitously and securely. With such potential use cases and scenarios, the role of MEC in health and social assistance industries becomes more evident~\cite{IntelWhitepaper_retail}.

Some research works have already been published about the cooperation between edge computing  and IoT in the healthcare sector. In~\cite{singh2017semantic}, authors describe a military healthcare service platform based on hierarchical IoT architecture and a semantic edge network model. The hierarchical IoT architecture can collect the vital health parameters of the soldiers, their weapon status, as well as their geographical locations. The control center of the battlefield performs the role of edge component which can process and store large amount of health data sent over an SDN-based network. The preliminary network architecture proposed in~\cite{nunna2015enabling} provides real-time context-aware collaboration for remote robotic tele-surgeries. Big data analytics performed by edge computing are also important in e-Healthcare applications~\cite{sharma2017live}. In ~\cite{rahmani2018exploiting}, Rahmani et. al. introduces the smart gateway concept for an IoT-based remote health monitoring system. Here they 
e
xploit edge computing nodes to update the centralized cloud based on the medical data generated by the IoT sensors. Their geo-distributed network of smart e-Health gateways provides local data processing for real-time notification for medical practitioners, secure and privacy preserved data gathering, patients' mobility, network interoperability, and energy efficient communication.

\subsection{Autonomous Vehicles/IoT Automotive}
5G is a key enabler of V2X (Vehicle to Everything) concept which covers Vehicle to Vehicle (V2V), vehicle to infrastructure, vehicle to device, vehicle to pedestrian, vehicle to home and vehicle to grid~\cite{SimallianceWhitepaper}. In the context of IoT Automotive, V2X requires critical communication infrastructure where reliability and ultra low latency are crucial factors~\cite{zakaria2017internet}. Use cases in these categories include autonomous and semi-autonomous driving, vehicle maintenance, and in vehicle infotainment. In order to operate an efficient and reliable vehicular network, several features have to be improved, these include real-time traffic monitoring \cite{balid2017intelligent, amini2017big},  continuous sensing in vehicles~\cite{yu2016senspeed, nawaz2016smart}, support for Infotainment applications~\cite{han2017software} and improved security~\cite{he2015efficient}. However, these features cannot be served by current mobile networks \cite{metis2016deliv}. In this vein, upcoming 5G mobile systems are expected to offer a higher level of flexibility, leveraging the emerging technologies related to network softwarization \cite{osseiran20165g}. In this context, V2X combined with MEC provides a viable and cost-effective solution that can accelerate development of V2X and IoT automotive systems~\cite{datta2017vehicles}.  
 
It is important to improve the performance of RAN technologies to enable IoT automatization. MEC will play a vital role here also. For instance, MEC technologies may fulfill the latency, reliability, and throughput requirements in V2X channel modeling of mmWave communication~\cite{frascolla5g}. Moreover, the placement of the MEC server  within the RAN provides flexible network services for the vehicle and to efficiently control the radio network resources\cite{li2017novel}. It is also possible to design a time-predicted handover mechanism for vehicles  by leveraging road side information at MEC server in order to meet the demand for high mobility and reliability in vehicular networks\cite{li2017novel}.

In addition, ICN-MEC integration can also tackle existing technical challenges such as  massive mobility of vehicles, scalability, deployment strategies, service orchestration,  massive data handling, fast big data  processing, as well as ensuring security and privacy\cite{grewe2017information}.
 
Unmanned aerial vehicles~(UAVs) or drones are another type of autonomous vehicles which are capable of sensing its environment and navigating without human inputs. UAV use cases include but not limited to, public safety, smart agriculture, surveillance, and environmental monitoring~~\cite{motlagh2017uav}. In order to maximize the flight time, the UAV battery life should be essentially conserved by minimizing the overhead onboard. When the required processing power exceeds the available resources on UAV, the application data can be offloaded to MEC. Accompanying the advanced RATs, MEC will facilitate the offloading process from UAV due to its expected wide deployment in the network~\cite{motlagh2017uav}.
 

\subsection{Gaming, AR and VR}

Mixed reality~(MR) combines virtual reality~(VR) and augmented reality~(AR) technologies thereby enabling humans to interact more naturally with the virtual worlds based on data aggregated by IoT devices~\cite{satyanarayanan2017emergence}. With IoT, AR technologies are able to benefit directly from the high end interconnection of objects that characterizes the IoT environment through which users can extend their interactions from the real world to the virtual world~\cite{hu2015mobile,baresi2017empowering}. Convergence of VR and IoT can  occur in many ways such as telepresence, tourism industry, smart transportation networks, and robotic assisted surgeries. Exclusive AR and VR experiences with the delivery of 360$^{\circ}$ navigable videos will be offered by enhanced mobile broadband connections with low latency and high reliability for mission-critical services. With present-day network standards, this might be impossible to achieve, however with the predicted characteristics of 5G such as 20~Gbps peak data rate and 1~ms round-trip over-the-air latency, this becomes more easily achievable. As identified by ETSI, MEC will be an ideal solution for low-latency offload services in AR and VR applications that combine computer generated data with physical reality~\cite{etsimec}. While operating VR devices over wireless links and deploying the VR control center at MEC server, the tracking accuracy can be increased with round trip latency of 1~ms and high reliability~\cite{chen2017virtual}. Migrating computationally intensive tasks to edge servers will increase the computational capacity of VR devices and save their battery-life. Furthermore, MEC will allow VR devices to access cloud resources in an on-demand fashion~\cite{bastug2017toward}.

MEC platforms provide high capacity and low latency wireless coverage for large venues like stadiums or smart cities with a massive density of users to enjoy the AR and VR experience. For instance, inside a smart building with a network of cameras, obtaining raw video frames and preparing the processed frames for display can be performed locally with the help of edge computing. Furthermore, tracking the local position of the user or object, building a model of the environment, and identifying known objects in the environment can be offloaded to the edge cloud. Similarly, in order to get absolute experience of VR glasses, the response time should be extremely low. When the user moves his head, he may experience delay if the glasses need to access remote data centers. Therefore, the expected interaction time between machines and humans needs to be less than 1ms. When the latency of a VR application is more than 1ms, the user will experience cyber sickness which will be interrupting the real VR experience. MEC servers in the nearest proximity will be able to serve such applications with ultra low latency. Future games will be played beyond the entertainment purposes on top of VR and AR applications which would require the minimum possible latency. Pokémon Go and Ingress are two examples of successful games that combine AR and sensor information such as user location. 

\subsection{Retail}
The second largest MEC use case is expected to be in the retail businesses~\cite{IntelWhitepaper_retail}. Currently, IoT has dominated retail market applications in many ways including digital signage, supply chain management, intelligent payment solutions, smart vending machines, shelves, doors, resource management, streaming, and safety. The high class retail stores which use facial recognition systems need high definition cameras that generate huge volumes of data requiring powerful servers within the premises. Therefore, the on-site MEC servers will assist to process these kind of large data sets produced by IoT devices in a retail market. Big data analytics in shopping centers can further exploit the collaborative processing between edge and cloud computing~\cite{sharma2017live}. Installation of MEC in a retail market also provides high speed mobile coverage throughout the store. Wi‐Fi access points that are maintained per store can be connected to the MEC server to provide Wi‐Fi connectivity for store customers as needed. The enabling of MEC will also omit load balancing, Wi-Fi controllers, or policy engines required in the wide area networks in the store. Although not many academic published research works are explicitly focusing on MEC and IoT~\cite{cheng2017fogflow}, they have become enormously reputed and commercialized technologies in the industry and the business sectors.

\subsection{Wearable IoT (WIoT)}
During the previous years, wearable technology has evolved tremendously from walkman to step trackers, smart watches to smart glasses. The development of low power wireless technologies such as  BLE (Bluetooth Low Energy) fuels the development of wearable devices. Present-day wearables span  from low-end devices such as health and fitness trackers to high-end devices such as VR/AR  helmets and smart watches. It is expected that wearables will become the world’s best-selling consumer electronics product after smartphones with a global availability of more than 929 million devices by 2021~\cite{CISCOVNI2017}. With the new application domains and enabling services, wearable devices will demand more sophisticated communication infrastructures.  For instance,  VR/AR wearables are demanding  gigabit/s throughput network connectivity to run their applications. On the other hand,  dense deployment of wearable devices in smart cities will increase the network traffic on communication networks. Thus, the next generation communication networks should be able to provide the gigabit experience for the anticipated ultra dense wearable devices\cite{sun2017challenges}.

Although cloud computing has enabled wide range of new networking services, it  cannot alone fulfill the upcoming requirements for the future wearable ecosystem. Mainly, the centralized cloud data centers fails due to long End-to-End~(E2E) latency. Delay-sensitive wearable applications such as VR perceptual stability requires ultra low delay. In this context, MEC has the potential to solve the limitations in current cloud based systems, by combining cloud and MEC infrastructures. This will enable providers deploy  storage, computing, and caching capabilities in close proximity with such wearable devices\cite{sun2017challenges}.


\subsection{IoT in Mechanized Agriculture}

In order to meet the demands for future food production, the agricultural sector will require some major evolution where IoT will be integrated in various production, management, and analytical processes~\cite{perera2015emerging,ferrandez2016developing}. The present-day agricultural sector has been slow to adopting the emerging Machine-to-Machine~(M2M) and IoT technologies when compared with other sectors like smart cities and the medical fields~\cite{SmartFarming}.  

Precision farming and smart agriculture can be achieved using autonomous vehicles (tractors), remote monitoring, and real-time analytics. It is reported that farmers are increasingly turning to agricultural drones and satellites to survey their lands and generate crop data. IoT sensors may provide information about crop yields, rainfall, pest infestation, and soil nutrition which are invaluable to production and can improve farming techniques over time. Although low latency is not a critical requirement in smart farming environment, management of large data sets will be a key requirement to consider. MEC servers located on-site can assist high tech farming by collecting and analyzing big data on agriculture in order to maximize efficiency.
Likewise, without moving everyday farming applications to a remote cloud, MEC platforms can benefit in terms of data access, synchronization, storage and other overhead costs the farmer might normally incur. 

The use of IoT-based automated data collection and monitoring systems in poultry houses can be used to increase work efficiency and service quality, and get a deeper understanding of chicken nurturing\cite{microsoftpoultry}. Sensing technologies can be used in carbon dioxide and luminosity sensing, these are important parameters in large scale poultry houses. Gas sensors can be used to get all necessary information to prevent chicken infertility due to problems such as low carbon dioxide levels. Luminosity senors can help to maintain the proper luminosity level for optimum productivity. Similar to smart farms, low latency is not a critical requirement in smart poultry houses\cite{microsoftpoultry}. However, it is critical to manage large data sets where on-site MEC servers can be used. In addition, sharing the data between poultry houses and storing legacy data in centralized servers are important in identifying abnormal incidents in the farm\cite{mahale2016smart}. With the use of MEC, poultry houses can work with intermittent connectivity to the centralized clouds. In that case, MEC servers can temporarily hold the data until farms are connected with the centralized clouds. 


\newcommand{\spheadingg}[2][7.5em]{
  \rotatebox{90}{\parbox{#1}{\raggedright #2}}}
\begin{table*}[hb]
  \centering
        \caption{The reviewed state-of-the-art MEC integration in different IoT applications.}
        \label{tab:mecIoTApp}
  \begin{tabular}{|c|p{9cm}|c|c|c|c|c|c|c|c|c|c|c|c|}
  \hline
      \textbf{Ref.}	
         & \textbf{Description}
         &\spheadingg{Smart Home}
         &\spheadingg{Smart City}
         &\spheadingg{Healthcare}
         &\spheadingg{IoT Automotive}
         &\spheadingg{Gaming, AR, VR}
         &\spheadingg{Retail}
       
         &\spheadingg{Wearable IoT}
         &\spheadingg{Smart Agriculture}
         &\spheadingg{Smart Energy}
         &\spheadingg{Industrial Internet}\\

  \hline

	\cite{vallati2016mobile}	& Preliminary design of deploying MEC server functionalities in a smart home to realize IoT gateway with direct M2M interaction in LTE networks & \checkmark & & & & &  & & & & \\
   \hline
  	\cite{morabito2016enabling}	& Introduce Gateway-as-a-Service for heterogeneous IoT devices on top of the virtualization technologies in edge computing. & \checkmark & & & & &  & & \checkmark & & \\ 

 \hline
 \cite{taleb2017mobile}	& Propose an autonomic creation of MEC services to enhance QoS of video streaming in smart cities. 		&	& \checkmark & & & &  & & & & \\
 \hline
  \cite{singh2017semantic} & Propose a semantic edge-based IoT architecture for military health services in battlefield.	 & & & \checkmark & & &  & & & & \\
    \hline
 \cite{nunna2015enabling} & Provide a conceptual MEC based architecture for mission-critical context aware collaboration in remote surgeries.	 & & & \checkmark & & &  & & & & \\
  \hline
 \cite{rahmani2018exploiting} & Describe and implement a smart e-Health gateway at the edge of the network suitable for ubiquitous healthcare systems. 	 & & & \checkmark & & &  & & & & \\
  \hline
 \cite{datta2017vehicles} & Analysis on research and engineering challenges co-existence of cloud, edge computing and data caching strategies at the edge for vehicular networks.  & & & & \checkmark  & &  & & & & \\
 \hline
 \cite{boban2016design} & Discuss the design aspects for the radio access in 5G V2X.  & & & & \checkmark  & &  & & & & \\
 \hline
  \cite{frascolla5g} & Discuss the benefits of merging MEC and
mmWave technologies for 5G applications.  & & & & \checkmark  & \checkmark &  & & & & \checkmark\\
 \hline
 \cite{li2017novel} & Propose a novel MEC-based architecture for future cellular vehicular networks.  & & & & \checkmark  &  &  & & & & \\
 \hline
  \cite{grewe2017information} & Discuss the benefits of combining ICN and MEC in the context of connected vehicle environments.  & & & & \checkmark  &  &  & & & & \\
 \hline
\cite{sharma2017live}   & Propose a framework for big data analytics between edge and cloud computing platforms. & 	& &\checkmark & &\checkmark &  & & & & \\
  \hline
 \cite{cheng2017fogflow} & 	Design and implement a fog computing based framework that support sharing and reusing contextual data across services in smart city and retail stores.	 & 				& \checkmark & & & & \checkmark  & & & & \\
  \hline
  
  
  \cite{braun2017study}  & Present a usecase of  MEC for Tactile Internet based 5G gaming application. & & & & &\checkmark & & & & & \\
  \hline
      \cite{pandi2017demonstration}  & A demonstration of MEC for Tactile Internet based 5G gaming application. & & & & &\checkmark & & & & & \\
  \hline
  \cite{sun2017challenges}  & Discuss the role of MEC in 5G WIoT communication and its challenges. & & & \checkmark  & & &  & \checkmark& & & \\
  \hline
  \cite{motlagh2017uav}& Propose an UAV-based IoT platform for a crowd surveillance use case. & & &  &  \checkmark & &  & &\checkmark & & \\
  \hline 
  
\cite{ferrandez2016developing}   & 	Develop and test a ubiquitous sensor network platform for crop lands automation maintenance in precision agriculture.	 & 				& & & & & & &\checkmark & & \\ \hline
  
 \cite{baresi2017empowering} & 	Present a serverless edge computing architecture that enables the offloading of mobile computation with low latency and high throughput, using a mobile AR application. 	 & 				& & & &\checkmark & & & & & \\
  \hline
\cite{satyanarayanan2017edge}  & 		Discuss the benefits of MEC and edge computing~(EC) to enhance the security of smart grids. 	 & 				& & & & &  & & & \checkmark & \\ \hline
\cite{kanzaki2017video}  & 	Present a method to optimize the EC based video streaming schemes for Industrial IoT. 	 & 				& & & & & & & &  & \checkmark\\
  \hline
  
 \cite{harper2016microdatabases}  & Present the use of edge computing to provide  elastic resources and services to enable microdatabases architecture for IIoT.	 & 				& & & & & & & & & \checkmark \\
  \hline
  
   \cite{peralta2017fog}  & Propose a fog-based communication architecture for Industry 4.0 applications.	 & 				& & & & &  & & & & \checkmark \\
   \hline
      \cite{bastug2017toward}  & Describe research directions and enablers of wireless interconnected VR systems. & & & & & \checkmark &  & & & &\\
  \hline
      \cite{chakareski2017vr}  & Design an optimization framework for VR/AR communication via small-cell cooperation. & & & & & \checkmark &  & & & &\\
  \hline
  \end{tabular}
  \end{table*}

\subsection{Smart Energy}

The smart grid system is an Information Communication Technology~(ICT)-enabled energy generation, transmission and distribution network. It has capabilities to continuously sense, analyze, and monitor both energy flow and energy transportation infrastructure. Such  features are enabled  by adding digital controls and enabling network monitoring and telecommunication capabilities. As a result, a smart grid does not only provide two-way flows of electrical power, but also  enables real-time, automated, bidirectional flow of information. Adding such smartness to the aging energy infrastructure will foster a more efficient energy system. 

IoT is considered as the foundation for realizing intelligence capabilities in  smart grid systems. IoT integrates the Internet-connectivity into all kinds of grid components such as transformers, breakers, switches, meters, relays, intelligent electronic devices, capacitor banks, voltage regulators, cameras and many more. These IoT devices are then used to capture the data required to enable automations. IoT-enabled smart grids provide several benefits such as reduced capital expenditure, optimized renewable capacity, lowered maintenance costs and enhanced customer engagement. On one hand, the transformation of an electrical grid into a smart system requires  nearly every device and piece of equipment to have built-in, secure, interconnected intelligence. On the other hand, an efficient system is required to manage the generated data, i.e. transferring, storing, and analyzing such huge amounts of data which are collected from these smart devices.  Therefore, cloud computing is a viable solution to these IoT-based smart grids\cite{carvallo2015advanced}.

Generally, smart grids are spanning over large geographical areas. They often confront bandwidth bottlenecks and communication delays due to poor network connectivity and vast number of devices generating data. Thus, the traditional centralized cloud architecture is not suitable for the domain of the smart grid since it relies heavily on centralized processing\cite{moghaddam2017performance}. Many delay sensitive smart grid applications, such as fault detection, isolation and service restoration or Volt/VAR optimization cannot tolerate round trip delay to access centralized cloud systems. MEC is identified as the viable cloud computing option to address these limitations. MEC allows  the computation to be performed closer to the data source. Moreover, the potential attack points for the grid is increasing with the growth of ubiquitous sensor deployment. Every smart IoT device can be vulnerable to potential attacks. MEC provides the opportunity to enforce security mechanism closer to the end devices. As such, even if an attacker gains access to an endpoint device, the attack  gets no further information beyond the local network segment since MEC has  capabilities to notice the intrusion and cease the  accessibility\cite{satyanarayanan2017edge}.


\subsection{Industrial Internet}
The Industrial Internet of Things (IIoT), also known as Industry 4.0\cite{lasi2014industry} is an application of IoT in the domain of manufacturing. IIoT incorporates numerous advanced communication and automation technologies such as M2M communication, machine learning and big data analytics to improve intelligence and the connectivity\cite{da2014internet}. For instance, IIoT networks can connect all of the employees data and processes from the factory floor and forward them to the executive offices. Thus, decision makers or employees can create a full and accurate view of their manufacturing process by using IIoT network, hence improving their ability to make more informed decisions. IIoT also helps the exploitation as well as implementation of new intelligent technologies to accelerate the innovation and transformation of the factory workforce\cite{lasi2014industry}.
 
Primarily, IIoT is seen as a way to improve operational efficiency. However, IIoT provides a wide range of other benefits such as improving connectivity, efficiency, scalability, time savings, as well as cost savings for manufacturing processes with the maximum use of smart machines\cite{lasi2014industry, perera2014survey}. In general, these smart machines operate with higher accuracy, greater efficiency and constant working capabilities than humans\cite{kehoe2015survey}. Thus, IIoT has great potential for improving quality control, sustainability and overall supply chain efficiency.

MEC will play a vital role in enabling future IIoT applications \cite{li2017industrial} by addressing the shortcomings of M2M communication (e.g. latency, resilience, cost, peer-to-peer, connectivity, security) in IIoT domain\cite{albano2017industrial, nelson2017smart}. Current market trends already show that edge computing will represent many implementation scenarios for IIoT. For instance, real-time edge analytics and enhanced edge security are two key drivers in the creation of new IIoT deployments. Thus, the addition of MEC in IIoT networks will fuel the evolution of IIoT as well as create new business applications\cite{steiner2016fog}.

One way to optimize the use of conventional edge computing in video streaming schemes for IIoT is presented in \cite{kanzaki2017video}. By using machine learning algorithms, edge computing can process the sensor data  before transmitting to the cloud. This mitigates against the degradation of service quality of the video streaming. Aggregation of all the sensor data to a single data center increases latency and raises performance concerns in IIoT domain. In order to solve this issue, a  microdatabase architecture is proposed for the Industrial Internet\cite{harper2016microdatabases}. It holds the data close to the industrial processes, but also makes it available near the applications that can benefit from the data. Edge computing also provides elastic resources and services to enable micro-database architecture\cite{harper2016microdatabases}. A fog-based communication architecture for Industry 4.0 applications is proposed in \cite{peralta2017fog}. This approach will substantially minimize the energy consumption of the IoT nodes. Edge computational capabilities are further used to predict future data measurements and reduce the throughput from IoT devices to the control unit.

%

\section{Technical Aspects of MEC Enabled IoT}
\label{sec:technical aspects}
To realize the MEC exploitation for IoT applications, the key value propositions are mostly seen from the technical parameters such as scalability, communication, computation offloading and resource allocation, mobility management, security, privacy, and trust management. This section describes the state-of-the-art of each of these technical parameters, hence giving a clear background against which the benefits of MEC can be envisioned.  

\subsection{Scalability}
\label{subsec:Scalability}

\subsubsection{Requirements}
When it comes to actual deployment of MEC platform for IoT systems, scalability is a key factor to consider. The compatibility of MEC servers to multiple network environments is one of the factors that will drive its large scale adoption in future networks \cite{liang2017mobile}. The IoT environment will consist of hundreds of billions of sensors, actuators, Radio-Frequency Identification (RFID)-tagged objects, software, vehicles, and embedded systems all interconnected in a huge network of cyber-physical systems. At a utility scale consideration, these devices will be working in close collaboration to deliver the expected services in technologies like the smart grids, virtual power plants, smart homes, intelligent transportation and smart cities. That being said, the role of scalability to the realization of such a hyper-connected IoT environment becomes more obvious. The IoT environment will require a dynamic range of capabilities in the network space if such large numbers of devices are to be supported effectively. 

\subsubsection{Related work}

Currently, MEC servers have been confirmed to be compatible with LTE macro base station (eNodeB) sites, 3G Radio Network Controller (RNC) site, multi-Radio Access Technology (RAT) cell aggregation site, and at the edge of the core network \cite{hu2015mobile}. Such multi-RAT cell aggregation schemes can be implemented indoor or outdoor settings depending on the requirements. This invariably enables MEC to be applied to many different possible scenarios. The larger the deployment scenarios for MEC the more the range of capabilities it can handle, this also translates to higher scalability for MEC-enable technologies like IoT.    

\begin{table*}[htbp]
  \centering
        \caption{Comparison of the reviewed state-of-the-art scalability feature in MEC enabled IoT.}
        \label{tab:scalability}
  \begin{tabular}{|p{0.8cm}|p{10cm}|p{2cm}|p{1.4cm}|p{1.2cm}|}
    \hline
   	\textbf{Ref.} 
    &  \textbf{Description}
    & \textbf{IoT application\newline/domain\newline/feature}	
    & \textbf{Addressing} 
    & \textbf{Search}
      \\
  \hline
  \cite{zhang2011searching} &  Discusses the challenges in searching imposed by the burgeoning field of IoT. & General IoT & & \checkmark \\ \hline
  \cite{perera2014survey} &  Examines a variety of popular and innovative IoT solutions in terms of context-aware technology perspectives, to serve as a conceptual framework for context-aware product development and research in the IoT paradigm & General IoT solutions. & \checkmark & \\ \hline
  \cite{bellavista2016towards} & Proposes an innovative distributed architecture combining machine-to-machine industry-mature protocols (i.e., MQTT and CoAP) in an original way to enhance the scalability of gateways for the efficient IoT-cloud integration & IoT cloud integration. & \checkmark & \checkmark \\ \hline
 \cite{ren2017serving} & Studies an implementation
of edge computing, which exploits transparent computing to build scalable IoT platforms using transparent computing.&Wearable IoT&\checkmark&\\ \hline
\cite{morabito2018legiot}
&Introduces a lightweight edge gateway for the IoT architecture using container-based virtualization techniques. &General IoT&&\checkmark \\ \hline
  \end{tabular}
  \end{table*}

Designing an edge cloud network implies that an optimal location for citing the cloud facility is first determined. In \cite{ceselli2017mobile}, authors present a design optimization scheme for the MEC architecture based on link-path formulation supported by heuristics in order to optimize the computation time for the scheme. In this approach, consideration is given to both users and VMs mobility. Hence, an optimal point to install the MEC server is determined through a tread-off between installation cost and the quality of service to be delivered. Table~\ref{tab:scalability} compares the reviewed state-of-the-art scalability feature in MEC enabled IoT.

%
%
\subsection{Communication}
\label{sec:Communication}
\subsubsection{Requirements}


There are three main categories for the communication concerns about MEC~\cite{LiangMECbook}: Wireless access while offloading to the mobile edge host; Backhaul access while offloading to a remote cloud server; Communication among IoT devices, mobile edge host, and remote cloud servers when they collaboratively execute multiple jobs. The first and the second categories are the most renowned on behalf of the MEC servers which are the small scale data centers deployed by the network operators and can be co-located with the Wireless Access Points~(WAPs). In the IoT supportive MEC systems, the consumer devices may communicate with the MEC servers either directly or with the support of neighboring devices using Device-to-Device~(D2D) communication. For the third category, WAPs enable access to the remote data centers in the central cloud through backhaul links.


In order to reap the maximum advantage of computation offloading leveraged at the edge servers, MEC systems need efficient communication channels. Unlike the wired connections in the conventional grid computing and cloud computing, the wireless access links between the mobile devices and cloud computing resources in the edge computing paradigm can be unstable. Sudden service outages may occur with the interruption of access links. The inherent challenges with wireless communication channels like multi-path fading, interference, and spectrum shortage should always be taken into account for the design of MEC systems to seamlessly integrate computation offloading and radio resource management~\cite{mao2017survey}. Moreover, both wireless and backhaul access links have limited capacities which should be properly shared among mobile devices in a similar way as sharing the computing resources of the MEC server. Hence, having a cooperative scheme for the joint allocation of communication and computation resources is important for the successful deployment of MEC~\cite{LiangMECbook}. Redesigning both communication and networking protocols to integrate communication infrastructures in MEC and IoT systems is a challenging task. The key focus should be on improving the computation efficiency with respect to data transmission.

Another major requirement is to maintain interoperability while addressing  heterogeneous communication technologies that have to be utilized in IoT and MEC paradigms in 5G. There are plenty of radio technologies that facilitate IoT Low-Power Wide Area Networks (LPWANs) (e.g., WCDMA, LTE, narrowband IoT~(NB-IoT), Wi-Fi, Bluetooth, Zigbee, SIGFOX and LoRA). The choice of these LPWAN technologies may create trade-offs among signal strength, operational range, throughput, and power consumption. With the arrival of 5G, the convergence of these communication technologies needs to be achieved since one network will not be fitting based on those trade-offs.

\subsubsection{Related work}
Recently, Fog-Radio Access Network~(F-RAN) was introduced by Peng et. al. to consolidate the heterogeneous networks into a single network architecture with 5G even though they do not operate in the same bands to gain high spectral and operating and energy efficiency~\cite{peng2016fog}. 
Well known Cloud Radio Access Network~(C-RAN) architecture can perform cooperative transmission across multiple edge nodes with centralized cloud computing servers via fronthaul links~\cite{tandon2016harnessing}. Although, C-RAN provides high spectral efficiencies due to the enhanced interference management capabilities with the centralized baseband  processing at the cloud, it has potentially large latencies. F-RAN is proposed for 5G MEC deployments as an advanced socially aware mobile networking architecture to provide high spectral efficiency while maintaining high energy efficiency and low latency~\cite{peng2016fog,tandon2016harnessing}. Precoding design, resource block allocation, user scheduling, and cell association are jointly designed for radio resource allocation in F-RANs in order to optimize spectral and energy efficiencies, and latency performances~\cite{peng2016recent}. In~\cite{rimal2017mobile}, Rimal et. al propose a unified Time-Division Multiple Access~(TDMA) based resource management scheme for offloading traffic over Fiber-enabled Wireless~(FiWi) access networks.

In the envisioned 5G systems and MEC architecture, both backhaul and wireless access links can be facilitated by millimeter-Wave~(mmW) spectrum~\cite{agiwal2016next}. The use of mmW spectrum will enable high data rate access to MEC functionalities with low latency. On the other hand, MEC provides local computation power usefully for optimizing the performance of mmW communications. In~\cite{barbarossa2017enabling,barbarossa2017overbooking}, the authors address the joint optimization of communication/computation resources with mmW communication. They have taken the advantage of blocking probabilities by considering intermittency of mmW multi-link communications.

An open source LPWAN infrastructure called OpenChirp is discussed in~\cite{OpenChirp7917625}. OpenChirp, which is developed using LoRWAN, allows multiple users to provision and to manage battery-powered transducers across large areas like campuses, industrial zones, or cities. As pointed out in~\cite{ansari2017mobile,baktir2017can}, SDN plays a vital role in improving MEC type technologies by removing the technical shortcomings in edge computing implementations. The authors summarize the work performed for implementing MEC based on NFV and SDN where the SDN controller manages the communication between MEC servers which form a data center at the edge. Table~\ref{tab:communication} summarizes the reviewed state-of-the-art communication issues and solutions in MEC enabled IoT.

\begin{table*}
  \centering
        \caption{Comparison of the reviewed state-of-the-art communication issues and solutions in MEC enabled IoT.}
        \label{tab:communication}
  \begin{tabular}{|p{0.8cm}|p{10cm}|p{2cm}|p{1.4cm}|p{1.2cm}|}
    \hline
   	\textbf{Ref.} 
    &  \textbf{Description}
    & \textbf{IoT application\newline/domain\newline/feature}	
    & \textbf{Comm. \newline network \newline architecture} 
    & \textbf{Comm. \newline resource \newline allocation}
      \\
  \hline
   \cite{peng2016recent} 
    & Performance analysis of radio resource allocation in F-RANs for edge cache and adaptive model selection to improve spectral efficiency and energy efficiency.
   & Low latency and \newline high reliability
   & 
   & \checkmark
  \\
    \hline
\cite{barbarossa2017enabling}, \newline{\cite{barbarossa2017overbooking}}  
&Use of mmWave spectrum for high data rate access to MEC servers and backhaul links.
&Low latency and \newline high reliability  
& 
&\checkmark
\\
   \hline
   \cite{OpenChirp7917625} 
      & An open source LPWAN infrastructure which allows multiple users to provision and manage battery-powered transducers across large areas.
   & LPWAN infrastructure 
   &\checkmark
   & 
 \\
   \hline
   \cite{sun2016edgeiot}
    &A virtualized edge computing architecture with a proxy VM migration scheme to minimize traffic in the core network.
   &IoT big data streams
   &\checkmark
   &
   \\
     \hline
     \cite{ansari2017mobile}
     & Proposed network architecture includes multi-interface wireless access network (e.g.,~FiWi), heterogeneous backhauling, distributed cloudlets, hierarchical structure of a cloudlet, and the SDN based mobile core network.
     & IoT big data streams
     &\checkmark
     &
     \\
       \hline
   \cite{rimal2017mobile}
   & A novel unified resource management scheme for Ethernet-based FiWi networks that jointly allocates bandwidth for transmissions of both conventional broadband traffic and MEC data in a TDMA fashion.
   & Mission-critical IoT
   & 
   & \checkmark
   \\
     \hline
      \cite{farris2017federations}
      & Introduce Mobile-IoT-Federation-as-a-Service~(MIFaaS) to enable dynamic cooperation among private/public local clouds of IoT devices at the edge of the cellular infrastructure. The selection of the best configuration of federated IoT cloud platforms are modeled as a coalition formation problem.
     &Cellular IoT
     & 
     &\checkmark
      \\
     \hline
     \cite{farris2017federated}
      &Allocation of radio resources in a joint LTE and NB-IoT system based of MIFaas paradigm~\cite{farris2017federations}. Discovered that in handling high-end IoT data traffic, a combination between NB-IoT and LTE is essential in providing the needed high data rate and low latency.
     & Mission-critical IoT
     & 
     &\checkmark
    \\
     \hline
    \cite{orsino2017exploiting}
      & Integration of D2D communications into edge computing environment reduce transmission delay and traffic load across the network.
     & Mission-critical IoT
     & 
     &\checkmark
    \\
     \hline
      \cite{ko2017wireless}
      & Use the theories of stochastic geometry, queueing, and parallel computing for provisioning and planning MEC networks. 
     & Communication latency
     &\checkmark 
     &
    \\
     \hline
  \end{tabular}
  \end{table*}

%
\subsection{Computation Offloading and Resource Allocation}
\label{sec:ComputationOffloading}

\subsubsection{Requirements}
Computation offloading is the most prominent and widely discussed feature of MEC that empowers resource-constrained IoT devices with augmented computational capabilities~\cite{wang2017survey,mach2017mobile}. This will not only prolong the battery life of the IoT sensor nodes, but also reduce E2E latency needed to run sophisticated applications. 
In the first place, UE has to decide whether to execute the relatively simple tasks locally or offload to the MEC servers (i.e., task model for binary offloading)~\cite{mao2017survey}. Secondly, the decision of computation offloading to the MEC servers can be performed fully or partially. In the partial offloading, a subset of computations is executed locally while the rest is offloaded to the MEC server by considering several factors such as users or application preferences (e.g., application buffer state), radio and backhaul connections quality (i.e., between UE and MEC servers), UE capabilities, or cloud capabilities, and availability~\cite{mach2017mobile}. 

The sole objective of the offloading policies need to be the minimization of execution delay. Other critical concerns are to define the dependency of offloadable components of the applications based on their ability to partition data (e.g., real-time user input has to be processed at UE without offloading) and to predict the execution time of multiple tasks. The execution order or routines have to be carefully formulated since certain outcomes can be the inputs of other tasks. As pointed out in~\cite{mao2017survey}, the task models for partial offloading can be represented by task-call graphs with sequential, parallel, and general dependencies. 

Although in MEC, computation offloading enables powerful cloud services at the edge level, the insufficient battery energy at the tiny IoT devices may incur new challenges. In applications like IoT surveillance or remote asset management, the nodes are typically hard to reach. Those applications may also require to offload data more frequently in small chunks by consuming more energy.  Therefore, it is necessary to consider not only the trade-off between energy consumption and execution delay in both full and partial offloading scenarios in MEC, but also the trade-off between computation energy and transmission energy consumption in order to extend battery life. 


The joint computation and communication resource allocation should be properly addressed in order to get the maximum utilization of available resources. Single MEC server will be allocated for the applications which cannot be partitioned. The resources in multiple MEC servers are allocated for the offloaded applications that can be split into several parts. When a job arrives at the MEC server, if there are enough resources, the scheduler has to allocate the VM for further processing. If there are no sufficient computation resources, it delegates the task to the centralized cloud. MEC servers also have to allocate computation and communication resources for user application jobs and MEC service jobs. User mobility, network topology, network scalability, and load balancing are some other factors to be considered in order to define fare resource utilization policies on MEC servers. Specifically when IoT gateways share limited bandwidth among multiple IoT devices which can handle video, audio or bio-medical signals,  the allocation of bandwidth will become challenging~\cite{samie2016computation}. The low power wireless technologies (e.g., BLE, ZigBee, low power Wi-Fi, and LPWAN standards like LoRA or SigFox) used in IoT networks have limited bandwidth. When the IoT devices access the MEC server, which is acting as the IoT gateway, they have to utilize either of those low-power wireless connections that have low bandwidth.

\subsubsection{Related work}
In the comprehensive survey presented in~\cite{mach2017mobile}, the existing work that addresses MEC computation offloading decisions have been nicely summarized based on full and partial offloading types. These solutions are proposed either to minimize the execution delay or to balance the trade-off between energy consumption and latency. Moreover,~\cite{mach2017mobile} provides an overview of the latest research works that address the allocation of computation resources for the data or application which it decides to offload in MEC systems. However, this analysis does not address the explicit applicability of computation offloading and resource allocation in IoT supportive MEC systems.

A preliminary study on how computation offloading and bandwidth allocation can be performed in MEC supportive IoT networks is presented in~\cite{samie2016computation}. Due to the discrete and coarse-grained offloading levels on the IoT end nodes, the gateway (i.e. MEC server) bandwidth will be under-utilized. This phenomenon is termed fragmentation. Based on the received transmission rates and power consumption parameters of IoT devices, the gateway runs an iterative algorithm to optimally allocate bandwidth in such a way as to optimize the battery life of the devices. The implementation of the algorithm for a health monitoring application shows more than 40\% improvement in using gateway bandwidth and up to 1.5 hour improvement in battery life of IoT devices. Replisom~\cite{abdelwahab2016replisom} designed by Abdelwahab et al., is a model for computation offloading for massive IoT applications where the replicated memory objects produced by IoT devices are offloaded to the LTE-aware edge cloud. Replisom protocol relies on D2D communication for effectively scheduling the memory replication occasions to resolve interference and scarcity in radio resources as a large number of devices simultaneously transmit their memory replicas.

\begin{table*} [hb]
  \centering
        \caption{Comparison of the reviewed state-of-the-art computation offloading and resource allocation features in MEC enabled IoT.}
        \label{tab:computation}
  \begin{tabular}{|p{1cm}|p{10cm}|p{2cm}|p{1.7cm}|p{1.2cm}|}
    \hline
   	 \textbf{Ref.} 
     & \textbf{Description}
     & \textbf{IoT application \newline /domain\newline /feature}	
     & \textbf{Computation Offloading}
     &\textbf{Comp. Resource \newline Allocation} 
       \\

     \hline
  \cite{samie2016computation}
  & Management of computation offloading in a local IoT network with the efficient utilization of IoT gateway bandwidth constraints.
  & IoT-gateway
  &\checkmark
  &
   \\
  \hline
  \cite{abdelwahab2016replisom}
   & Replicated memory objects produced by IoT devices are offloaded to the LTE-aware edge cloud based on D2D communication.
  & Massive-IoT
  & \checkmark
  &
 \\
  \hline
  \cite{yu2017sdlb}
  & Proposes a portable MEC load balancer which is scalable, software based, memory efficient and adaptive to device heterogeneity. The design takes the advantages of SDN and POG data structure.
  & IoT big data streams
  & 
  & \checkmark
    \\
  \hline
  \cite{vilalta2017telcofog}
  &Defines an architecture to allocate cloud and edge resources for deploying NFV, MEC, and IoT services on top of a telecom operator's network.
  &Low latency
  &
  &\checkmark 
  \\
  \hline
  \cite{bouet2017geo}
   &Propose a MEC clustering algorithm to consolidate the maximum communications at the edge which stands for the spatial temporal dynamics of the traffic.
  &IoT big data streams
  &
  &\checkmark 
 \\
  \hline
  \cite{flores2017large}
   &Defines a scalable offloading architecture and a simulator  with  multi-tenancy  ability and dynamic horizontal scaling based on Amazon Autoscale service-oriented architecture.
  & Massive-IoT
  & \checkmark 
  & \checkmark
 \\
  \hline
     \cite{wang2017computation}
     & Formulate the computation offloading decision, resource allocation, and content caching in wireless cellular networks with mobile edge computing as an optimization problem and solve it applying alternating direction method of multipliers based distributed algorithm.
     & Cellular IoT
     & \checkmark 
     & \checkmark 
     \\
     \hline
     \cite{lyu2017optimal}
     & Introduces asymptotically optimal offloading schedules, which are tolerant to partial out-of-date network knowledge and stochastically maximize a time-average network utility balancing system throughput and fairness.
     & Massive IoT
     & \checkmark 
     & \checkmark
     \\ \hline
       \cite{gupta2017ifogsim}
     & Develop a toolkit for modeling and simulation of resource management techniques in the IoT, edge and fog computing environments
     & General IoT
     &  
     & \checkmark 
     \\
     \hline
         
  \end{tabular}
  \end{table*}
  
Furthermore, with the advent of mobile device performance and D2D communication technologies, computation offloading can be performed at the mobile devices. As shown in~\cite{habak2015femto}, a collection of co-located mobile devices can be utilized to provide cloud services at the edge instead of using MEC servers. Such an offloading mechanism will allow the very constrained tiny IoT devices to outsource the computation intensive tasks to the high performing mobile devices in the closest proximity. Few research efforts were performed to derive computation offloading strategies in MEC that support user mobility. In~\cite{chen2016mobility}, the authors propose a hybrid computation offloading mechanism for edge computing considering the hardware heterogeneity of the mobile devices, various user’s requirements on Quality of Experience~(QoE) and the heterogeneity status of the network.
  
The requests for computation offloading generated by end devices have to be handled by the software load balancer according to the availability of the MEC servers and resources. Yu et. al. proposes a softwarized load balancer technique called SDLB for edge computing based on the minimal perfect hashing algorithm~\cite{yu2017sdlb}. Their scalable and dynamic load balancer SDLB is derived based on POG data structure and able to support about one million update requests per second. In~\cite{vilalta2017telcofog}, the authors propose a virtualized network architecture with intelligent resource allocation capabilities for NFV, MEC and IoT services. This so called TelcoFog architecture provides seamless and unified control for the complete visibility, computation, and allocation of both cloud and network resources through different network segments (access, aggregation, and transport) assuming heterogeneous access and transport technologies (e.g., Wi-Fi, packet switching, optical transmission).

The game theoretic approach is also designed for selecting the most appropriate wireless channels to transmit offloading data in a multi-user multi-channel MEC systems~\cite{chen2016efficient,sardellitti2015joint}. In~\cite{wang2017joint}, the MEC server makes the offloading decisions and physical resource block allocation to the UEs using the graph coloring method. Furthermore, in~\cite{bouet2017geo}, authors propose a graph-based algorithm that takes into account, the maximum MEC server capacity, provides a partition of geographic area, and consolidates as many communications as possible at the edge. The offloading architecture proposed in~\cite{flores2017large}, addresses the scaling of offloading support to large-scale IoT environments. Their application level task scheduler uses horizontal scaling to allocate the available resources in the edge cloud. Moreover, content caching strategy is also considered in some work for the optimized joint computation and communication resource allocation~\cite{wang2017computation}. Table~\ref{tab:computation} summarizes the reviewed state-of-the-art computation offloading and resource allocation features in MEC enabled IoT.


\subsection{Mobility Management}
\label{sec:mobility}

\subsubsection{Requirements}
A more general concept in cellular and IP networks is mobility management for moving users. Since earlier generations of mobile cellular networks, mobility management has been the ultimate way of ensuring that mobile services are delivered to subscribers wherever they are within the coverage areas of the service provider. The cellular network is a radio network that consists of multiple base stations; each base station is designated to provide mobile services within a particular cell, and hence combining several base stations enables the service provider to cover wider geographical locations. In LTE, mobility management advanced significantly through the introduction of moving networks, seamless roaming, and vertical handovers which is enabled when the UE changes the serving eNB/SCeNB.

In the case of MEC, mobility management is particularly crucial, given that when mobile UEs move far away from the computing node, then there is the possibility of degrading the QoS due to latency. A severe degradation could lead to a complete disconnection of a UE from the MEC network. In MEC-enabled IoT, a large majority of the nodes will be mobile nodes, hence the goal is to exploit MEC services to offer an ultra-reliable mobility management scheme for IoT applications. In traditional mobile networks, the key issues with mobility management are mainly connectivity, location management, routing group formation, seamless mobility, mobility context management, and migration among others. Among these issues, seamless mobility tends to be the most trivial. There is a need for mobile devices to have uninterrupted access to information, communication, monitoring and control – when, where and how they want, regardless of the device, service, network or location. For the MEC architecture, using such traditional approach to mobility management will certainly lead to a degraded performance in the overall MEC network; one key reason for this shortfall is due to the co-provision of radio access and computing services of the MEC-enabled base stations.

\subsubsection{Related Work}

Several mobility management policies have been proposed for the MEC architecture~\cite{sun2017emm, mach2017mobile, liu2016delay, you2017energy}.
In \cite{sun2017emm} authors developed a novel user-centric Energy-aware Mobility Management (EMM) scheme based on Lyapunov optimization and multi-armed bandit theories. The EMM scheme works in an online fashion without using future system state information is hence able to manage the imperfect system state information. The goal of EMM is to optimize the offloading delay that results from both radio access and computation, under the long-term energy consumption constraint of the user. Here, the experiment results showed that the proposed algorithms can optimize the delay performance while approximately satisfying the energy consumption budget of the user. However a major issue with this algorithm is that it will not be effective for a high mobility scenario where a connected node will move in a great deal during the processing of a task, and such high mobility scenario is a typical feature of the IoT networks.



\begin{figure}[ht]
  \centering
  \includegraphics[width=0.49\textwidth]{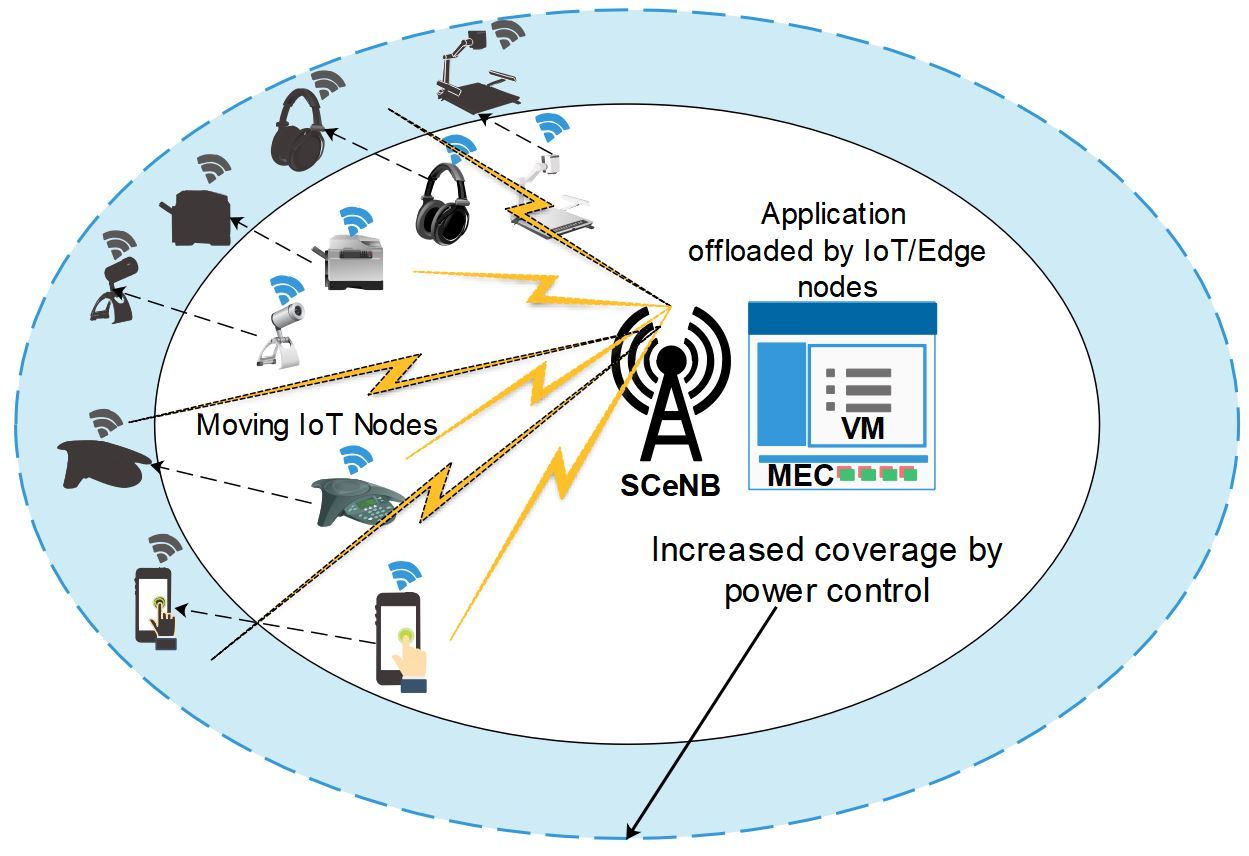}
  \caption{CaPC Power Control Principle\cite{mach2017mobile}.}
  \label{fig_SCeNB}
\end{figure}

In \cite{mach2017mobile}, authors presented a user-oriented use case of MEC from the perspective of computational offloading and mobility management. They first discuss the power control approach where the mobility management entity regulates the transmission power of the eNB/SCeNB, which is mostly used in scenarios where the UEs mobility is confined within a given space such as an office room~\cite{mach2017mobile, mach2014cloud, mach2016cloud}. The principle of this approach is depicted in Figure~\ref{fig_SCeNB}. Accordingly, the MEC services are extended to slowly moving IoT devices within a given space by adjusting the transmission power of the serving and/or neighboring SCeNBs. This Cloud-aware Power Control (CaPC) algorithm is mostly suitable for managing the offloading of real-time applications where delay requirements are strict. It allows the MEC system to handle higher amounts of offloaded applications within specific latency constraint. Typically, increasing the transmission power of SCeNB will momentarily increase the coverage region of MEC signals, hence allowing IoT nodes to move beyond the default coverage region for the duration of the power boost. This will help to avoid the need for handover as much as possible, especially in cases where the moving distance of the IoT device is relatively small. The moving IoT devices are able to roam certain distance away from the coverage region of MEC services just by adapting the transmission power of the eNB/SCeNB, without discontinuity in service and handovers. 

Another scenario is when the IoT node decides to initiate an offload either within the coverage region increased by power control or as it roams beyond. Two possible procedures could be used in this case; one is by performing a VM migration, i.e. migrating a VM from the less effective to a more effective computing node, and two is by path selection, i.e. selecting a new path for communication between the computing node and the IoT device. The need for VM migration arises when the IoT node roams beyond the region extended by the power control mechanism. In that case, the risk of service discontinuity and poor QoS factors tend to be higher, hence there is a need to 
strategically design the VM migration process. Analysis of the influence of such migration on the performance of a typical IoT node is described in \cite{taleb2013analytical}, using the Markov chain analytical models. Based on the outcome of the analysis, when VM migration is not implemented, the probability that the edge device will connect to the optimal MEC decreases with the increase in hops between the eNB and the UE. Meanwhile there is also an additional delay that occurs in when VM migration is not used.  In addition to the literature mentioned in~\cite{mach2017mobile}, Table~\ref{tab:mobility} summarizes the reviewed state-of-the-art mobility management in MEC enabled IoT.

\begin{table*}
  \centering
        \caption{Summary of the reviewed state-of-the-art mobility management in MEC enabled IoT.}
        \label{tab:mobility}
  \begin{tabular}{|p{0.8cm}|p{10cm}|p{2cm}|p{1.4cm}|p{1.2cm}|}
    \hline
   	\textbf{Ref.} 
    &  \textbf{Description}
    & \textbf{IoT application\newline /domain\newline/feature}	
    & \textbf{Mobility Management} 
    & \textbf{Flow Scheduling}
      \\
  \hline
  \cite{sun2017emm} & Develop a user-centric energy-aware mobility management (EMM) scheme, to optimize the delay due to both radio access and computation, under the long-term energy consumption constraint of the user. & General IoT & \checkmark &  \\ \hline
  \cite{wu2015ubiflow} &  Present UbiFlow, the first software-defined IoT system for ubiquitous flow control and mobility management base don distributed controller in multinetworks. & Software defined IoT & \checkmark & \checkmark\\
     \hline
  \cite{shang2016named} & Explores how Named Data Networking, a proposed future Internet architecture, can address the challenges of interoperability in IoT networks. & IoT applications & & \checkmark \\ \hline
\cite{giust2015distributed} & Analyzed distributed mobility management for future IoT sensor networks. & IoT sensors & \checkmark & \checkmark \\ \hline
\cite{le2017location} & Propose a location-aware load prediction at edge data centers which supports user mobility. &General IoT & \checkmark& \\\hline  
  \end{tabular}
  \end{table*}


\subsection{Security}
\label{sec:Security}

\subsubsection{Requirements}
Integrating MEC capabilities to the IoT systems come with an assurance of better performance in terms of quality of service and ease of implementation. This however, raises concerns in both research and the industry first on the heterogeneity of connected devices, and second on the potential repercussions of such architectural modification on the overall security of MEC-enabled systems. Typical security threats in these areas are Denial of Service (DoS) attacks, Man-in-the-Middle (MitM) attacks and malicious node problems \cite{vassilakis2016security, roman2016mobile}. More detailed descriptions of these threats are presented in \cite{roman2016mobile}.  

\par IoT systems in general inherit most of the security vulnerabilities commonly found on sensor networks, mobile communication networks and the Internet as a whole. Thus making security one of the application challenges of IoT in present and future networks. Such security vulnerabilities in IoT networks include DoS/Distributed DoS~(DDoS) attacks, forgery/middle attack, heterogeneous network attacks, application risk of IPv6, Wireless Local Area Networking~(WLAN) application conflicts also affect the transport security of IoT \cite{jing2014security}.

Here we define the possible security attacks in the context of MEC-enabled IoT environment. Security threats are mostly targeted towards the MEC nodes, e.g. MEC server and other IoT nodes. In DoS attacks, the adversaries tend to attack critical networking or computing resources by sending requests at rates that are beyond the handling capacity of such networking or computing equipment, hence inundating such facility and preventing other users or nodes from getting access to the resources offered. DoS attacks could happen in the form of DDoS or wireless jamming and could be launched on both the virtualization and network infrastructures. 

\par MitM happens when an adversary interposes between two nodes or entities and secretly relaying or altering the communication between such parties, common example is the MitM attack between a server and a client. For the MEC-enabled IoT scenario, the most vulnerable location for MitM attack is the infrastructure layer where the malicious attacker tries to hijack certain segments of the network and begins to launch attacks like eavesdropping and phishing on connected devices. As claimed in~\cite{stojmenovic2016overview} MitM attacks can be launched between 3G and WLAN networks. Such attacks would be even more threatening for the MEC-enabled IoT scenario, given that MEC relies heavily on virtualization, hence launching a MitM attack on multiple VMs could very easily affect all other elements on both sides of the attack.  

\par VM Manipulation is a typical attack for all virtualized and edge computing systems. In MEC-enabled IoT system, VM manipulation is mainly targeted towards the virtualization infrastructures. In this case, the attacker is more likely to be a malicious insider with enough privileges or a VM that has escalated privileges.  The adversary in such attack begins to launch multiple attacks to the VMs running inside it. When VM manipulation attack is launched, the affected VMs are further exposed to numerous other potential attacks like logic bombs. 

\subsubsection{Related Work}
On the application layer, security threats are mostly in the context of information access and user authentication. Others include possibility of tracking and destroying data streams, tampering with the stability of the IoT platform, attacking the middleware layer and/or management platform\cite{wan2013cloud,wan2014vcmia}. Given that IoT will further converge people’s everyday life activities and devices on the network, the need for faster access to data which is largely addressed by introducing MEC to the IoT system, must be balanced by a robust and highly reliable security technology in addition to creating more security awareness for users and application developers. 



The architecture proposed in~\cite{jing2014security} has three key layers namely perception, transportation and application. The authors have identified different potential security vulnerabilities on each layer. For the perception layer, potential security vulnerabilities are mainly on the RFID, the wireless sensor networks, and the RFID sensor networks. For the transport layer, security vulnerabilities are mainly found at the access network, the core network, and the local network. Here, vulnerabilities can also be unique to the different access technologies, i.e. for 3G access network, Ad-Hoc network, and Wi-Fi. On the application layer, vulnerabilities exist for the application support layer as well as for specific IoT applications.



\subsection{Privacy}
\label{sec:Privacy}

\subsubsection{Issues and challenges} 

The early designs of IoT systems were largely closed, homogeneous and single-purpose with limited functionality, geographic scope and scale. In contrast, the present-day IoT systems are much larger and spanning across countries or continents, making them to comply with the varying rules and regulations. Similarly, in health care\cite{varga2016network} type of applications, which invade personal spaces, privacy is becoming a significant concern~\cite{porambage2016quest, sicari2015security}. Governing organizations like European Commission have recognized that privacy in the processing of personal data and the confidentiality of communications as fundamental rights that should be protected~\cite{taylor2006eu}. In an IoT application, when the data sharing principle is leveraged by a cloud based system, that could raise a lot of privacy concerns. The potential use of data for unpredicted future applications may compromise privacy. 

MEC enables caching, data processing and analytics to be done closer the source of the data and reduces the burden on centralized cloud servers and core networks\cite{garcia2015edge}. Importantly, this will support differentiated privacy since raw, unprocessed data does not have to be stored or processed by a centralized cloud systems which can be located in distance. Only the processed and selected data are needed to reach the centralize cloud for further processing \cite{porambage2016quest, sicari2015security}.  For instance, the image processing of car number plate recognition  can be done in the edge without transferring the location information to the centralized cloud servers. Such MEC based local processing protects the privacy of data without leaving the jurisdiction of the user. Moreover, the decentralized approach reduces the impact of data breaches such as Sony breach \cite{haggard2015north} and OPM (Office of Personnel Management) breach \cite{german2016new} . MEC approach also enable the possibility to implement specific or local privacy policies\cite{roman2013features}, contrary to the uniform privacy policies applied in centrally managed public cloud. In some IoT applications such eHealth services (for instance, mental and abortion clinics) local privacy polices with edge intelligence is required to meet the required privacy protection which cannot be met by only using a centralized approach\cite{roman2013features}.

The requirements in privacy protection are identified based on the generic and the regulatory objectives. First, it is required to harmonize the privacy of digital services at global level by promoting the digital single market. All relevant directives and legislative instruments should be encouraged to enable cross border policies. Then, it is necessary to balance the interests in protecting privacy and in fostering the global use of services.  

Second, the privacy legislation should be done at a global level to ensure their compatibility with new technologies such as MEC.  Different jurisdictions should cooperate together to develop inter-operable privacy requirements and  facilitate the flow of information with the required level of privacy protection. For instance, the "Safe Harbor" agreement between US and EU, requires US companies  to obey  EU regulations so that  EU companies can  store and process data in US data centers\cite{kemmer2016software}.

Third, it is necessary to foster interoperability and data portability to support the adaptation of new technologies. For instance, it can be done by avoiding mandated standards or preferences which could prevent interoperability. Moreover, it is necessary to  promote the on-going interoperability efforts in the industries, this will be useful in defining uniform and global privacy policies. Finally, it is required to define one framework with a set of data protection laws which can be used across the border and they should be simple enough to be set up globally. This framework should be based on the concept of accountability and the laws should also support self-regulatory codes and mechanisms.

\subsubsection{Related work}
Security and privacy challenges in MEC like edge computing paradigms are surveyed in \cite{roman2018mobile} and \cite{yi2015security}. A partially distributed approach that allows edge intelligence that can meet the privacy requirements of IoT use cases such as eHealth services is presented in \cite{roman2013features}. The possibility of exploiting edge computing to solve the problem of loss of privacy by releasing personal and social data to centralized services such as e-commerce sites, rating services, search engines, social networks, and location services are presented in \cite{garcia2015edge}. Possibilities of improving the data privacy of IoT data by using edge computing is presented in \cite{shi2016edge}.

%

\subsection{Trust management}
\subsubsection{Requirements}

Trust is a rather complex property to define, it is closely associated with the overall security of any network or platform. Trust is significant in  critical 5G use cases like remote surgeries, emergency autonomous vehicles, factory automation, and tele-operated driving (e.g. drones). In these scenarios, latency and reliability are highly regarded. Although trust is an equally important property similar to security and privacy in IoT and MEC, it is hardly addressed lately in research works~\cite{roman2018mobile}. The need to implement the appropriate trust management scheme is very essential when it comes to IoT technologies. This is because IoT devices offload their delay critical applications to the edge cloud which is normally out of the direct control of the client. 

According to Yan et. al., the key challenges of trust management in IoT are not only limited to system security robustness and privacy preservation~\cite{yan2014survey}. Trust relationships have to be sustained among all IoT system entities including the enabling technologies such as MEC. Data perception trust determines the reliability of data sensing and collection in the IoT perception layer. Data fusion and mining trust explains the efficiency and trustworthiness of big data handling in the IoT network layer. Enabling secure data transmission and communication while maintaining the quality of IoT services and identity trust are other important aspects of IoT trust. It is equally important to apply a more generic trust management framework for IoT since it is a collaboration of multiple technologies and systems. The utilization of tamper resistive secure elements will enable the trust in the end user devices with physical protections to prevent the compromising of cryptographic security parameters. However, due to limited resources in many tiny IoT devices, the integration of such trust enabling devices will also be challenging. Above all, the most significant is the realization of human-computer trust interaction which requires more attention to the subjective properties of IoT users at the application layer. 

In cloud computing, trust is targeted towards long-term underlying properties or infrastructure (persistent trust), and such trust can be specific to context-based social and technological mechanisms (dynamic trust). Moreover, when edge cloud computing is collaborating with IoT, it introduces more trust related objectives such as maintaining the trust for computation offloading IoT services or collected data to the edge cloud and the cooperative trust among edge servers. The edge servers should ensure the trustworthiness of end users and IoT devices, which acquire the resources from the edge cloud. Likewise, the edge servers should also assure their reliability and trustworthiness to the end users/devices and other edge servers for providing guaranteed services. More importantly, the efficient resource sharing among the edge servers has to be accomplished based on a proper trust management framework.

\begin{table*}[hb]
  \centering
        \caption{Comparison of the reviewed state-of-the-art security, privacy, and trust management in MEC enabled IoT.}
        \label{tab:security}
  \begin{tabular}{|p{1cm}|p{9cm}|p{2.2cm}|p{0.9cm}|p{0.9cm}|p{0.9cm}|}
    \hline
   	 \textbf{Ref.} & \textbf{Description} &  \textbf{IoT application\newline/domain/feature} & \textbf{Security}	& \textbf{Privacy}& \textbf{Trust}  \\
  \hline
  
  \cite{vassilakis2016security} & Proposed a security framework for virtualized Small Cell Networks, with the aim of further extending MEC in the broader 5G environment & Cloud-enabled IoT & \checkmark & \checkmark &\\ \hline
   \cite{abedin2015fog} &  Addresses the utility based matching or pairing problem within the same domain of IoT nodes by using Irving's matching algorithm under the node specified preferences to endure a stable IoT node pairing & IoT node pairing services & \checkmark & \checkmark & \checkmark \\ \hline
   \cite{jing2014security} &  Analyzes the cross-layer heterogeneous integration issues and security issues in detail and discusses the security issues of IoT as a whole and tries to find solutions to them & General IoT & \checkmark & \checkmark & \\ \hline
   \cite{garcia2015edge}  & Presents the research challenges associated with security, privacy and trust management in Edge-centric Computing   & General IoT & \checkmark & \checkmark & \checkmark\\ \hline
\cite{de2012proposed}  & Holistically analyses the security and privacy threats, challenges, and mechanisms inherent in all edge paradigms including MEC.   & General IoT & \checkmark & \checkmark &\\ \hline
\cite{roman2018mobile}  & Holistically analyses the security and privacy threats, challenges, and mechanisms inherent in all edge paradigms including MEC.   & General IoT & \checkmark & \checkmark &\\ \hline
\cite{yi2015security}  & A survey on security and privacy  challenge in fog computing   & General IoT & \checkmark & \checkmark &\\ \hline
\cite{roman2013features}  & Present a edge computing based distributed approach to satisfy the security and privacy requirements of IoT   & General IoT & \checkmark & \checkmark &\\ \hline
\cite{shi2016edge}  & Discuss the methods of improving security and  privacy of IoT data by using edge computing   & General IoT & \checkmark & \checkmark & \\ \hline

\cite{ziegler2017anastacia}         & Introduce the preliminary design of a holistic framework for enabling trust
and security by-design for cyber physical systems (CPS) based on  IoT  and  edge cloud  architectures.   & IoT architecture &\checkmark & \checkmark &\checkmark\\ \hline
\cite{dang2017data} & Propose a trust translation model for fog nodes and a privacy-aware model for access control at fog nodes.  &IoT big data streams& &\checkmark &\checkmark\\
     \hline
         
  \end{tabular}
  \end{table*}
  
\subsubsection{Related work}
The comprehensive literature surveys in~\cite{yan2014survey,sicari2015security} summarize the recent research works on IoT trust. Accordingly, the researchers have addressed IoT trust in multiple perspectives including trust evaluation, trust framework, data perception trust, identity trust and privacy preservation, transmission and communication trust, secure multi-party computation, user trust, and application trust. Existing IoT trust evaluation mechanisms are mathematically formed and have considered different trust metrics like social trust and QoS trust using both direct observations and indirect recommendations. Most of the trust frameworks proposed in IoT address security and privacy in IoT data transmission and communications. In~\cite{ziegler2017anastacia}, a preliminary design of a holistic solution with trust and security-by-design for cyber physical systems based on IoT and cloud architectures is presented. They have taken the initiative to develop and demonstrate a trustworthy-by-design autonomic security framework based on SDN/NFV and IoT networks. 

In many previous literatures, data perception trust is addressed in the context of security and privacy, mainly by mitigating security attacks on data aggregation and processing, as well as exploiting some key management techniques~\cite{yan2014survey}. Some recent literatures have also addressed data protection and performance improvement at the edge computing servers by trust management among fog servers~\cite{dang2017data}. Furthermore, trust is paramount to the effectiveness of node interaction in SIoT where the objects are building up a social network and becoming more autonomous~\cite{atzori2012social}. Table~\ref{tab:security} summarizes the reviewed state-of-the-art security, privacy, and trust management in MEC enabled IoT.
%

\section{Integration Technologies}
\label{sec:Technologies}

The realization of MEC for IoT is fueled by several integrating technologies such as SDN, NFV, ICN and Network Slicing. This section provides a high level overview of the role of each technology in MEC-IoT environment and the related works. 

\subsection{Network Function Virtualization}

NFV is a network concept which proposes to use virtualization technologies to manage core networking functions using a software based approach\cite{mijumbi2016network}. NFV has been proven as one of the key enablers for not only the development of 5G but also MEC-IoT integration\cite{gupta2016mobile}. Specifically, MEC reuses the NFV virtualization infrastructure and the NFV infrastructure management to the largest extent possible\cite{yang2016seamless}.

Both MEC and NFV technologies can be used together in environments such as 5G mobile networks to elevate computing capacity to meet the increased networking demands. MEC architecture is also based on a virtualized platform quite similar to NFV architecture. Both technologies feature stackable components and each has a virtualization layer. 

According to ESTI\cite{hu2015mobile}, it is beneficial to reuse the infrastructure and infrastructure management of NFV to the largest extent possible, by hosting both Virtual Network Functions~(VNFs) and MEC applications on the same platform, computing experience is enhanced. The use of NFV will equally increase the scalability of MEC application. NFV can improve the scalability  by dynamically scaling up/down the network resources depending on demand.

Several NFV-MEC ingratiation research works have been proposed recently. In \cite{yang2016seamless}, NFV-enabled MEC scheme is proposed to optimize the placement of resources among NFV-enabled nodes to support low latency mobile multimedia applications. A novel MEC and NFV integrated network architecture is presented in \cite{li2017mec}, this can be used to enhance the mobile game experience, optimized high speed HD video streaming and local content caching for AR. The double-tier MEC-NFV architecture in~\cite{sciancalepore2016double} aligns and integrates the MEC system with the NFV Management and Orchestration (MANO) by introducing a management subsystem that enriches the MANO with application-oriented orchestration capabilities. To support the deployment of container-based network services at the edge of the network, an architecture based on the Open Baton MANO framework is proposed by combining the NFV and MEC within a single orchestration environment\cite{carella2017prototyping}.

\subsection{Software Defined Networking}
SDN is another 5G enabling technology which will help to design dynamic, manageable, cost-effective, and adaptable networks. SDN has fuel the advancement of network softwarization by proposing to transfer the control functionality to software based entities, i.e. network controllers. SDN eliminates the use of vendor specific black-box hardware, thereby promoting the use of commodity servers and switches over proprietary appliances. 

Notwithstanding, the transfer of network control functionalities to software based centralized entities, demands the data plane devices to communicate frequently with the SDN controllers. Thus, SDN controllers are located closer to the data plane to reduce the latency in packet processing. MEC offers the opportunity to locate control functions closer to data plane devices. Moreover, MEC complements the SDN advancement of the transformation of the mobile-broadband network into a programmable world, ensuring highly efficient network operation and service delivery\cite{blanco2017technology}. Thus, the popularity of SDN  in different domains including 5G, IoT will fuel the adaption of MEC concept as well.

Many recent research works justify the added benefits of the combine use of SDN and MEC in IoT systems\cite{peng2017qoe, ali2017real, farristowards, huang2017low, nguyen2017simeca, hossain2017impact, liu2017high, phemius2016bringing}. The role of NFV and SDN in  MEC ecosystem is discussed in \cite{peng2017qoe}. SDN can be also used to  make MEC more flexible and cost-effective for 5G applications. The  real-time heart attack mobile detection service proposed in \cite{ali2017real}, is a novel e-health IoT service that employs SDN-powered MEC in a Vehicular Ad-hoc Network~(VANET) architecture for reliable performance. In~\cite{farristowards}, a novel SDN/NFV-based security framework is presented to enable integrated protection for IoT systems and in MEC applications. An SDN-based MEC framework has been proposed to provide the required data-plane flexibility,  programmability and reduced latency for applications such as VR and Vehicular IoT\cite{huang2017low}. 

In addition, a conceptual approach to  providing security for  IoT systems by using SDN and edge computing is presented in \cite{aggarwal2016securing}. The SDN-based IoT mobile edge cloud architecture (SIMECA) proposed in \cite{nguyen2017simeca} can deploy diverse IoT services at the mobile edge by leveraging distributed, lightweight control and data planes optimized for IoT communications. In~\cite{hossain2017impact}, the utilization of SDN and MEC to overcome the challenges of network densification of IoT–cloud integration over a smart home is presented. Likewise, the MEC-SDN framework presented in \cite{phemius2016bringing} guarantees the QoS requirement satisfaction and efficient use of the wireless resources in tactical network applications.

\subsection{Information Centric Networking}

To address the ever increasing traffic volume in the Internet applications such as HD mobile video, AR/VR, 3D gaming and cloud computing, a new set of network architectures and networking technologies  are developed over the past few decades. These technologies employ caching, replication and content distribution in optimum ways. Among them, ICN has become one of the main approaches to addressing this demand\cite{vasilakos2015information, piro2014information}. ICN is an Internet architecture that puts information at the center where it needs to be and replaces the client-server model by proposing  a new publish-subscribe model. The key benefits of ICN include fast and efficient data delivery and improved reliability. Thus, ICN is considered one of the promising networking models for IoT ecosystem.


MEC and ICN are complementary concepts which can be deployed independently \cite{grewe2017information}. However, both could add value to 5G and IoT domains in a complementary fashion. Certain synergies can be exploited when these two technologies are deployed cooperatively.  For example, ICN can be used for content distribution over an unreliable radio links and  transparent mobility among multiple technologies\cite{lloret2017enabling}, while MEC can be used  to reduce the latency for delay critical applications such as tactile Internet\cite{maier2016tactile} and AR/VR applications, or to perform distributed data-reduction and security functions for an IoT network.

In addition, the use of MEC with ICN can further improve the performance of edge computing. It can solve some of the existing challenges in MEC ecosystem. For instance, MEC is facing a challenge of application level reconfiguration, since it requires a re-initialization of the session whenever a session is being served by a non-optimal service instance. Such application level reconfiguration will increase the delay in session migration. However, the natural support for service-centric networking in ICN can minimize the network related configuration for applications. It will reduce the reconfiguration delay and  allow fast resolution for named service instances\cite{ravindran2017realizing}. 

ICN can also improve the edge storage and caching features of MEC enabled networks. ICN allows location independent data replication and opportunistic caching at strategic points in the network. These features benefit both real-time and non-realtime IoT applications where a set of IoT devices or users share the same content\cite{ravindran2017realizing}.


Opportunities and challenges of MEC and ICN integration for IoT are presented in \cite{5GAmericaICNMEC}. Here, the authors highlight the synergies that can be exploited when the two technologies are deployed cooperatively for IoT applications. In addition, several research works have also verified  the importance of ICN and MEC cooperation\cite{zhou2017video, grewe2017information, zhou2017resource, huo2016software, ge2016qoe}.  A novel HetNets virtualization architecture with ICN and MEC techniques is proposed for video trans-coding, caching, and multi-cast in \cite{zhou2017video}. A virtual multi-resources allocation scheme is used in the designed framework to  maximize the utility of computing, caching, and communication to support the massive content delivery. The vision of combining ICN and MEC in the context of connected vehicle environments is presented in ~\cite{grewe2017information}. It shows how ICN in combination with MEC can address the challenges of futuristic vehicular application scenarios. A novel information-centric heterogeneous networks framework is proposed in \cite{zhou2017resource} to optimize the virtual resource allocation at the edge. Authors formulate the virtual resource allocation strategy as a joint optimization problem by considering both virtualization and  caching and computing at the edge. A novel framework which jointly considers networking, caching, and computing techniques to support energy-efficient information retrieval and computing services is presented in \cite{huo2016software}. This framework integrates SDN, MEC and ICN  to enable the dynamic orchestration of different resources in next generation green wireless networks. A MEC-enabled ICN-based content handling framework at the mobile network edge is presented in \cite{ge2016qoe}. The proposed framework realizes context-aware content localization in order to enhance user QoE in video distribution applications. 

\subsection{Network Slicing}
Network slicing proposes a way of separating the network into different network segments. Thus, it allows multiple logical network segments to be created on top of a common shared physical infrastructure\cite{samdanis2016network}. Future IoT will enable a wide range of different types of connections and services. These connections and services will need performance guarantees as well as security. Network slicing can satisfy these requirements. Moreover, 5G mobile network will support both MEC and network slicing technologies\cite{alliance2016description}. 

Network slicing can be used in different IoT domains. One of such application domain is massive IoT\cite{nikaein2015network}. In order to support massive IoT systems, the network should be able to satisfy requirements such as massive cost reduction in communication, network scalability and edge analytics. The integration of MEC with Network slicing can be used to satisfy some of these requirements such as scalability and edge analytics. Another use case is critical communications for delay critical applications such healthcare, autonomous driving and industrial Internet. The key requirements to enable such critical communications are reduced latency and traffic prioritization. While MEC can be used to reduce latency, network slicing can support traffic prioritization. 

\begin{figure}[ht]
  \centering
  \includegraphics[width=0.45\textwidth]{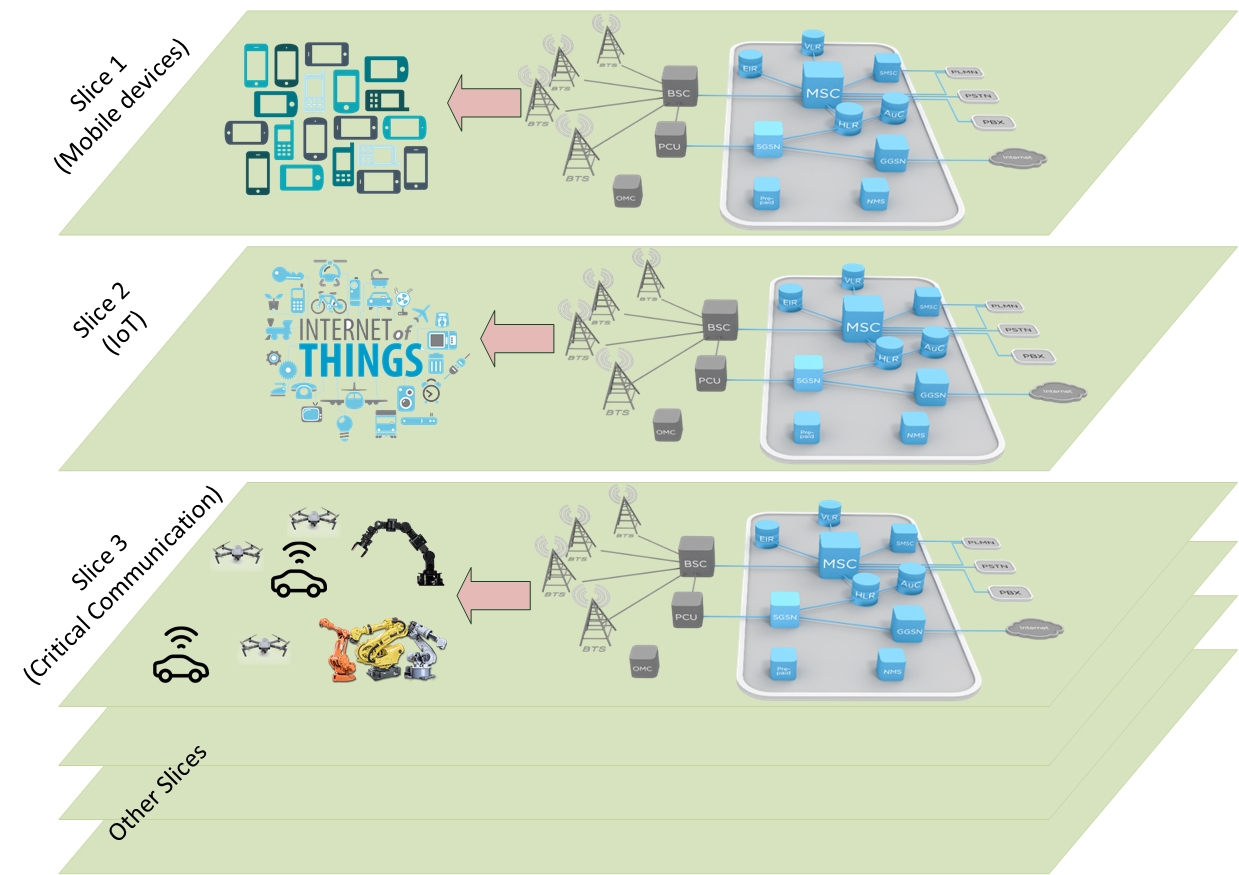}
  \caption{Use of Network Slicing in different applications\cite{5gamericasNS}.}
  \label{fig_SlicingUC}
\end{figure}

Figure \ref{fig_SlicingUC} illustrates the utilization of network slicing in different applications. Here, network slicing can be use to divide the MEC resources in to different slices dynamically. It will improve the efficiency of using MEC resources in different IoT applications.

Several research articles  already presented the possibility of using Network slicing with MEC to provide improved services for IoT and other 5G applications. 

An overview of the Third Generation Partnership Project~(3GPP) standard evolution from network sharing principles, mechanisms, and architectures to future on-demand multi-tenant systems is presented in \cite{samdanis2016network}. MEC is identified as one of the key  attributes to realize the aforementioned network slicing extensions in 3GPP toward full multi-tenancy. A logical architecture for network-slicing-based 5G systems is presented in \cite{zhang2017network}. Here, authors show the evolution of network slicing in network architecture and the synergy with SDN, NFV and MEC technologies. The work presented in \cite{katsalis2017network} discusses the design challenges of network slicing  with other concepts such as cloud-RAN and MEC.  A SDN/NFV packet/optical transport network and edge/core cloud platform for E2E 5G and IoT services is presented in ADRENALINTE testbed\cite{munoz2017adrenaline}. It demonstrates the use of SDN/NFV control system to provide the global orchestration of the multi-layer (packet/optical) network resources and network slicing based distributed cloud infrastructure for multi-tenancy.

Table~\ref{tab:Integrationtech} summarizes the reviewed state-of-the-art MEC-IoT integration technologies.

\begin{table*}
  \centering
        \caption{Comparison of the reviewed state-of-the-art MEC-IoT Integration Technologies.}
        \label{tab:Integrationtech}
  \begin{tabular}{|p{1cm}|p{8.1cm}|p{2cm}|p{0.9cm}|p{0.9cm}|p{0.9cm}|p{0.9cm}|}
    \hline
   	 \textbf{Ref.} & \textbf{Description} &  \textbf{IoT application OR domain} & \textbf{NFV}	& \textbf{SDN}&  \textbf{ICN} & \textbf{Network Slicing}   \\
  \hline
  
\cite{li2017mec}    & Present a NFV-enabled MEC  architecture for video streaming, gaming and AR & Gaming and AR. & \checkmark & & &\\ 
     \hline
     
     \cite{yang2016seamless}    & Present an double-tier MEC-NFV integrated architecture for 5G applications. & Gaming & \checkmark & & &\\ 
     \hline
     
      \cite{carella2017prototyping}    & Present an integrated orchestration solution by combining the NFV and MEC use cases within a single orchestration environment. & General IoT & \checkmark & & &\\ 
     \hline

\cite{sciancalepore2016double}   & Present a NFV-enabled MEC  framework for low latency mobile applications. & General IoT & \checkmark & & &\\ 
     \hline

     \cite{van2017unavoidable}   & Preset an network architecture to addresses some of the central convergence  challenges  of NFV, 5G/MEC, IoT, and fog.  & General IoT & \checkmark & & &\\ 
     \hline

\cite{peng2017qoe}   & Discuss the role of NFV and SDN in  MEC and IoT ecosystem.  & General IoT & \checkmark &  \checkmark & &\\ 
     \hline

\cite{ali2017real}   & Present SDN-MEC based  Real-time Heart Attack Mobile Detection Service (RHAMDS) by using smart watches.  & ehealth, WIoT & &  \checkmark & &\\ 
     \hline
     
     \cite{farristowards}   & Present SDN-NFV based  security  framework which can integration with existing IoT security mechanisms.  & General IoT & \checkmark &  \checkmark & &\\ 
     \hline
     
     \cite{huang2017low}   & Present SDN-based MEC framework for low latency applications  & VR and IoT Automotives. &  &  \checkmark & &\\ 
     \hline
     
      \cite{vilalta2016sdn}   & Present an SDN/NFV architecture to delivery of future 5G services across multiple technological and administrative networks.  & General IoT & \checkmark &  \checkmark & &\\ 
     \hline
     
       \cite{aggarwal2016securing}   & Present an conceptual approach to  provide security for  IoT systems by using SDN and edge computing.  & General IoT &  &  \checkmark & &\\ 
     \hline    
     
            \cite{nguyen2017simeca}    & Presnet an SDN-based IoT Mobile Edge Cloud Architecture (SIMECA) for future IoT applications.  & General IoT &  \checkmark &  \checkmark & &\\ 
     \hline   
     
            \cite{liu2017high}   & Present a four-tier architecture assisted by MEC and SDN for VANETs.  & IoT Automotive &  &  \checkmark & &\\ 
     \hline

\cite{hossain2017impact}   & Discuss the utilization of SDN and MEC to overcome the challenges of network densification.    & Smart homes &  &  \checkmark & &\\ 
  \hline 
\cite{phemius2016bringing}   & Present  an MEC and SDN based framework for efficient and flexible service delivery.     & Tactile Internet &  &  \checkmark & &\\ 
     \hline 
     
\cite{5GAmericaICNMEC}   & A white paper on opportunities and challenges of MEC and ICN integration for IoT.      & General IoT &  &   & \checkmark &\\ 
     \hline      
\cite{zhou2017video}   & Present an novel HetNets virtualization architecture for video trans-coding, caching, and multi-cast.     & VR,AR, Gaming, WIoT &  &   & \checkmark &\\ 
     \hline 
     
\cite{grewe2017information}   & Present the vision of combining ICN and MEC in the context of connected vehicle environments. & IoT Automotive &  &   & \checkmark &\\ 
     \hline
     
\cite{zhou2017resource}   & Present a novel information-centric heterogeneous networks framework for virtual resource allocation at the edge. & General IoT  &  &   & \checkmark &\\ 
     \hline
     
 \cite{huo2016software}   & Present a novel framework which jointly considers networking, caching, and computing techniques to support energy-efficient information retrieval and computing services. & General IoT  &  & \checkmark  & \checkmark &\\ 
     \hline
     
     \cite{ge2016qoe}   & Present a  content handling framework which realizes context-aware content localization  to enhance user QoE in video distribution applications. & VR,AR, Gaming, WIoT  &  &   & \checkmark &\\ 
     \hline
     
          \cite{ravindran20175g}   & Propose an 5G-ICN architecture  to realize an ICN-based service delivery  for future IoT applications. & General IoT  & \checkmark &  \checkmark & \checkmark &\\ 
     \hline
     
       \cite{nikaein2015network}   &  A discussion on use of network slicing for Massive IoT services.   & General IoT  &  &  &  &\checkmark\\ 
     \hline  
     
\cite{katsalis2017network}   & Propose an   novel network slicing architecture for integrated 5G  communications including IoT   & General IoT  &  &  &  &\checkmark\\ 
     \hline
     
           \cite{munoz2017adrenaline}   & Propose an  packet/optical transport network and edge/core cloud platform and testbed implementation for E2E 5G and IoT services. & General IoT  & \checkmark &  \checkmark &  &\checkmark\\ 
     \hline

  \end{tabular}
  \end{table*}


\section{Projects}
\label{sec:projects}


The European 5G PPP (5G Infrastructure Public Private Partnership) is one of the key layers on efforts to leverage MEC and IoT technologies to support the evolution towards 5G networks. In this section, we discuss some renowned ongoing EU research projects which are explicitly contributing to MEC and IoT technologies. These projects along with their technological aspects and the key research areas are summarized in Table~\ref{tab:projects}. Since the concept of MEC was initiated by ETSI, all of these projects are EU based. However, they have other non-EU partners as Japan, Taiwan, and China. The recent Horizon 2020~(H2020) funding scheme has fueled the MEC related research in Europe with the cooperation of other parts of the globe. Although, non-EU international level projects are hardly found on integrating MEC and IoT, the other countries have projects on different edge technologies including MCC, fog and cloudlets. We have excluded these projects from our survey since they are out of scope from the mainstream of the paper.

\subsubsection{SESAME: Small cEllS coordinAtion for Multi-tenancy and Edge services (June 2015 - Dec. 2017)}

SESAME~\cite{SESAME} is one of the front-line EU H2020 projects which focuses on the development and demonstration of an innovative architecture, capable of providing Small Cell~(SC) coverage to multiple virtual operators as-a-Service. This is a pioneering project that uses MEC and NFV technologies to realize the cloud-enabled small cell~(CESC) concept by supporting powerful self-x (x stands for organizing, optimizing, or healing) management features and executing novel applications and services inside the access network infrastructure. SESAME is expected to deliver the small cell concept in high dense 5G scenarios. Moreover, it intends to consolidate multi-tenancy in communications infrastructures. This allows several operators or service providers to engage in new sharing models of both access capacity and edge computing capabilities.

\subsubsection{ANASTACIA: Advanced Networked Agents for Security and Trust Assessment in CPS / IOT Architectures (Jan 2017 - Dec 2019)}
ANASTACIA~\cite{Anastacia}, an EU H2020 funded project which promises to develop and demonstrate a holistic solution enabling trust and security by-design for heterogeneous, distributed and dynamically evolving CPS based on IoT and virtualised cloud architectures. The security framework, with self-protection, self-healing, and self-repair capabilities,  will be designed in full compliance to SDN/NFV standards. This will include the security development paradigm, distributed trust and security enabler, and dynamic security and privacy seal. In particular ANASTACIA will address the security challenges in two use cases on the deployment of MEC server and  smart buildings.

\subsubsection{5G-MiEdge: Millimeter-wave Edge Cloud as an Enabler for 5G Ecosystem (July 2016 - June 2019)}
5G-MiEdge~\cite{miEdge} is a publicly supported research project bringing Millimeter-Wave (mmWave) technology and MEC into the mobile radio world. It was co-funded by EU H2020 and Japanese government. It combines mmW access/backhauling with MEC to enable enhanced mobile broadband~(eMBB) services and mission critical low-latency applications using cost-efficient RANs. The project is composed of three key technologies; naming the protocols of mmWave access/backhaul links, ultra-lean and inter-operable control signaling mechanism (liquid RAN C-plane) over 3GPP LTE, and user or application centric orchestration algorithms for edge resource allocation. 5G-MiEdge intends to develop transmission schemes and protocols of mmWave access/backhauling which can assist the mobile edge cloud with caching/prefetching. This will be useful in realizing ultra-high speed and low latency service delivery which will be resilient to network bottlenecks such as backhaul congestion, users' density, and mission-critical service deployments. The targeted use cases are mostly stadiums, offices, and train stations.

\subsubsection{5G!Pagoda}
5G!Pagoda project~\cite{5GPagoda} aims at creating a virtual mobile network that can be deployed upon request, dedicated to an application, to be used during the Tokyo Olympic Games in 2020. 5G!Pagoda intends to develop a scalable 5G slicing architecture and a highly programmable network control and data path supporting mechanism for use cases in IoT and human communication. This would be achievable through the development of a scalable network slice management and orchestration frameworks.  These frameworks would serve distributed, edge dominated network infrastructures and convergent software functionality for lightweight control plane and data plane programmability.

\subsubsection{Inter-IoT (Jan. 2016 - Dec. 2018)}
Horizon 2020 EU project INTER-IoT project~\cite{InterIoT} aims to design, implement and test an open framework that will allow interoperability among different IoT platforms. The project uses a layer-oriented approach for the interoperability framework in four application domains: smart grid, e-health, smart factories, and transport-logistics. The final goal is to integrate different IoT devices, networks, platforms, services and applications that will allow a global continuum of data, infrastructures and services which can enable different IoT use cases.

\subsubsection{5G-MoNArch: 5G Mobile Network Architecture for diverse services, use cases, and applications in 5G and beyond (July 2017 - June 2019)}
5G-MoNArch~\cite{5g-monarch} is another project funded by EU Horizon 2020 programme and it will evolve 5G-PPP Phase 1 concepts to a fully-fledged architecture, develop prototype implementations and apply these prototypes to representative use cases. 5G-MoNArch’s specific technical goal is to use network slicing, which capitalizes on the capabilities of SDN, NFV, orchestration of access network and core network functions, and analytics, to support a variety of use cases in vertical industries such as automotive, healthcare, and media. The devised 5G-MoNArch architecture will be deployed in two test beds: a sea port and a tourist city.

\subsubsection{5G-ESSENSE: Embedded Network Services for 5G Experiences (June 2017 - June 2019)}
5G ESSENCE~\cite{5gEssence} is an EU H2020 funded project that  proposes a highly flexible and scalable 5G small cell platform leveraging the paradigms of edge cloud computing and Small-Cell-as-a-Service. ESSENCE builds virtualization techniques on the distributed and network-integrated cloud inherited by 5G-PPP Phase 1 SESAME project that provides processing power at the edge of the network. The project will explicitly address two use cases including in-flight entertainment and connectivity systems and mission critical applications for public safety. 

\subsubsection{MATILDA (June 2017 - June 2019)}
The EU H2020 funded 5G-PPP Phase 2 project, MATILDA~\cite{Matilda}, aims to design and implement a holistic 5G framework for the design, development and orchestration of 5G-ready applications and 5G network services over a sliced, programmable infrastructure using VNFs. Intelligent and unified orchestration mechanisms will be applied for the automated placement of the 5G-ready applications and the creation and maintenance of the required network slices. The management of the cloud/edge computing and IoT resources is supported by a multi-site virtualized infrastructure manager.

\subsubsection{5GCITY (June 2017 - June 2019)}
5GCity~\cite{5GCity} is also an EU H2020 funded 5G-PPP Phase 2 project which demonstrates how to empower the city infrastructure and transform them into a hyper-connected, distributed 5G-enabled edge virtualization domain. The project targets three different cities (Barcelona, Bristol and Lucca), and would benefit telecommunication infrastructure providers, municipalities, and a number of different vertical sectors utilizing the city infrastructure. It will leverage the virtualization platform in order to enable the cities to create dynamic E2E slices containing both virtualized edge and network resources and lease to third-party operators.

\subsubsection{MONICA: Management Of Networked IoT Wearables – Very Large Scale Demonstration of Cultural and Societal Applications (Jan 2017 - Dec 2019)}

MONICA~\cite{MONICA} is an EU H2020 funded large scale pilot project which aims to provide a very large scale demonstration of multiple existing and new IoT technologies for smarter living.
It demonstrates a large scale IoT ecosystem that uses innovative wearable and portable IoT sensors and actuators with closed-loop back-end services integrated into an interoperable, cloud-based platform capable of offering a multitude of simultaneous, targeted applications. The key objectives of this project are to strengthen crowd safety and security at at big, cultural, open-air events, and improve user experience. Given these goals, the final solution should be compatible with many different IoT sensors, open source, with cost effective wearables, and strengthened with data security, privacy, and trust.

\subsubsection{AUTOPILOT: AUTOmated driving Progressed by Internet Of Things (Jan 2017 - Dec 2019)}
Another large scale pilot project funded by EU H2020, AUTOPILOT~\cite{Autopilot} will deploy, test and demonstrate IoT-based automated driving use cases comprising urban driving, highway pilot, automated valet parking, and platooning. The project will integrate into vehicle IoT sensors and use cloud and MEC type IoT platforms (e.g., Brainport pilot site in Netherlands) to share sensor data and create new autonomous mobility services. The AUTOPILOT project will create and deploy new business products and services for fully automated driving vehicles used at the pilot sites. This project will feature innovations such as driving route optimization, vulnerable road user sensing and dynamically updating an IoT based HD map.

\subsubsection{5G-CORAL: A 5G Convergent Virtualised Radio Access Network Living at the Edge (Sep. 2017 - Aug. 2019) }
The newly initiated EU H2020 project, 5G-CORAL~\cite{5Gcoral} leverages on the pervasiveness of edge and fog computing in RAN to create a unique opportunity for access convergence. This is envisioned by the means of virtualised networking and computing solution where virtualised functions, context-aware services, and user and third-party applications are blended together to offer enhanced connectivity and better quality of experience. The proposed solution considers two major building blocks, namely the edge and fog computing system and the orchestration and control system. 5G-CORAL project will be validated in three testbeds; a shopping mall, high-speed train, and connected cars. 

\begin{table*}
  \centering
        \caption{Contribution of global level ongoing projects on MEC and IoT. Todo: Shall we remove mmWave here. We did not discuss that a lot}
        \label{tab:projects}
  \begin{tabular}{|l|c|c|c|c|c|c|c|c|c|c|c|c|}
  \hline
      	\textbf{Project}
         &{\rotatebox[origin=c]{90}{~SESAME~\cite{SESAME}~}}
         &{\rotatebox[origin=c]{90}{~ANASTACIA~\cite{Anastacia}~}}
         &{\rotatebox[origin=c]{90}{5G Mi-Edge~\cite{miEdge}}}
         &{\rotatebox[origin=c]{90}{5G!Pagoda~\cite{5GPagoda}}}
         &{\rotatebox[origin=c]{90}{Inter-IoT~\cite{InterIoT}}}
         &{\rotatebox[origin=c]{90}{5G-MoNArch~\cite{5g-monarch}}}
         &{\rotatebox[origin=c]{90}{5G-ESSENSE~\cite{5gEssence}}}
         &{\rotatebox[origin=c]{90}{MATILDA~\cite{Matilda}}}
         &{\rotatebox[origin=c]{90}{5GCITY~\cite{5GCity}}}
         &{\rotatebox[origin=c]{90}{MONICA~\cite{MONICA}}}
         &{\rotatebox[origin=c]{90}{AUTOPILOT~\cite{Autopilot}}}
         &{\rotatebox[origin=c]{90}{5G-CORAL~\cite{5Gcoral}}}\\

  \hline
    \textbf{Technologies}   & & &	& & & & & & & & & \\
\hline
~~MEC                  & \checkmark  & \checkmark & \checkmark	&\checkmark &  &\checkmark & \checkmark & \checkmark & \checkmark & \checkmark & \checkmark & \checkmark\\
\hline
~~IoT                   & & \checkmark & \checkmark &\checkmark & \checkmark & \checkmark & \checkmark & \checkmark & \checkmark & \checkmark & \checkmark & \checkmark \\
\hline
~~SDN                   &\checkmark & \checkmark &	\checkmark &\checkmark & &\checkmark & \checkmark & \checkmark & \checkmark & & & \\
\hline
~~NFV                    &\checkmark & \checkmark &	&\checkmark & &\checkmark & \checkmark & \checkmark & \checkmark & & & \checkmark \\
\hline
     ~~Network slicing        &\checkmark & &	&\checkmark &  &\checkmark & \checkmark & \checkmark & \checkmark & & & \\
\hline
~~mmWave             			&& &\checkmark	& & & & & & & &  & \\
\hline
     \textbf{Research focus} &  & &	& & & & & & & & & \\
\hline
~~Network architecture OR framework		& \checkmark & \checkmark  & & \checkmark & \checkmark & \checkmark & \checkmark & \checkmark & \checkmark & \checkmark & \checkmark & \\
\hline
    ~~Communication and network infrastructure  	& \checkmark &  & \checkmark 	& & & & \checkmark &\checkmark & & & & \checkmark \\
\hline
~~Computation offloading &\checkmark & 			 & 				& & & & & & & & & \checkmark \\
\hline
~~Resource management  & \checkmark & & & & & \checkmark & \checkmark & \checkmark & \checkmark & & & \\
\hline
    ~~Mobility   			&	& 			 & 				& & & & & & \checkmark & \checkmark & \checkmark & \\
\hline
~~Scalability   		&	& 			 & 				& \checkmark & & & & & \checkmark & \checkmark & & \\
\hline
~~Interoperability   	&		& \checkmark	& 	&  & \checkmark & & & & & & & \\
\hline
~~Security   			&	& \checkmark & 	&  & & \checkmark & & & \checkmark & \checkmark& & \\
\hline
~~Privacy   			&	& \checkmark &		& & & & & & \checkmark & \checkmark & & \\
\hline
~~Trust   			&	& \checkmark &				& & & & & & & \checkmark & & \\
  \hline
  \end{tabular}
  \end{table*}


\section{Lessons Learned and Future Research Directions}
\label{sec:lessons_learned}
In this section, we present the lessons learned and the future research directions with respect to MEC-IoT integration. In particular, we focus on MEC-IoT application paradigms, technical aspects (i.e., scalability, communication, computation offloading and resource allocation, mobility management, security, privacy, and trust management), and standardization efforts. 

\subsection{Applications}
\subsubsection{Lessons learned}
MEC is an ideal solution that supports the increased demand for bandwidth consumption and ultra low latency requirements of IoT applications. MEC resources can be utilized for the pre-processing of massive IoT data which will reduce bandwidth consumption, provide network scalability, and ensure a fast response to user requests. However, in order to reap the maximum benefits of MEC for IoT, there needs to be more in dept research on how to efficiently distribute and manage data storage and computing resources at the network edge. Since MEC is still not well established, there can be myriad of technical challenges that need to be addressed. Moreover, due to much unprecedented user expectations, the requirements for designing MEC systems may vary upon the IoT application area.  

\subsubsection{Future research directions}
The applications described in Section~\ref{sec:IoT and MEC applications} are overlapping in several ways. For instance, AR and VR may explicitly support autonomous driving by exchanging information derived from multi-resolution maps created using the local sensors of the vehicles. This will extend the visibility of the vehicle. The edge servers are expected to perform pro-actively in such AR and VR systems.
Tele-surgery is another domain that takes advantage of AR and VR exploitation. In the ideal situation, VR should have no distinction between real and virtual worlds. In order to achieve this goal, the concepts of MEC in VR applications might be merged with concepts like quantum computing.  It is reported that ETSI and Virtual Reality/Augmented Reality Association (VRARA) intend to collaborate on interactive VR and AR technologies delivered over emerging 5G networks and hosted on MEC sites~\cite{ETSIVRARA}. VRARA will encourage common member companies to pursue VR/AR-focused use cases and requirements for ETSI MEC Phase 2.

The adoption of machine learning techniques in 5G networks has increasingly attracted the attention of the research community. This will provide adaptive learning and decision-making approaches to meet the requirements of different verticals. The integration of Artificial Intelligence~(AI) algorithms and machine learning at the edge of the networks will further assist the data-intensive requirements of the IoT applications. Particularly, AI techniques can be exploited for adaptive, optimal, and pro-active action on instantaneous networking demand in vehicular communications, in the context of self-driving vehicles. However, more efforts are needed to adopt machine learning techniques such as recursive neural networks, reservoir computing and deep learning in autonomous vehicles kind of applications due to their complex network architecture and enormous data sets. More importantly there is no unifying theories to define how such a network will behave.

\subsection{Scalability}

\subsubsection{Lessons learned}
Several aspects of the present-day scalability schemes and data management paradigms will need substantial refinement in order to be able to handle the changes that are expected in future MEC-enabled IoT networks. IoT devices like sensors and RFID capturing devices are expected to keep capturing objects almost in real-time, hence generating a huge amount of readings. Timeliness is another factor in such scenarios since generated data usually have very short life-span of about 2 seconds. Obviously, the present-day approach to information search and data management cannot handle this expectation in a scalable manner. For this reason a more refined search and indexing algorithm will be required for both MEC-enabled IoT applications and IoT systems in general.


\subsubsection{Future research directions}
The adoption of the IPv6 is a significant move that will further advance scalability in MEC-enabled IoT applications going forward. In \cite{liu2014concert}, authors proposed the idea of CONCERT, a term coined from the combination of cloud and cellular system. The CONCERT solution exploits the principles of NFV and SDN to enhance scalability in future networks. Since scalability is a huge factor to determine where the MEC server gets deployed, and since the devices exploiting the MEC server located in the core network will inevitably experience longer latencies, then there could be a major hindrance to the use of real-time applications in such MEC settings. Regarding control signaling in MEC, the proposed CONCERT approach also adopts either a fully centralized control or a hierarchical control for better scalability and flexibility.  


\subsection{Communication}
\subsubsection{Lessons learned}
As MEC is still at its infancy, defining a solid communication model for the entire MEC architecture is an open research question that paves many opportunities to the academia, industry and the standardization entities. Advanced wireless communication techniques are required to design for interference cancellation and adaptive power control at the MEC servers in order to reduce the offloading energy consumption in a significant manner. The tight alliance between MEC and IoT may also create new research challenges in communication perspective. 

\subsubsection{Future research directions}
As pointed out by Raza et. al. in~\cite{raza2017low}, interoperability among various IoT LPWAN technologies encountered in IoT is still an open research question to address. There are still insufficient testbeds and open-source tool chains for LPWAN technologies. Massive connectivity and high data rate requirements of IoT devices (e.g., wearables) can be fulfilled by accompanying new radio access technologies such as Non-Orthogonal Multiple Access~(NOMA) and massive Multiple-Input-Multiple-Output~(MIMO)~\cite{sun2017challenges}.

Moreover, many research efforts on edge caching are underway to achieve the trade-off between the transmission rate and storage at the MEC hosts~\cite{peng2016fog}. The co-existence of different wireless communication technologies available for IoT may still create many challenges for edge level accessibility, since the IoT applications are diversified in versatile areas, where each has a unique set of requirements. Furthermore, they have conflicting goals such as energy efficiency, high throughput, and wide coverage. Therefore, system-level research is required to reap out the maximum benefit on exploiting such communication technologies. 

Implementing MEC over FiWi access networks are investigated due to their low costs, wide deployments, and high capacity~\cite{rimal2017mobile}. These fiber-wireless broadband access networks may provide a single communication platform for MEC and centralized cloud services over the wired and wireless networking technologies. ICN in combination with MEC is identified as another promising way of establishing a communication model for vehicular networks~\cite{grewe2017information} where moving vehicles may incur frequent disconnects and re-connects to different network access points. 

\subsection{Computation Offloading and Resource Allocation}
\subsubsection{Lessons learned}
Decision making for data offloading at the user-end devices and the resource allocation for those offloaded data/application at the edge clouds are two highly regarded topics discussed among the research community, especially  those who engaged in MEC and IoT eras. Most of the prior works were focused on the offloading mechanisms for latency critical applications while minimizing energy consumption at the UE. In contrary, IoT permits a platform that has both delay sensitive and delay tolerant applications. Although, most of the proposed solutions are evaluated by means of theoretical analysis or simulations, there is still no proper formation of standard offloading mechanism for IoT and MEC systems. 

\subsubsection{Future research directions}
Mobility is a principal feature of IoT devices which are either being transported by humans (e.g. wearable sensor) or by another carrier (e.g. vehicular networks), or being mobile by itself (e.g. robots). Mobility-aware resource management and computation offloading strategies need to be precisely investigated in the era of IoT supportive MEC systems. Scalability is the other equally important feature to consider in large scale IoT deployments where edge computing needs seamless offloading and resource allocation policies. Other accelerating tendencies towards future research efforts in the field of MEC and IoT may include server cooperation in MEC, dependency-aware offloading, and dynamic resource allocation.

The exploitation of Knowledge-Defined Networking~(KDN) to make intelligent predictions about offload costs can be leveraged for efficient resource allocation at MEC servers as well as the offloading decision making at IoT devices~\cite{mestres2017knowledge}. The new paradigm of KDN is composed of Network Analytics~(NA), SDN, and AI techniques.
The introductory work in~\cite{crutcher2017hyperprofile} proposes an intelligent computation offloading framework based on user dynamics and historical data. 

\subsection{Mobility Management}
\subsubsection{Lessons learned}
Mobility management in MEC-enabled IoT has attracted a lot of attention in both research and the industry. This comes natural, given that mobile nodes are expected to dominate the future IoT networks. An optimal offloading decision will be necessary for effective integration of MEC with IoT. Thus far, most of the works on mobility management in the context of MEC are solely focusing on optimizing the energy consumption at IoT nodes. However, designing efficient and optimal MEC-enabled IoT systems will require energy optimization at the MEC end also. This includes energy consumed on computation and energy consumed on communication. 

Furthermore, most works on offloading decisions are based on static scenarios where the IoT device moves from one MEC eNB to another and remains in one steady location during the offload, which is not necessarily the situation in most cases. 

\subsubsection{Future research directions}
The energy required for offloading or handover could vary substantially based on the movement factor during the offload\cite{taleb2013analytical}. For this reason, there will be a need for more advanced decision making algorithms. They will leverage on various prediction techniques to determine when offloading is in fact necessary, what the channel quality will be like during the offloading and what the entire offloading process will cost for each offloading condition.     


For advancing the VM migration techniques, a crucial step moving forward is to optimize the migration process by minimizing the time required to complete a full migration. This will mostly dependent on the protocol design of the migration process.
Hence an optimal solution is required for a collaborative effort on the side of individuals and organizations. That notwithstanding, still the VM migration scheme might not be suitable for highly delay-sensitive real-time applications. In general, to achieve an efficient and highly optimized mobility management scheme for MEC-enabled IoT applications, there will be a need for a more holistic approach. Such a solution will encompass power control, VM migration, data compression, and path selection\cite{mach2017mobile}. 

\subsection{Security}
\subsubsection{Lessons learned}
Notwithstanding the closed paradigm of MEC, it is important to realize that the whole ecosystem of MEC will not be controlled by one single owner or service provider. MEC data centers are capable of providing services without relying on centralized infrastructures. Thus, it is certain that all MEC relevant assets, such as the network infrastructure, the service infrastructure (e.g. edge data centers, core infrastructure), the virtualization infrastructure, and the user devices will not be controlled by a single entity. The scale of this effect is further confounded by the diversity that exists in IoT applications. Consequently, every element of MEC and IoT infrastructure should be targeted towards global networking environment. As discussed in \cite{roman2016mobile}, the \textquotedblleft anything, anytime" principle should be the underlying building blocks and application scenarios for MEC-enabled IoT systems\cite{roman2013features}. Conversely, the \textquotedblleft anywhere" principle also implies that attacks can be performed from anywhere, making the edge paradigms a double-edged sword and hence the need for security measures that span the entire global networking paraphernalia.

\subsubsection{Future research directions}
The future of MEC-enabled IoT systems will revolve mostly around developing universal standard security mechanisms that can adequately protect the whole ecosystem against security threats. Such universal standards will enable both service providers and developers to understand the particularities of every edge paradigm, as they have subtle differences that will affect the implementation and deployment of the security mechanisms\cite{roman2016mobile}. Currently, the absence of such global perimeters is seen as one of the banes to the security of the edge paradigms.
\par One notable effect of the lack of a global perimeter is the nature of the different attacker profiles that will target edge paradigms~\cite{ahmed2017bringing}. In the present day networks, adversaries are mostly external entities with no stake in control of network elements. However, with the advent of MEC-enabled IoT, there exist many adversaries that will control one or more elements of the infrastructure such as user devices, VMs, servers, sections of the network, and in the worst case, an entire edge data center\cite{roman2013features}. Adopting deep-learning-based models at the edge level to detect malicious applications will be another interesting research area. Applying reinforcement learning techniques to develop edge security solutions can be exploited for anomaly detection and lightweight authentication.

\subsection{Privacy}
\subsubsection{Lessons learned}The rise of  new architecture, new technologies and new network services will open up new challenges to privacy protection. On the one hand, the existing privacy objectives are outdated and are not compatible with current technologies such as MEC, IoT and 5G. Therefore, these privacy directives have to be updated. Governing organizations have already started redefining the privacy objectives. For instance, the European commission adopted  a General Data Protection Regulation~(GDPR), in April 2016. It will be superseded by the data protection directive and is planned to be enforceable starting on 25 May 2018. On the other hand, privacy awareness is significantly increasing among the general public users\cite{sorensen20155g}. Therefore, the future networks require to provide an extra level of privacy than the earlier generation of networks. 

\subsubsection{Future research directions}The future research work should be focused on addressing above privacy challenges. New privacy protection mechanisms such as Software Defined Privacy (SDP)\cite{kemmer2016software}, Privacy by Design (PbD)\cite{cavoukian2014regulartor} and SDN based privacy-aware routing\cite{pitt2013trust} can be used to provide the required level of privacy while or after the integration of MEC to IoT systems. SDP\cite{kemmer2016software} allows easy orchestrations of existing tools for enforcing privacy requirements of an Infrastructure as a Service (IaaS) cloud customer. This concept can further be extended to provide privacy protection for MEC enabled IoT systems. PbD is an approach in system engineering, which promotes the integration of privacy throughout the whole design process\cite{cavoukian2014regulartor}. PbD approach can be used during the MEC integration in IoT systems. If SDN is used in MEC-IoT systems, which is highly likely, user data packets containing privacy information that should not cross local spaces or even  country borders could be identified. Then, the SDN controller could define flow rules so that these packets are routed only via the links and routers with high security. More sophisticated routing protocols can be designed by increasing the number of such qualifiers. 

\subsection{Trust Management}
\subsubsection{Lessons learned}
Trust management in MEC systems is still a barely investigated area. In order to strengthen the user ecosystem in centralized cloud environment, a flexible trust manager can be shared among the cloud infrastructure providers~\cite{aikat2017rethinking}. Likewise, the mutual trust should be incorporated among the MEC servers to enhance the secure sharing of IoT datasets. 

\subsubsection{Future research directions}
Context-aware trust relationships based on social computing are yet to be investigated in the paradigm of IoT and edge computing. A comprehensive trust framework is still lacking for holistic trust management in IoT with the context of MEC which is capable of achieving all the objectives listed above and fulfills the requirements from different trust levels. Future research needs to focus on data collecting at IoT perception layer and processing at edge servers in order to improve the IoT and MEC service quality. Complex and resource consuming trust management algorithms are not affordable by the tiny IoT devices. Furthermore, device and network heterogeneity in IoT raises further challenges. There are also some open research trends for making light-weight trust management mechanisms suitable for heterogeneous IoT.

\subsection{Standardization}
The standardization of the MEC technology is relatively recent and currently ongoing. The goal is to bring together all experts and industry players in consensus to define the characteristics and rules that will govern the implementation and interconnection of the MEC technology globally. Just like other standardized technologies, the standardization of MEC will open up an infinite avenue for developers and innovators to harness the benefits of MEC in designing cutting-edge technologies and innovative solutions that will drive 5G and future networks. On the side of the customers, such standardization would by no small measure affirm their trust in MEC and other related products and services.

\subsubsection{Future research directions}
The standardization processes of MEC along with the coordination and management tasks are lead by an ETSI ISG~\cite{etsimec}. The MEC ISG group aims at creating an open standardized and efficient platform for the seamless integration of enterprise applications from different vendors and service providers into the MEC platform. Most recently, the 3GPP has shown a growing interest in incorporating MEC into its 5G standard and has identified functionality supports for edge computing in a recent technical specification contribution. 
 
The standardization entities are required to ensure that MEC architecture works harmoniously with the heterogeneous IoT echo systems and related technologies. Moreover, since there are numerous third-party partners such as application developers, content providers and network device vendors, the complexity of the services and the management of very large scale environment becomes challenging~\cite{li2016mobile}.

It is also important to do security and privacy legislation and standardization in a global context. Different jurisdictions should cooperate together to develop inter-operable security and privacy requirements to  facilitate the flow of information with the required level of  protection. Thus, the security and privacy regulations will play a vital role to promote the adaptations new technologies such as MEC. Regulatory entities such as governments and standardization organizations have to work together with industry to define and/or update the regulations according to the new technologies.




\section{Conclusions}
\label{sec:Discussion}
The advancements of MEC and IoT technologies will be contributing immensely to the realization of the highly anticipated game-changing vision of 5G and future generations of mobile networks. The propounders of MEC; which is relatively a recent technology, have identified IoT as one of the important use cases of MEC. MEC server performs as a gateway between the latency critical and massive IoT networks and the core network where it can provide edge-cloud computing and networking functionalities. IoT application domains are empowered with MEC technology by extending some intelligence to the edge of the network. Although MEC will provide on-site cloud computing services for IoT networks, there are still challenges in terms of device and network heterogeneity, scalability, mobility, and security. In addition to the possible future works discussed in Section~\ref{sec:lessons_learned}, there are few other research topics including but not limited to MEC service level congestion control, latency aware routing, and dynamic application routing. In all essence, MEC and IoT are two complementary technologies that if well harnessed have the potential of advancing the course of the 5G networks and beyond. 

\section*{Acknowledgment}

This work has been performed under the framework of the Infotech Doctoral Program of UniOGS and the three projects, 6Genesis Flagship (grant 318927), SECUREConnect (Secure Connectivity of Future Cyber-Physical Systems) and Towards Digital Paradise. This research is funded by Academy of Finland and TEKES, Finland.


%





\ifCLASSOPTIONcaptionsoff
  \newpage
\fi

\bibliographystyle{IEEEtran} 
\bibliography{bibtex/bib/MECIoTSurvey}

\begin{thebibliography}{100}
\providecommand{\url}[1]{#1}
\csname url@samestyle\endcsname
\providecommand{\newblock}{\relax}
\providecommand{\bibinfo}[2]{#2}
\providecommand{\BIBentrySTDinterwordspacing}{\spaceskip=0pt\relax}
\providecommand{\BIBentryALTinterwordstretchfactor}{4}
\providecommand{\BIBentryALTinterwordspacing}{\spaceskip=\fontdimen2\font plus
\BIBentryALTinterwordstretchfactor\fontdimen3\font minus
  \fontdimen4\font\relax}
\providecommand{\BIBforeignlanguage}[2]{{%
\expandafter\ifx\csname l@#1\endcsname\relax
\typeout{** WARNING: IEEEtran.bst: No hyphenation pattern has been}%
\typeout{** loaded for the language `#1'. Using the pattern for}%
\typeout{** the default language instead.}%
\else
\language=\csname l@#1\endcsname
\fi
#2}}
\providecommand{\BIBdecl}{\relax}
\BIBdecl

\bibitem{IoTdefinition}
\BIBentryALTinterwordspacing
P.~Guillemin and P.~Friess, ``{The Industrial Internet of Things Volume G1:
  Reference Architecture,},'' The Cluster of European Research Projects, Tech.
  Rep.,, September 2009. [Online]. Available:
  \url{http://www.internet-of-things-research.eu/pdf/IoT_Cluster_Strategic_Research_Agenda_2009.pdf}
\BIBentrySTDinterwordspacing

\bibitem{hu2015mobile}
Y.-C. Hu, M.~Patel, D.~Sabella, N.~Sprecher, and V.~Young, ``{Mobile Edge
  Computing—A Key Technology Towards 5G},'' \emph{ETSI White Paper}, vol.~11,
  no.~11, pp. 1--16, 2015.

\bibitem{schulz2017latency}
P.~Schulz, M.~Matthe, H.~Klessig, M.~Simsek, G.~Fettweis, J.~Ansari, S.~A.
  Ashraf, B.~Almeroth, J.~Voigt, I.~Riedel \emph{et~al.}, ``{Latency Critical
  IoT Applications in 5G: Perspective on the Design of Radio Interface and
  Network Architecture},'' \emph{IEEE Communications Magazine}, vol.~55, no.~2,
  pp. 70--78, 2017.

\bibitem{taleb2017multi}
T.~Taleb, K.~Samdanis, B.~Mada, H.~Flinck, S.~Dutta, and D.~Sabella, ``{On
  Multi-Access Edge Computing: A Survey of the Emerging 5G Network Edge
  Architecture \& Orchestration},'' \emph{Communications Surveys \& Tutorials},
  vol.~19, no.~3, pp. 1657--1681, 2017.

\bibitem{atzori2010internet}
L.~Atzori, A.~Iera, and G.~Morabito, ``{The Internet of Things: A Survey},''
  \emph{Computer networks}, vol.~54, no.~15, pp. 2787--2805, 2010.

\bibitem{gubbi2013internet}
J.~Gubbi, R.~Buyya, S.~Marusic, and M.~Palaniswami, ``{Internet of Things
  (IoT): A Vision, Architectural Elements, and Future Directions},''
  \emph{Future generation computer systems}, vol.~29, no.~7, pp. 1645--1660,
  2013.

\bibitem{al2015internet}
A.~Al-Fuqaha, M.~Guizani, M.~Mohammadi, M.~Aledhari, and M.~Ayyash, ``{Internet
  of Things: A Survey on Enabling Technologies, Protocols, and Applications},''
  \emph{IEEE Communications Surveys \& Tutorials}, vol.~17, no.~4, pp.
  2347--2376, 2015.

\bibitem{gazis2017survey}
V.~Gazis, ``{A Survey of Standards for Machine-to-Machine and the Internet of
  Things},'' \emph{IEEE Communications Surveys \& Tutorials}, vol.~19, no.~1,
  pp. 482--511, 2017.

\bibitem{weyrich2016reference}
M.~Weyrich and C.~Ebert, ``{Reference Architectures for the Internet of
  Things},'' \emph{IEEE Software}, vol.~33, no.~1, pp. 112--116, 2016.

\bibitem{sicari2015security}
S.~Sicari, A.~Rizzardi, L.~A. Grieco, and A.~Coen-Porisini, ``{Security,
  Privacy and Trust in Internet of Things: The Road Ahead},'' \emph{Computer
  Networks}, vol.~76, pp. 146--164, 2015.

\bibitem{granjal2015security}
J.~Granjal, E.~Monteiro, and J.~S. Silva, ``{Security for the Internet of
  Things: A Survey of Existing Protocols and Open Research Issues},''
  \emph{IEEE Communications Surveys \& Tutorials}, vol.~17, no.~3, pp.
  1294--1312, 2015.

\bibitem{porambage2016quest}
P.~Porambage, M.~Ylianttila, C.~Schmitt, P.~Kumar, A.~Gurtov, and A.~V.
  Vasilakos, ``{The Quest for Privacy in the Internet of Things},'' \emph{Cloud
  Computing}, vol.~3, no.~2, pp. 36--45, 2016.

\bibitem{yan2014survey}
Z.~Yan, P.~Zhang, and A.~V. Vasilakos, ``{A Survey on Trust Management for
  Internet of Things},'' \emph{Journal of network and computer applications},
  vol.~42, pp. 120--134, 2014.

\bibitem{atzori2012social}
L.~Atzori, A.~Iera, G.~Morabito, and M.~Nitti, ``{The Social Internet of Things
  (SIoT)--when social networks meet the internet of things: Concept,
  architecture and network characterization},'' \emph{Computer networks},
  vol.~56, no.~16, pp. 3594--3608, 2012.

\bibitem{raza2017low}
U.~Raza, P.~Kulkarni, and M.~Sooriyabandara, ``{Low Power Wide Area Networks:
  An Overview},'' \emph{Communications Surveys \& Tutorials}, vol.~19, no.~2,
  pp. 855--873, 2017.

\bibitem{perera2014context}
C.~Perera, A.~Zaslavsky, P.~Christen, and D.~Georgakopoulos, ``{Context Aware
  Computing for the Internet of Things: A Survey},'' \emph{Communications
  Surveys \& Tutorials}, vol.~16, no.~1, pp. 414--454, 2014.

\bibitem{stankovic2014research}
J.~A. Stankovic, ``{Research Directions for the Internet of Things},''
  \emph{Internet of Things Journal}, vol.~1, no.~1, pp. 3--9, 2014.

\bibitem{lin2017survey}
J.~Lin, W.~Yu, N.~Zhang, X.~Yang, H.~Zhang, and W.~Zhao, ``{A Survey on
  Internet of Things: Architecture, Enabling Technologies, Security and
  Privacy, and Applications},'' \emph{Internet of Things Journal}, 2017.

\bibitem{miorandi2012internet}
D.~Miorandi, S.~Sicari, F.~De~Pellegrini, and I.~Chlamtac, ``{Internet of
  Things: Vision, Applications and Research Challenges},'' \emph{Ad Hoc
  Networks}, vol.~10, no.~7, pp. 1497--1516, 2012.

\bibitem{atzori2017understanding}
L.~Atzori, A.~Iera, and G.~Morabito, ``{Understanding the Internet of Things:
  Definition, Potentials, and Societal Role of a Fast Evolving Paradigm},''
  \emph{Ad Hoc Networks}, vol.~56, pp. 122--140, 2017.

\bibitem{sabella2016mobile}
D.~Sabella, A.~Vaillant, P.~Kuure, U.~Rauschenbach, and F.~Giust,
  ``{Mobile-edge Computing Architecture: The Role of MEC in the Internet of
  Things},'' \emph{Consumer Electronics Magazine}, vol.~5, no.~4, pp. 84--91,
  2016.

\bibitem{garcia2015edge}
P.~Garcia~Lopez, A.~Montresor, D.~Epema, A.~Datta, T.~Higashino, A.~Iamnitchi,
  M.~Barcellos, P.~Felber, and E.~Riviere, ``{Edge-centric computing: Vision
  and challenges},'' \emph{SIGCOMM Computer Communication Review}, vol.~45,
  no.~5, pp. 37--42, 2015.

\bibitem{shahzadi2017multi}
S.~Shahzadi, M.~Iqbal, T.~Dagiuklas, and Z.~U. Qayyum, ``{Multi-access Edge
  Computing: Open Issues, Challenges and Future Perspectives},'' \emph{Journal
  of Cloud Computing}, vol.~6, no.~1, p.~30, 2017.

\bibitem{ahmed2017mobile}
E.~Ahmed and M.~H. Rehmani, ``{Mobile Edge Computing: Opportunities, Solutions,
  and Challenges},'' \emph{Future Generation Computer Systems}, 2017.

\bibitem{ai2017edge}
Y.~Ai, M.~Peng, and K.~Zhang, ``{Edge Cloud Computing Technologies for Internet
  of Things: A Primer},'' \emph{Digital Communications and Networks}, 2017.

\bibitem{abbas2017mobile}
N.~Abbas, Y.~Zhang, A.~Taherkordi, and T.~Skeie, ``{Mobile Edge Computing: A
  Survey},'' \emph{IEEE Internet of Things Journal}, vol.~5, no.~1, pp.
  450--465, 2018.

\bibitem{ahmed2016survey}
A.~Ahmed and E.~Ahmed, ``{A Survey on Mobile Edge Computing},'' in \emph{10th
  IEEE International Conference on Intelligent Systems and Control (ISCO)},
  2016, pp. 1--8.

\bibitem{beck2014mobile}
M.~T. Beck, M.~Werner, S.~Feld, and T.~Schimper, ``{Mobile Edge Computing: A
  Taxonomy},'' in \emph{Proc. of the Sixth International Conference on Advances
  in Future Internet}, 2014, pp. 48--55.

\bibitem{mach2017mobile}
P.~Mach and Z.~Becvar, ``{Mobile Edge Computing: A Survey on Architecture and
  Computation Offloading},'' \emph{IEEE Communications Surveys \& Tutorials},
  vol.~19, no.~3, pp. 1628--1656, 2017.

\bibitem{baktir2017can}
A.~C. Baktir, A.~Ozgovde, and C.~Ersoy, ``{How Can Edge Computing Benefit from
  Software-Defined Networking: A Survey, Use Cases \& Future Directions},''
  \emph{Communications Surveys \& Tutorials}, vol.~19, no.~4, pp. 2359--2391,
  2017.

\bibitem{afolabi2018network}
I.~Afolabi, T.~Taleb, K.~Samdanis, A.~Ksentini, and H.~Flinck, ``{Network
  Slicing \& Softwarization: A Survey on Principles, Enabling Technologies \&
  Solutions},'' \emph{IEEE Communications Surveys \& Tutorials}, 2018.

\bibitem{mao2017survey}
Y.~Mao, C.~You, J.~Zhang, K.~Huang, and K.~B. Letaief, ``{A Survey on Mobile
  Edge Computing: The Communication Perspective},'' \emph{Communications
  Surveys \& Tutorials}, vol.~19, no.~4, pp. 2322--2358, 2017.

\bibitem{wang2017survey}
S.~Wang, X.~Zhang, Y.~Zhang, L.~Wang, J.~Yang, and W.~Wang, ``{A Survey on
  Mobile Edge Networks: Convergence of Computing, Caching and
  Communications},'' \emph{Access}, vol.~5, pp. 6757--6779, 2017.

\bibitem{roman2018mobile}
R.~Roman, J.~Lopez, and M.~Mambo, ``{Mobile Edge Computing, Fog et al.: A
  Survey and Analysis of Security Threats and Challenges},'' \emph{Future
  Generation Computer Systems}, vol.~78, pp. 680--698, 2018.

\bibitem{perez2016impact}
M.~Perez, S.~Xu, S.~Chauhan, A.~Tanaka, K.~Simpson, H.~Abdul-Muhsin, and
  R.~Smith, ``{Impact of Delay on Telesurgical Performance: Study on the
  Robotic Simulator dV-Trainer},'' \emph{International journal of computer
  assisted radiology and surgery}, vol.~11, no.~4, pp. 581--587, 2016.

\bibitem{ar_vr}
\BIBentryALTinterwordspacing
``{Unlocking Commercial Opportunities From 4G Evolution to 5G.}'' GSMA Network
  Tech. Report, accessed on 21.03.2018. [Online]. Available:
  \url{https://www.gsma.com/futurenetworks/wpcontent/uploads/2016/02/704_GSMA_unlocking_comm_opp_report_v5.pdf}
\BIBentrySTDinterwordspacing

\bibitem{IntelWhitepaper_retail}
\BIBentryALTinterwordspacing
``{The Business Case for MEC in Retail: A TCO Analysis and its Implications in
  the 5G Era},'' {Intel technical White paper}, June 2017, accessed on
  14.03.2018. [Online]. Available:
  \url{https://builders.intel.com/docs/networkbuilders/the-business-case-for-mec-in-retail-a-tco-analysis-and-its-implications-in-the-5g-era.pdf}
\BIBentrySTDinterwordspacing

\bibitem{iiot}
\BIBentryALTinterwordspacing
``{Putting Sensors to Work in the Factory Environment: Data to Information to
  Wisdom},'' accessed on 29.04.2018. [Online]. Available:
  \url{https://itpeernetwork.intel.com/putting-sensors-to-work-in-the-factory-environment/}
\BIBentrySTDinterwordspacing

\bibitem{stojkoska2017review}
B.~L.~R. Stojkoska and K.~V. Trivodaliev, ``{A Review of Internet of Things for
  Smart Home: Challenges and Solutions},'' \emph{Journal of Cleaner
  Production}, vol. 140, pp. 1454--1464, 2017.

\bibitem{vallati2016mobile}
C.~Vallati, A.~Virdis, E.~Mingozzi, and G.~Stea, ``{Mobile-Edge Computing Come
  Home Connecting Things in Future Smart Homes Using LTE Device-to-Device
  Communications},'' \emph{Consumer Electronics Magazine}, vol.~5, no.~4, pp.
  77--83, 2016.

\bibitem{morabito2016enabling}
R.~Morabito, R.~Petrolo, V.~Loscr{\'\i}, and N.~Mitton, ``{Enabling a
  Lightweight Edge Gateway-as-a-Service for the Internet of Things},'' in
  \emph{7th International Conference on the Network of the Future (NOF)}.\hskip
  1em plus 0.5em minus 0.4em\relax IEEE, 2016, pp. 1--5.

\bibitem{sun2016edgeiot}
X.~Sun and N.~Ansari, ``{Edgeiot: Mobile Edge Computing for the Internet of
  Things},'' \emph{Communications Magazine}, vol.~54, no.~12, pp. 22--29, 2016.

\bibitem{nguyen2016virtual}
K.-K. Nguyen and M.~Cheriet, ``Virtual edge-based smart community network
  management,'' \emph{Internet Computing}, vol.~20, no.~6, pp. 32--41, 2016.

\bibitem{taleb2017mobile}
T.~Taleb, S.~Dutta, A.~Ksentini, M.~Iqbal, and H.~Flinck, ``{Mobile Edge
  Computing Potential in Making Cities Smarter},'' \emph{Communications
  Magazine}, vol.~55, no.~3, pp. 38--43, 2017.

\bibitem{shi2016edge}
W.~Shi, J.~Cao, Q.~Zhang, Y.~Li, and L.~Xu, ``{Edge Computing: Vision and
  Challenges},'' \emph{Internet of Things Journal}, vol.~3, no.~5, pp.
  637--646, 2016.

\bibitem{hossain2016cloud}
M.~S. Hossain and G.~Muhammad, ``{Cloud-assisted Industrial Internet of Things
  (IIoT)--Enabled Framework for Health Monitoring},'' \emph{Computer Networks},
  vol. 101, pp. 192--202, 2016.

\bibitem{islam2015internet}
S.~R. Islam, D.~Kwak, M.~H. Kabir, M.~Hossain, and K.-S. Kwak, ``{The Internet
  of Things for Health Care: A Comprehensive Survey},'' \emph{Access}, vol.~3,
  pp. 678--708, 2015.

\bibitem{shi2016promise}
W.~Shi and S.~Dustdar, ``{The Promise of Edge Computing},'' \emph{Computer},
  vol.~49, no.~5, pp. 78--81, 2016.

\bibitem{tran2017collaborative}
T.~X. Tran, A.~Hajisami, P.~Pandey, and D.~Pompili, ``{Collaborative Mobile
  Edge Computing in 5G Networks: New Paradigms, Scenarios, and Challenges},''
  \emph{Communications Magazine}, vol.~55, no.~4, pp. 54--61, 2017.

\bibitem{singh2017semantic}
D.~Singh, G.~Tripathi, A.~M. Alberti, and A.~Jara, ``{Semantic Edge Computing
  and IoT Architecture for Military Health Services in Battlefield},'' in
  \emph{14th Annual Consumer Communications Networking Conference
  (CCNC)}.\hskip 1em plus 0.5em minus 0.4em\relax IEEE, 2017, pp. 185--190.

\bibitem{nunna2015enabling}
S.~Nunna, A.~Kousaridas, M.~Ibrahim, M.~Dillinger, C.~Thuemmler, H.~Feussner,
  and A.~Schneider, ``{Enabling Real-time Context-aware Collaboration Through
  5G and Mobile Edge Computing},'' in \emph{12th International Conference on
  Information Technology-New Generations (ITNG)}.\hskip 1em plus 0.5em minus
  0.4em\relax IEEE, 2015, pp. 601--605.

\bibitem{sharma2017live}
S.~K. Sharma and X.~Wang, ``{Live Data Analytics With Collaborative Edge and
  Cloud Processing in Wireless IoT Networks},'' \emph{Access}, vol.~5, pp.
  4621--4635, 2017.

\bibitem{rahmani2018exploiting}
A.~M. Rahmani, T.~N. Gia, B.~Negash, A.~Anzanpour, I.~Azimi, M.~Jiang, and
  P.~Liljeberg, ``{Exploiting Smart e-Health Gateways at the Edge of Healthcare
  Internet of Things: A Fog Computing Approach},'' \emph{Future Generation
  Computer Systems}, vol.~78, pp. 641--658, 2018.

\bibitem{SimallianceWhitepaper}
\BIBentryALTinterwordspacing
``{5G Security – Making the Right Choice to Match your Needs},'' {SIMalliance
  5GWG technical White paper}, Feb 2016, accessed on 12.02.2018. [Online].
  Available: \url{http://simalliance.org/}
\BIBentrySTDinterwordspacing

\bibitem{zakaria2017internet}
O.~Zakaria, J.~Britt, and H.~Forood, ``{Internet of Things (IoT) Automotive
  Device, System, and Method},'' Jul.~25 2017, uS Patent 9,717,012.

\bibitem{balid2017intelligent}
W.~Balid, H.~Tafish, and H.~H. Refai, ``{Intelligent Vehicle Counting and
  Classification Sensor for Real-Time Traffic Surveillance},''
  \emph{Transactions on Intelligent Transportation Systems}, 2017.

\bibitem{amini2017big}
S.~Amini, I.~Gerostathopoulos, and C.~Prehofer, ``{Big Data Analytics
  Architecture for Real-time Traffic Control},'' in \emph{5th International
  Conference on Models and Technologies for Intelligent Transportation Systems
  (MT-ITS)}.\hskip 1em plus 0.5em minus 0.4em\relax IEEE, 2017, pp. 710--715.

\bibitem{yu2016senspeed}
J.~Yu, H.~Zhu, H.~Han, Y.~J. Chen, J.~Yang, Y.~Zhu, Z.~Chen, G.~Xue, and M.~Li,
  ``{Senspeed: Sensing Driving Conditions to Estimate Vehicle Speed in Urban
  Environments},'' \emph{Transactions on Mobile Computing}, vol.~15, no.~1, pp.
  202--216, 2016.

\bibitem{nawaz2016smart}
S.~Nawaz, C.~Efstratiou, and C.~Mascolo, ``{Smart Sensing Systems for the Daily
  Drive},'' \emph{Pervasive Computing}, vol.~15, no.~1, pp. 39--43, 2016.

\bibitem{han2017software}
G.~Han, M.~Guizani, Y.~Bi, T.~H. Luan, K.~Ota, H.~Zhou, W.~Guibene, and
  A.~Rayes, ``{Software-Defined Vehicular Networks: Architecture, Algorithms,
  and Applications: Part 1},'' \emph{Communications Magazine}, vol.~55, no.~7,
  pp. 78--79, 2017.

\bibitem{he2015efficient}
D.~He, S.~Zeadally, B.~Xu, and X.~Huang, ``{An Efficient Identity-based
  Conditional Privacy-preserving Authentication Scheme for Vehicular Ad Hoc
  Networks},'' \emph{Transactions on Information Forensics and Security},
  vol.~10, no.~12, pp. 2681--2691, 2015.

\bibitem{metis2016deliv}
``{Deliverable D1.1 Refined scenarios and requirements, consolidated use cases,
  and qualitative techno-economic feasibility assessment},''
  \url{https://metis-ii.5g-ppp.eu/wp-content/uploads/deliverables/METIS-II_D1.1_v1.0.pdf},
  2016, accessed on 18.04.2018.

\bibitem{osseiran20165g}
A.~Osseiran, J.~F. Monserrat, and P.~Marsch, \emph{{5G Mobile and Wireless
  Communications Technology}}.\hskip 1em plus 0.5em minus 0.4em\relax Cambridge
  University Press, 2016.

\bibitem{datta2017vehicles}
S.~K. Datta, J.~Haerri, C.~Bonnet, and R.~F. Da~Costa, ``{Vehicles as Connected
  Resources: Opportunities and Challenges for the Future},'' \emph{Vehicular
  Technology Magazine}, vol.~12, no.~2, pp. 26--35, 2017.

\bibitem{frascolla5g}
V.~Frascolla, F.~Miatton, G.~K. Tran, K.~Takinami, A.~De~Domenico,
  E.~Calvanese, K.~K. Strinati, T.~Haustein, K.~Sakaguchi, S.~Barbarossa
  \emph{et~al.}, ``{5G-MiEdge: Design, Standardization and Deployment of 5G
  Phase II Technologies},'' in \emph{Conference on Standards for Communications
  \& Networking}.\hskip 1em plus 0.5em minus 0.4em\relax IEEE, 2017, pp. 1--6.

\bibitem{li2017novel}
L.~Li, Y.~Li, and R.~Hou, ``{A Novel Mobile Edge Computing-Based Architecture
  for Future Cellular Vehicular Networks},'' in \emph{Wireless Communications
  and Networking Conference (WCNC)}.\hskip 1em plus 0.5em minus 0.4em\relax
  IEEE, 2017, pp. 1--6.

\bibitem{grewe2017information}
``{Information-Centric Mobile Edge Computing for Connected Vehicle
  Environments: Challenges and Research Directions}, author={Grewe, Dennis and
  Wagner, Marco and Arumaithurai, Mayutan and Psaras, Ioannis and Kutscher,
  Dirk}, booktitle={Proceedings of the Workshop on Mobile Edge Communications},
  pages={7--12}, year={2017}, organization={ACM}.''

\bibitem{motlagh2017uav}
N.~H. Motlagh, M.~Bagaa, and T.~Taleb, ``{UAV-based IoT platform: A crowd
  surveillance use case},'' \emph{IEEE Communications Magazine}, vol.~55,
  no.~2, pp. 128--134, 2017.

\bibitem{satyanarayanan2017emergence}
M.~Satyanarayanan, ``The emergence of edge computing,'' \emph{Computer},
  vol.~50, no.~1, pp. 30--39, 2017.

\bibitem{baresi2017empowering}
L.~Baresi, D.~F. Mendon{\c{c}}a, and M.~Garriga, ``{Empowering Low-Latency
  Applications Through a Serverless Edge Computing Architecture},'' in
  \emph{European Conference on Service-Oriented and Cloud Computing}.\hskip 1em
  plus 0.5em minus 0.4em\relax Springer, 2017, pp. 196--210.

\bibitem{etsimec}
``{ETSI} executive briefing - mobile edge computing (mec) initiative,'' \url{
  https://portal.etsi.org/portals/0/tbpages/mec/docs/
  mec%20executive%20brief%20v1%2028-09-14.pdf}, accessed on 01.02.2018.

\bibitem{chen2017virtual}
M.~Chen, W.~Saad, and C.~Yin, ``{Virtual Reality Over Wireless Networks:
  Quality-of-Service Model and Learning-based Resource Management},''
  \emph{arXiv preprint arXiv:1703.04209}, 2017.

\bibitem{bastug2017toward}
E.~Bastug, M.~Bennis, M.~M{\'e}dard, and M.~Debbah, ``{Toward Interconnected
  Virtual Reality: Opportunities, Challenges, and Enablers},'' \emph{IEEE
  Communications Magazine}, vol.~55, no.~6, pp. 110--117, 2017.

\bibitem{cheng2017fogflow}
B.~Cheng, G.~Solmaz, F.~Cirillo, E.~Kovacs, K.~Terasawa, and A.~Kitazawa,
  ``{FogFlow: Easy Programming of IoT Services Over Cloud and Edges for Smart
  Cities},'' vol.~5, no.~2, pp. 696--707, 2018.

\bibitem{CISCOVNI2017}
\BIBentryALTinterwordspacing
``{Cisco Visual Networking Index: Forecast and Methodology, 2016–2021},''
  Cisco White Paper, June 2017. [Online]. Available:
  \url{https://www.cisco.com/c/en/us/solutions/collateral/service-provider/visual-networking-index-vni/complete-white-paper-c11-481360.pdf}
\BIBentrySTDinterwordspacing

\bibitem{sun2017challenges}
H.~Sun, Z.~Zhang, R.~Q. Hu, and Y.~Qian, ``{Challenges and Enabling
  Technologies in 5G Wearable Communications},'' \emph{arXiv preprint
  arXiv:1708.05410}, 2017.

\bibitem{perera2015emerging}
C.~Perera, C.~H. Liu, and S.~Jayawardena, ``{The Emerging Internet of Things
  Marketplace from an Industrial Perspective: A Survey},'' \emph{Transactions
  on Emerging Topics in Computing}, vol.~3, no.~4, pp. 585--598, 2015.

\bibitem{ferrandez2016developing}
F.~J. Ferr{\'a}ndez-Pastor, J.~M. Garc{\'\i}a-Chamizo, M.~Nieto-Hidalgo,
  J.~Mora-Pascual, and J.~Mora-Mart{\'\i}nez, ``{Developing Ubiquitous Sensor
  Network Platform Using Internet of Things: Application in Precision
  Agriculture},'' \emph{Sensors}, vol.~16, no.~7, p. 1141, 2016.

\bibitem{SmartFarming}
\BIBentryALTinterwordspacing
``{Smart Farming: The sustainable way to food},'' {Beecham Research Report},
  May 2017, accessed on 04.04.2018. [Online]. Available:
  \url{http://www.beechamresearch.com/}
\BIBentrySTDinterwordspacing

\bibitem{microsoftpoultry}
\BIBentryALTinterwordspacing
``{Building an IoT solution with PeakUp to improve management of poultry
  houses},'' Microsoft Technical Case Studies, March 2017. [Online]. Available:
  \url{https://microsoft.github.io/techcasestudies/iot/2017/03/30/PeakUp.html}
\BIBentrySTDinterwordspacing

\bibitem{mahale2016smart}
R.~B. Mahale and S.~Sonavane, ``{Smart Poultry Farm Monitoring Using IoT and
  Wireless Sensor Networks},'' \emph{International Journal of Advanced Research
  in Computer Science}, vol.~7, no.~3, 2016.

\bibitem{boban2016design}
M.~Boban, K.~Manolakis, M.~Ibrahim, S.~Bazzi, and W.~Xu, ``{Design aspects for
  5G V2X physical layer},'' in \emph{Conference on Standards for Communications
  and Networking (CSCN)}.\hskip 1em plus 0.5em minus 0.4em\relax IEEE, 2016,
  pp. 1--7.

\bibitem{braun2017study}
P.~J. Braun, S.~Pandi, R.-S. Schmoll, and F.~H. Fitzek, ``{On the Study and
  Deployment of Mobile Edge Cloud for Tactile Internet using a 5G Gaming
  Application},'' in \emph{14th Consumer Communications \& Networking
  Conference (CCNC)}.\hskip 1em plus 0.5em minus 0.4em\relax IEEE, 2017, pp.
  154--159.

\bibitem{pandi2017demonstration}
S.~Pandi, R.~S. Schmoll, P.~J. Braun, and F.~H. Fitzek, ``{Demonstration of
  Mobile Edge Cloud for Tactile Internet using a 5G Gaming Application},'' in
  \emph{14th Annual Consumer Communications \& Networking Conference
  (CCNC)}.\hskip 1em plus 0.5em minus 0.4em\relax IEEE, 2017, pp. 607--608.

\bibitem{satyanarayanan2017edge}
M.~Satyanarayanan, ``{Keynotes: Edge Computing: Vision and Challenges},'' in
  \emph{2nd International Conference on Collaboration and Internet Computing
  (CIC)}, 2016.

\bibitem{kanzaki2017video}
H.~Kanzaki, K.~Schubert, and N.~Bambos, ``{Video Streaming Schemes for
  Industrial IoT},'' in \emph{26th International Conference on Computer
  Communication and Networks (ICCCN)}.\hskip 1em plus 0.5em minus 0.4em\relax
  IEEE, 2017, pp. 1--7.

\bibitem{harper2016microdatabases}
K.~E. Harper, T.~de~Gooijer, J.~O. Schmitt, and D.~Cox, ``Microdatabases for
  the industrial internet,'' \emph{arXiv preprint arXiv:1601.04036}, 2016.

\bibitem{peralta2017fog}
G.~Peralta, M.~Iglesias-Urkia, M.~Barcelo, R.~Gomez, A.~Moran, and J.~Bilbao,
  ``{Fog Computing Based Efficient IoT Scheme for the Industry 4.0},'' in
  \emph{International Workshop of Electronics, Control, Measurement, Signals
  and their Application to Mechatronics (ECMSM)}.\hskip 1em plus 0.5em minus
  0.4em\relax IEEE, 2017, pp. 1--6.

\bibitem{chakareski2017vr}
J.~Chakareski, ``{VR/AR Immersive Communication: Caching, Edge Computing, and
  Transmission Trade-Offs},'' in \emph{Proceedings of the Workshop on Virtual
  Reality and Augmented Reality Network}.\hskip 1em plus 0.5em minus
  0.4em\relax ACM, 2017, pp. 36--41.

\bibitem{carvallo2015advanced}
A.~Carvallo and J.~Cooper, \emph{{The Advanced Smart Grid: Edge Power Driving
  Sustainability}}.\hskip 1em plus 0.5em minus 0.4em\relax Artech House, 2015.

\bibitem{moghaddam2017performance}
M.~H.~Y. Moghaddam, A.~Leon-Garcia, and M.~Moghaddassian, ``{On the Performance
  of Distributed and Cloud-Based Demand Response in Smart Grid},''
  \emph{Transactions on Smart Grid}, 2017.

\bibitem{lasi2014industry}
H.~Lasi, P.~Fettke, H.-G. Kemper, T.~Feld, and M.~Hoffmann, ``{Industry 4.0},''
  \emph{Business \& Information Systems Engineering}, vol.~6, no.~4, pp.
  239--242, 2014.

\bibitem{da2014internet}
L.~Da~Xu, W.~He, and S.~Li, ``{Internet of Things in Industries: A Survey},''
  \emph{Transactions on industrial informatics}, vol.~10, no.~4, pp.
  2233--2243, 2014.

\bibitem{perera2014survey}
C.~Perera, C.~H. Liu, S.~Jayawardena, and M.~Chen, ``{A Survey on Internet of
  Things from Industrial Market Perspective},'' \emph{Access}, vol.~2, pp.
  1660--1679, 2014.

\bibitem{kehoe2015survey}
B.~Kehoe, S.~Patil, P.~Abbeel, and K.~Goldberg, ``{A Survey of Research on
  Cloud Robotics and Automation},'' \emph{Transactions on automation science
  and engineering}, vol.~12, no.~2, pp. 398--409, 2015.

\bibitem{li2017industrial}
J.-q. Li, F.~R. Yu, G.~Deng, C.~Luo, Z.~Ming, and Q.~Yan, ``{Industrial
  Internet: A Survey on the Enabling Technologies, Applications, and
  Challenges},'' \emph{Communications Surveys \& Tutorials}, 2017.

\bibitem{albano2017industrial}
M.~Albano, J.~B. Silva, and L.~Lino~Ferreira, ``{The Industrial Internet of
  Things},'' \emph{22{\textordmasculine} Semin{\'a}rio da Rede Tem{\'a}tica de
  Comunica{\c{c}}{\~o}es M{\'o}veis}, 2017.

\bibitem{nelson2017smart}
R.~Nelson, ``{Smart Factories Leverage Cloud, Edge Computing},''
  \emph{EE-Evaluation Engineering}, vol.~56, no.~6, pp. 14--18, 2017.

\bibitem{steiner2016fog}
W.~Steiner and S.~Poledna, ``{Fog Computing as Enabler for the Industrial
  Internet of Things},'' \emph{e \& i Elektrotechnik und Informationstechnik},
  vol. 133, no.~7, pp. 310--314, 2016.

\bibitem{liang2017mobile}
B.~Liang, \emph{{Mobile Edge Computing}}.\hskip 1em plus 0.5em minus
  0.4em\relax Cambridge University Press, 2017.

\bibitem{zhang2011searching}
D.~Zhang, L.~T. Yang, and H.~Huang, ``{Searching in Internet of Things: Vision
  and Challenges},'' in \emph{9th International Symposium on Parallel and
  Distributed Processing with Applications (ISPA)}.\hskip 1em plus 0.5em minus
  0.4em\relax IEEE, 2011, pp. 201--206.

\bibitem{bellavista2016towards}
P.~Bellavista and A.~Zanni, ``{Towards Better Scalability for IoT-cloud
  Interactions via Combined Exploitation of MQTT and CoAP},'' in \emph{2nd
  International Forum on Research and Technologies for Society and Industry
  Leveraging a better tomorrow (RTSI)}.\hskip 1em plus 0.5em minus 0.4em\relax
  IEEE, 2016, pp. 1--6.

\bibitem{ren2017serving}
J.~Ren, H.~Guo, C.~Xu, and Y.~Zhang, ``{Serving at the Edge: A Scalable IoT
  Architecture Based on Transparent Computing},'' \emph{IEEE Network}, vol.~31,
  no.~5, pp. 96--105, 2017.

\bibitem{morabito2018legiot}
R.~Morabito, R.~Petrolo, V.~Loscri, and N.~Mitton, ``{LEGIoT: a Lightweight
  Edge Gateway for the Internet of Things},'' \emph{Future Generation Computer
  Systems}, vol.~81, pp. 1--15, 2018.

\bibitem{ceselli2017mobile}
A.~Ceselli, M.~Premoli, and S.~Secci, ``{Mobile Edge Cloud Network Design
  Optimization},'' \emph{IEEE/ACM Transactions on Networking}, vol.~25, no.~3,
  pp. 1818--1831, 2017.

\bibitem{LiangMECbook}
B.~Liang, \emph{{Mobile edge computing}}.\hskip 1em plus 0.5em minus
  0.4em\relax Cambridge University Press, 2017.

\bibitem{peng2016fog}
M.~Peng, S.~Yan, K.~Zhang, and C.~Wang, ``Fog-computing-based radio access
  networks: issues and challenges,'' \emph{Network}, vol.~30, no.~4, pp.
  46--53, 2016.

\bibitem{tandon2016harnessing}
R.~Tandon and O.~Simeone, ``{Harnessing Cloud and Edge Synergies: Toward an
  Information Theory of Fog Radio Access Networks},'' \emph{Communications
  Magazine}, vol.~54, no.~8, pp. 44--50, 2016.

\bibitem{peng2016recent}
M.~Peng and K.~Zhang, ``{Recent Advances in Fog Radio Access Networks:
  Performance Analysis and Radio Resource Allocation},'' \emph{Access}, vol.~4,
  pp. 5003--5009, 2016.

\bibitem{rimal2017mobile}
B.~P. Rimal, D.~P. Van, and M.~Maier, ``{Mobile-Edge Computing vs. Centralized
  Cloud Computing over a Converged FiWi Access Network},'' \emph{Transactions
  on Network and Service Management}, vol.~14, no.~3, pp. 498--513, 2017.

\bibitem{agiwal2016next}
M.~Agiwal, A.~Roy, and N.~Saxena, ``{Next Generation 5G Wireless Networks: A
  Comprehensive Survey},'' \emph{Communications Surveys \& Tutorials}, vol.~18,
  no.~3, pp. 1617--1655, 2016.

\bibitem{barbarossa2017enabling}
S.~Barbarossa, E.~Ceci, M.~Merluzzi, and E.~Calvanese-Strinati, ``{Enabling
  Effective Mobile Edge Computing Using millimeterwave Links},'' in
  \emph{International Conference on Communications Workshops (ICC
  Workshops)}.\hskip 1em plus 0.5em minus 0.4em\relax IEEE, 2017, pp. 367--372.

\bibitem{barbarossa2017overbooking}
S.~Barbarossa, E.~Ceci, and M.~Merluzzi, ``{Overbooking Radio and Computation
  Resources in mmW-Mobile Edge Computing to Reduce Vulnerability to Channel
  Intermittency},'' in \emph{European Conference on Networks and Communications
  (EuCNC)}.\hskip 1em plus 0.5em minus 0.4em\relax IEEE, 2017, pp. 1--5.

\bibitem{OpenChirp7917625}
A.~Dongare, C.~Hesling, K.~Bhatia, A.~Balanuta, R.~L. Pereira, B.~Iannucci, and
  A.~Rowe, ``{OpenChirp: A Low-Power Wide-Area Networking Architecture},'' in
  \emph{IEEE International Conference on Pervasive Computing and Communications
  Workshops (PerCom Workshops)}, 2017, pp. 569--574.

\bibitem{ansari2017mobile}
N.~Ansari and X.~Sun, ``{Mobile Edge Computing Empowers Internet of Things},''
  \emph{IEICE Transactions on Communications}, vol. 101, no.~3, pp. 604--619,
  2018.

\bibitem{farris2017federations}
I.~Farris, A.~Orsino, L.~Militano, M.~Nitti, G.~Araniti, L.~Atzori, and
  A.~Iera, ``{Federations of Connected Things for Ddelay-sensitive IoT Services
  in 5G Environments},'' in \emph{International Conference on Communications
  (ICC)}.\hskip 1em plus 0.5em minus 0.4em\relax IEEE, 2017, pp. 1--6.

\bibitem{farris2017federated}
I.~Farris, A.~Orsino, L.~Militano, A.~Iera, and G.~Araniti, ``{Federated IoT
  Services Leveraging 5G Technologies at the Edge},'' \emph{Ad Hoc Networks},
  vol.~68, pp. 58--69, 2018.

\bibitem{orsino2017exploiting}
A.~Orsino, I.~Farris, L.~Militano, G.~Araniti, S.~Andreev, I.~Gudkova,
  Y.~Koucheryavy, and A.~Iera, ``{Exploiting D2D Communications at the Network
  Edge for Mission-Critical IoT Applications},'' in \emph{Proceedings of 23th
  European Wireless Conference}.\hskip 1em plus 0.5em minus 0.4em\relax VDE,
  2017, pp. 1--6.

\bibitem{ko2017wireless}
``{{Wireless Networks for Mobile Edge Computing: Spatial Modeling and Latency
  Analysis (Extended version)}}, author={Ko, Seung-Woo and Han, Kaifeng and
  Huang, Kaibin}, journal={arXiv preprint arXiv:1709.01702}, year={2017}.''

\bibitem{samie2016computation}
F.~Samie, V.~Tsoutsouras, L.~Bauer, S.~Xydis, D.~Soudris, and J.~Henkel,
  ``{Computation Offloading and Resource Allocation for Low-power IoT Edge
  Devices},'' in \emph{3rd World Forum on Internet of Things (WF-IoT)}.\hskip
  1em plus 0.5em minus 0.4em\relax IEEE, 2016, pp. 7--12.

\bibitem{abdelwahab2016replisom}
S.~Abdelwahab, B.~Hamdaoui, M.~Guizani, and T.~Znati, ``{Replisom: Disciplined
  Tiny Memory Replication for Massive IoT Devices in LTE Edge Cloud},''
  \emph{Internet of Things Journal}, vol.~3, no.~3, pp. 327--338, 2016.

\bibitem{yu2017sdlb}
Y.~Yu, X.~Li, and C.~Qian, ``{SDLB: A Scalable and Dynamic Software Load
  Balancer for Fog and Mobile Edge Computing},'' in \emph{Proceedings of the
  Workshop on Mobile Edge Communications}.\hskip 1em plus 0.5em minus
  0.4em\relax ACM, 2017, pp. 55--60.

\bibitem{vilalta2017telcofog}
R.~Vilalta, V.~Lopez, A.~Giorgetti, S.~Peng, V.~Orsini, L.~Velasco,
  R.~Serral-Gracia, D.~Morris, S.~De~Fina, F.~Cugini \emph{et~al.},
  ``{TelcoFog: A Unified Flexible Fog and Cloud Computing Architecture for 5G
  Networks},'' \emph{Communications Magazine}, vol.~55, no.~8, pp. 36--43,
  2017.

\bibitem{bouet2017geo}
M.~Bouet and V.~Conan, ``Geo-partitioning of mec resources,'' in
  \emph{Proceedings of the Workshop on Mobile Edge Communications}.\hskip 1em
  plus 0.5em minus 0.4em\relax ACM, 2017, pp. 43--48.

\bibitem{flores2017large}
H.~Flores, X.~Su, V.~Kostakos, A.~Y. Ding, P.~Nurmi, S.~Tarkoma, P.~Hui, and
  Y.~Li, ``{Large-scale offloading in the Internet of Things},'' in
  \emph{International Conference on Pervasive Computing and Communications
  Workshops (PerCom Workshops)}.\hskip 1em plus 0.5em minus 0.4em\relax IEEE,
  2017, pp. 479--484.

\bibitem{wang2017computation}
C.~Wang, C.~Liang, F.~R. Yu, Q.~Chen, and L.~Tang, ``Computation offloading and
  resource allocation in wireless cellular networks with mobile edge
  computing,'' \emph{IEEE Transactions on Wireless Communications}, vol.~16,
  no.~8, pp. 4924--4938, 2017.

\bibitem{lyu2017optimal}
X.~Lyu, W.~Ni, H.~Tian, R.~P. Liu, X.~Wang, G.~B. Giannakis, and A.~Paulraj,
  ``{Optimal Schedule of Mobile Edge Computing for Internet of Things Using
  Partial Information},'' \emph{Journal on Selected Areas in Communications},
  vol.~35, no.~11, pp. 2606--2615, 2017.

\bibitem{gupta2017ifogsim}
H.~Gupta, A.~Vahid~Dastjerdi, S.~K. Ghosh, and R.~Buyya, ``{iFogSim: A toolkit
  for modeling and simulation of resource management techniques in the Internet
  of Things, Edge and Fog computing environments},'' \emph{Software: Practice
  and Experience}, vol.~47, no.~9, pp. 1275--1296, 2017.

\bibitem{habak2015femto}
K.~Habak, M.~Ammar, K.~A. Harras, and E.~Zegura, ``{Femto Clouds: Leveraging
  Mobile Devices to Provide Cloud Service at the Edge},'' in \emph{8th
  International Conference on Cloud Computing (CLOUD)}.\hskip 1em plus 0.5em
  minus 0.4em\relax IEEE, 2015, pp. 9--16.

\bibitem{chen2016mobility}
M.~Chen, Y.~Hao, M.~Qiu, J.~Song, D.~Wu, and I.~Humar, ``{Mobility-aware
  Caching and Computation Offloading in 5G Ultra-dense Cellular Networks},''
  \emph{Sensors}, vol.~16, no.~7, p. 974, 2016.

\bibitem{chen2016efficient}
X.~Chen, L.~Jiao, W.~Li, and X.~Fu, ``{Efficient multi-user computation
  offloading for mobile-edge cloud computing},'' \emph{IEEE/ACM Transactions on
  Networking}, vol.~24, no.~5, pp. 2795--2808, 2016.

\bibitem{sardellitti2015joint}
S.~Sardellitti, G.~Scutari, and S.~Barbarossa, ``{Joint Optimization of Radio
  and Computational Resources for Multicell Mobile-edge Computing},''
  \emph{Transactions on Signal and Information Processing over Networks},
  vol.~1, no.~2, pp. 89--103, 2015.

\bibitem{wang2017joint}
C.~Wang, F.~R. Yu, C.~Liang, Q.~Chen, and L.~Tang, ``{Joint Computation
  Offloading and Interference Management in Wireless Cellular Networks With
  Mobile Edge Computing},'' \emph{Transactions on Vehicular Technology},
  vol.~66, no.~8, pp. 7432--7445, 2017.

\bibitem{sun2017emm}
Y.~Sun, S.~Zhou, and J.~Xu, ``{EMM: Energy-Aware Mobility Management for Mobile
  Edge Computing in Ultra Dense Networks},'' \emph{IEEE Journal on Selected
  Areas in Communications}, vol.~35, no.~11, pp. 2637--2646, 2017.

\bibitem{liu2016delay}
J.~Liu, Y.~Mao, J.~Zhang, and K.~B. Letaief, ``{Delay-optimal Computation Task
  Scheduling for Mobile-edge Computing Systems},'' in \emph{International
  Symposium on Information Theory (ISIT)}.\hskip 1em plus 0.5em minus
  0.4em\relax IEEE, 2016, pp. 1451--1455.

\bibitem{you2017energy}
C.~You, K.~Huang, H.~Chae, and B.-H. Kim, ``{Energy-efficient Resource
  Allocation for Mobile-edge Computation Offloading},'' \emph{Transactions on
  Wireless Communications}, vol.~16, no.~3, pp. 1397--1411, 2017.

\bibitem{mach2014cloud}
P.~Mach and Z.~Becvar, ``{Cloud-aware Power Control for Cloud-enabled Small
  Cells},'' in \emph{Globecom Workshops}.\hskip 1em plus 0.5em minus
  0.4em\relax IEEE, 2014, pp. 1038--1043.

\bibitem{mach2016cloud}
{P. Mach and Z. Becvar}, ``{Cloud Aware Power Control for Real-time Application
  Offloading in Mobile Edge Computing},'' \emph{Transactions on Emerging
  Telecommunications Technologies}, vol.~27, no.~5, pp. 648--661, 2016.

\bibitem{taleb2013analytical}
T.~Taleb and A.~Ksentini, ``{An Analytical Model for Follow Me Cloud},'' in
  \emph{Global Communications Conference (GLOBECOM)}.\hskip 1em plus 0.5em
  minus 0.4em\relax IEEE, 2013, pp. 1291--1296.

\bibitem{wu2015ubiflow}
D.~Wu, D.~I. Arkhipov, E.~Asmare, Z.~Qin, and J.~A. McCann, ``{UbiFlow:
  Mobility Management in Urban-scale Software Defined IoT},'' in
  \emph{Conference on Computer Communications (INFOCOM)}.\hskip 1em plus 0.5em
  minus 0.4em\relax IEEE, 2015, pp. 208--216.

\bibitem{shang2016named}
W.~Shang, A.~Bannis, T.~Liang, Z.~Wang, Y.~Yu, A.~Afanasyev, J.~Thompson,
  J.~Burke, B.~Zhang, and L.~Zhang, ``{Named Data Networking of Things},'' in
  \emph{First International Conference on Internet-of-Things Design and
  Implementation (IoTDI)}.\hskip 1em plus 0.5em minus 0.4em\relax IEEE, 2016,
  pp. 117--128.

\bibitem{giust2015distributed}
F.~Giust, L.~Cominardi, and C.~J. Bernardos, ``{Distributed mobility Management
  for Future 5G Networks: Overview and Analysis of Existing Approaches},''
  \emph{Communications Magazine}, vol.~53, no.~1, pp. 142--149, 2015.

\bibitem{le2017location}
C.~N. Le~Tan, C.~Klein, and E.~Elmroth, ``{Location-aware load prediction in
  Edge Data Centers},'' in \emph{Second International Conference on Fog and
  Mobile Edge Computing (FMEC)}.\hskip 1em plus 0.5em minus 0.4em\relax IEEE,
  2017, pp. 25--31.

\bibitem{vassilakis2016security}
V.~Vassilakis, I.~P. Chochliouros, A.~S. Spiliopoulou, E.~Sfakianakis,
  M.~Belesioti, N.~Bompetsis, M.~Wilson, C.~Turyagyenda, and A.~Dardamanis,
  ``{Security Analysis of Mobile Edge Computing in Virtualized Small Cell
  Networks},'' in \emph{IFIP International Conference on Artificial
  Intelligence Applications and Innovations}.\hskip 1em plus 0.5em minus
  0.4em\relax Springer, 2016, pp. 653--665.

\bibitem{roman2016mobile}
R.~Roman, J.~Lopez, and M.~Mambo, ``{Mobile Edge Computing, Fog et al.: A
  Survey and Analysis of Security Threats and Challenges},'' \emph{Future
  Generation Computer Systems}, 2016.

\bibitem{jing2014security}
Q.~Jing, A.~V. Vasilakos, J.~Wan, J.~Lu, and D.~Qiu, ``{Security of the
  Internet of Things: Perspectives and Challenges},'' \emph{Wireless Networks},
  vol.~20, no.~8, pp. 2481--2501, 2014.

\bibitem{stojmenovic2016overview}
I.~Stojmenovic, S.~Wen, X.~Huang, and H.~Luan, ``{An Overview of Fog Computing
  and its Security Issues},'' \emph{Concurrency and Computation: Practice and
  Experience}, vol.~28, no.~10, pp. 2991--3005, 2016.

\bibitem{wan2013cloud}
J.~Wan, C.~Zou, S.~Ullah, C.-F. Lai, M.~Zhou, and X.~Wang, ``{Cloud-enabled
  Wireless Body Area Networks for Pervasive Healthcare},'' \emph{Network},
  vol.~27, no.~5, pp. 56--61, 2013.

\bibitem{wan2014vcmia}
J.~Wan, D.~Zhang, Y.~Sun, K.~Lin, C.~Zou, and H.~Cai, ``{VCMIA: a Novel
  Architecture for Integrating Vehicular Cyber-physical Systems and Mobile
  Cloud Computing},'' \emph{Mobile Networks and Applications}, vol.~19, no.~2,
  pp. 153--160, 2014.

\bibitem{varga2016network}
N.~Varga, L.~Bokor, and E.~Piri, ``{A Network-assisted Flow Mobility
  Architecture for Optimized Mobile Medical Multimedia Transmission},''
  \emph{Annals of Telecommunications}, vol.~71, no. 9-10, pp. 489--502, 2016.

\bibitem{taylor2006eu}
M.~Taylor, ``{The EU Data Retention Directive},'' \emph{Computer Law \&
  Security Review}, vol.~22, no.~4, pp. 309--312, 2006.

\bibitem{haggard2015north}
S.~Haggard and J.~R. Lindsay, ``North korea and the sony hack: exporting
  instability through cyberspace,'' 2015, accessed on 02.05.2018.

\bibitem{german2016new}
P.~German, ``{A New Month, a New Data Breach},'' \emph{Network Security}, vol.
  2016, no.~3, pp. 18--20, 2016.

\bibitem{roman2013features}
R.~Roman, J.~Zhou, and J.~Lopez, ``{On the Features and Challenges of Security
  and Privacy in Distributed Internet of Things},'' \emph{Computer Networks},
  vol.~57, no.~10, pp. 2266--2279, 2013.

\bibitem{kemmer2016software}
F.~Kemmer, C.~Reich, M.~Knahl, and N.~Clarke, ``{Software Defined Privacy},''
  in \emph{International Conference on Cloud Engineering Workshop
  (IC2EW)}.\hskip 1em plus 0.5em minus 0.4em\relax IEEE, 2016, pp. 25--29.

\bibitem{yi2015security}
S.~Yi, Z.~Qin, and Q.~Li, ``{Security and Privacy Issues of Fog Computing: A
  Survey},'' in \emph{International Conference on Wireless Algorithms, Systems,
  and Applications}.\hskip 1em plus 0.5em minus 0.4em\relax Springer, 2015, pp.
  685--695.

\bibitem{abedin2015fog}
S.~F. Abedin, M.~G.~R. Alam, N.~H. Tran, and C.~S. Hong, ``{A Fog Based System
  Model for Cooperative IoT Node Pairing Using Matching Theory},'' in
  \emph{17th Asia-Pacific Network Operations and Management Symposium
  (APNOMS)}.\hskip 1em plus 0.5em minus 0.4em\relax IEEE, 2015, pp. 309--314.

\bibitem{de2012proposed}
P.~De~Hert and V.~Papakonstantinou, ``{The Proposed Data Protection Regulation
  replacing Directive 95/46/EC: A Sound System for the Protection of
  Individuals},'' \emph{Computer Law \& Security Review}, vol.~28, no.~2, pp.
  130--142, 2012.

\bibitem{ziegler2017anastacia}
S.~Ziegler, A.~Skarmeta, J.~Bernal, E.~E. Kim, and S.~Bianchi, ``{ANASTACIA:
  Advanced Networked Agents for Security and Trust Assessment in CPS IoT
  Architectures},'' in \emph{Global Internet of Things Summit (GIoTS)}.\hskip
  1em plus 0.5em minus 0.4em\relax IEEE, 2017, pp. 1--6.

\bibitem{dang2017data}
T.~D. Dang and D.~Hoang, ``{A Data Protection Model for Fog Computing},'' in
  \emph{Fog and Mobile Edge Computing (FMEC)}.\hskip 1em plus 0.5em minus
  0.4em\relax IEEE, 2017, pp. 32--38.

\bibitem{mijumbi2016network}
R.~Mijumbi, J.~Serrat, J.-L. Gorricho, N.~Bouten, F.~De~Turck, and R.~Boutaba,
  ``{Network Function Virtualization: State-of-the-art and Research
  Challenges},'' \emph{Communications Surveys \& Tutorials}, vol.~18, no.~1,
  pp. 236--262, 2016.

\bibitem{gupta2016mobile}
\BIBentryALTinterwordspacing
L.~Gupta, R.~Jain, and H.~A. Chan, ``Mobile edge computing--an important
  ingredient of 5g networks,'' \emph{IEEE Software Defined Networks,
  Newsletter}, 2016. [Online]. Available: \url{http://sdn. ieee.
  org/newsletter/march-2016/mobile-edge-computing-an-important-ingredient-of-5g-network}
\BIBentrySTDinterwordspacing

\bibitem{yang2016seamless}
B.~Yang, W.~K. Chai, G.~Pavlou, and K.~V. Katsaros, ``{Seamless Support of Low
  Latency Mobile Applications with NFV-Enabled Mobile Edge-Cloud},'' in
  \emph{International Conference on Cloud Networking (Cloudnet)}.\hskip 1em
  plus 0.5em minus 0.4em\relax IEEE, 2016, pp. 136--141.

\bibitem{li2017mec}
B.~LI, Y.~ZHANG, and L.~XU, ``{An MEC and NFV Integrated Network
  Architecture},'' \emph{ZTE COMMUNICATIONS}, vol.~15, no.~2, p.~1, 2017.

\bibitem{sciancalepore2016double}
V.~Sciancalepore, F.~Giust, K.~Samdanis, and Z.~Yousaf, ``{A Double-tier
  MEC-NFV Architecture: Design and Optimisation},'' in \emph{Conference on
  Standards for Communications and Networking (CSCN)}.\hskip 1em plus 0.5em
  minus 0.4em\relax IEEE, 2016, pp. 1--6.

\bibitem{carella2017prototyping}
G.~A. Carella, M.~Pauls, T.~Magedanz, M.~Cilloni, P.~Bellavista, and
  L.~Foschini, ``{Prototyping NFV-based Multi-access Edge Computing in 5G Ready
  Networks with Open Baton},'' in \emph{Conference on Network Softwarization
  (NetSoft)}.\hskip 1em plus 0.5em minus 0.4em\relax IEEE, 2017, pp. 1--4.

\bibitem{blanco2017technology}
B.~Blanco, J.~O. Fajardo, I.~Giannoulakis, E.~Kafetzakis, S.~Peng,
  J.~P{\'e}rez-Romero, I.~Trajkovska, P.~S. Khodashenas, L.~Goratti, M.~Paolino
  \emph{et~al.}, ``{Technology Pillars in the Architecture of Future 5G Mobile
  Networks: NFV, MEC and SDN},'' \emph{Computer Standards \& Interfaces},
  vol.~54, pp. 216--228, 2017.

\bibitem{peng2017qoe}
S.~Peng, J.~O. Fajardo, P.~S. Khodashenas, B.~Blanco, F.~Liberal, C.~Ruiz,
  C.~Turyagyenda, M.~Wilson, and S.~Vadgama, ``{QoE-Oriented Mobile Edge
  Service Management Leveraging SDN and NFV},'' \emph{Mobile Information
  Systems}, vol. 2017, 2017.

\bibitem{ali2017real}
S.~Ali and M.~Ghazal, ``{Real-time Heart Attack Mobile Detection Service
  (RHAMDS): An IoT use case for Software Defined Networks},'' in \emph{30th
  Canadian Conference on Electrical and Computer Engineering (CCECE)}.\hskip
  1em plus 0.5em minus 0.4em\relax IEEE, 2017, pp. 1--6.

\bibitem{farristowards}
I.~Farris, J.~Bernabe, N.~Toumi, D.~Garcia-Carrillo, T.~Taleb, A.~Skarmeta, and
  B.~Sahlin, ``{Towards Provisioning of SDN/NFV-based Security Enablers for
  Integrated Protection of IoT Systems},'' in \emph{Conference on Standards for
  Communications \& Networking (CSCN)}.\hskip 1em plus 0.5em minus 0.4em\relax
  IEEE, 2017, pp. 1--6.

\bibitem{huang2017low}
A.~Huang, N.~Nikaein, T.~Stenbock, A.~Ksentini, and C.~Bonnet, ``{Low Latency
  MEC Framework for SDN-based LTE/LTE-A Networks},'' in \emph{International
  Conference on Communications (ICC)}.\hskip 1em plus 0.5em minus 0.4em\relax
  IEEE, 2017, pp. 1--6.

\bibitem{nguyen2017simeca}
B.~Nguyen, N.~Choi, M.~Thottan, and J.~Van~der Merwe, ``{SIMECA: SDN-based IoT
  Mobile Edge Cloud Architecture},'' in \emph{IFIP Symposium on Integrated
  Network and Service Management (IM)}.\hskip 1em plus 0.5em minus 0.4em\relax
  IEEE, 2017, pp. 503--509.

\bibitem{hossain2017impact}
M.~S. Hossain, C.~Xu, Y.~Li, A.-S.~K. Pathan, J.~Bilbao, W.~Zeng, and
  A.~El~Saddik, ``{Impact of Next-Generation Mobile Technologies on IoT-Cloud
  Convergence},'' \emph{Communications Magazine}, vol.~55, no.~1, pp. 18--19,
  2017.

\bibitem{liu2017high}
J.~Liu, J.~Wan, D.~Jia, B.~Zeng, D.~Li, C.-H. Hsu, and H.~Chen,
  ``{High-Efficiency Urban Traffic Management in Context-Aware Computing and 5G
  Communication},'' \emph{Communications Magazine}, vol.~55, no.~1, pp. 34--40,
  2017.

\bibitem{phemius2016bringing}
K.~Phemius, J.~Seddar, M.~Bouet, H.~Khalif{\'e}, and V.~Conan, ``{Bringing SDN
  to the Edge of Tactical Networks},'' in \emph{Military Communications
  Conference (MILCOM)}.\hskip 1em plus 0.5em minus 0.4em\relax IEEE, 2016, pp.
  1047--1052.

\bibitem{aggarwal2016securing}
C.~Aggarwal and K.~Srivastava, ``{Securing IoT devices using SDN and edge
  computing},'' in \emph{2nd International Conference on Next Generation
  Computing Technologies (NGCT)}.\hskip 1em plus 0.5em minus 0.4em\relax IEEE,
  2016, pp. 877--882.

\bibitem{vasilakos2015information}
A.~V. Vasilakos, Z.~Li, G.~Simon, and W.~You, ``{Information centric network:
  Research challenges and opportunities},'' \emph{Journal of Network and
  Computer Applications}, vol.~52, pp. 1--10, 2015.

\bibitem{piro2014information}
G.~Piro, L.~A. Grieco, G.~Boggia, and P.~Chatzimisios, ``{Information-centric
  Networking and Multimedia Services: Present and Future Challenges},''
  \emph{Transactions on Emerging Telecommunications Technologies}, vol.~25,
  no.~4, pp. 392--406, 2014.

\bibitem{lloret2017enabling}
E.~Ahmed, M.~Imran, M.~Guizani, A.~Rayes, J.~Lloret, G.~Han, and W.~Guibene,
  ``{Enabling Mobile and Wireless Technologies for Smart Cities},'' \emph{IEEE
  Communications Magazine}, vol.~55, no.~1, pp. 74--75, 2017.

\bibitem{maier2016tactile}
M.~Maier, M.~Chowdhury, B.~P. Rimal, and D.~P. Van, ``{The Tactile Internet:
  Vision, Recent Progress, and Open Challenges},'' \emph{Communications
  Magazine}, vol.~54, no.~5, pp. 138--145, 2016.

\bibitem{ravindran2017realizing}
R.~Ravindran, A.~Chakraborti, S.~O. Amin, A.~Azgin, and G.~Wang, ``{Realizing
  ICN in 3GPP's 5G NextGen Core Architecture},'' \emph{arXiv preprint
  arXiv:1711.02232}, 2017.

\bibitem{5GAmericaICNMEC}
\BIBentryALTinterwordspacing
``{Understanding Information-Centric Networking and Mobile Edge Computing},''
  5G Americas, December 2016, accessed on 12.01.2018. [Online]. Available:
  \url{http://www.5gamericas.org/files/1214/8175/3330/Understanding_Information_Centric_Networking_and_Mobile_Edge_Computing.pdf}
\BIBentrySTDinterwordspacing

\bibitem{zhou2017video}
{Y. Zhou and F. R. Yu and J. Chen and Y. Kuo}, ``{Video Transcoding, Caching,
  and Multicast for Heterogeneous Networks over Wireless Network
  Virtualization},'' \emph{Communications Letters}, vol.~22, no.~1, pp.
  141--144, 2018.

\bibitem{zhou2017resource}
Y.~Zhou, F.~R. Yu, J.~Chen, and Y.~Kuo, ``{Resource Allocation for Information
  Centric Virtualized Heterogeneous Networks with In-Network Caching and Mobile
  Edge Computing},'' \emph{Transactions on Vehicular Technology}, vol.~66,
  no.~12, pp. 11\,339--11\,351, 2017.

\bibitem{huo2016software}
R.~Huo, F.~R. Yu, T.~Huang, R.~Xie, J.~Liu, V.~C. Leung, and Y.~Liu,
  ``{Software Defined Networking, Caching, and Computing for Green Wireless
  Networks},'' \emph{Communications Magazine}, vol.~54, no.~11, pp. 185--193,
  2016.

\bibitem{ge2016qoe}
C.~Ge, N.~Wang, S.~Skillman, G.~Foster, and Y.~Cao, ``{QoE-Driven DASH Video
  Caching and Adaptation at 5G Mobile Edge},'' in \emph{Proceedings of 3rd ACM
  Conference on Information-Centric Networking}.\hskip 1em plus 0.5em minus
  0.4em\relax ACM, 2016, pp. 237--242.

\bibitem{samdanis2016network}
K.~Samdanis, X.~Costa-Perez, and V.~Sciancalepore, ``{From Network Sharing to
  Multi-tenancy: The 5G Network Slice Broker},'' \emph{Communications
  Magazine}, vol.~54, no.~7, pp. 32--39, 2016.

\bibitem{alliance2016description}
N.~Alliance, ``{Description of Network Slicing Concept},'' \emph{NGMN 5G P},
  vol.~1, 2016.

\bibitem{nikaein2015network}
N.~Nikaein, E.~Schiller, R.~Favraud, K.~Katsalis, D.~Stavropoulos, I.~Alyafawi,
  Z.~Zhao, T.~Braun, and T.~Korakis, ``{Network Store: Exploring Slicing in
  Future 5G Networks},'' in \emph{Proceedings of the 10th International
  Workshop on Mobility in the Evolving Internet Architecture}.\hskip 1em plus
  0.5em minus 0.4em\relax ACM, 2015, pp. 8--13.

\bibitem{5gamericasNS}
\BIBentryALTinterwordspacing
``{Network Slicing for 5G Networks \& Services},'' 5G Americas White Paper –
  Network Slicing for 5G and Beyond, November 2015, accessed on 03.01.2018.
  [Online]. Available:
  \url{http://www.5gamericas.org/files/3214/7975/0104/5G_Americas_Network_Slicing_11.21_Final.pdf}
\BIBentrySTDinterwordspacing

\bibitem{zhang2017network}
H.~Zhang, N.~Liu, X.~Chu, K.~Long, A.-H. Aghvami, and V.~C. Leung, ``{Network
  Slicing Based 5G and Future Mobile Networks: Mobility, Resource Management,
  and Challenges},'' \emph{Communications Magazine}, vol.~55, no.~8, pp.
  138--145, 2017.

\bibitem{katsalis2017network}
K.~Katsalis, N.~Nikaein, E.~Schiller, A.~Ksentini, and T.~Braun, ``{Network
  Slices toward 5G Communications: Slicing the LTE Network},''
  \emph{Communications Magazine}, vol.~55, no.~8, pp. 146--154, 2017.

\bibitem{munoz2017adrenaline}
R.~Mu{\~n}oz, L.~Nadal, R.~Casellas, M.~S. Moreolo, R.~Vilalta, J.~M.
  F{\`a}brega, R.~Mart{\'\i}nez, A.~Mayoral, and F.~J. V{\'\i}lchez, ``{The
  ADRENALINE testbed: An SDN/NFV packet/optical transport network and edge/core
  cloud platform for end-to-end 5G and IoT services},'' in \emph{European
  Conference on Networks and Communications (EuCNC)}.\hskip 1em plus 0.5em
  minus 0.4em\relax IEEE, 2017, pp. 1--5.

\bibitem{van2017unavoidable}
F.~van Lingen, M.~Yannuzzi, A.~Jain, R.~Irons-Mclean, O.~Lluch, D.~Carrera,
  J.~L. Perez, A.~Gutierrez, D.~Montero, J.~Marti \emph{et~al.}, ``{The
  Unavoidable Convergence of NFV, 5G, and Fog: A Model-Driven Approach to
  Bridge Cloud and Edge},'' \emph{Communications Magazine}, vol.~55, no.~8, pp.
  28--35, 2017.

\bibitem{vilalta2016sdn}
R.~Vilalta, A.~Mayoral, R.~Casellas, R.~Mart{\'\i}nez, and R.~Mu{\~n}oz,
  ``{SDN/NFV Orchestration of Multi-technology and Multi-domain Networks in
  Cloud/Fog Architectures for 5G Services},'' in \emph{21st OptoElectronics and
  Communications Conference (OECC) held jointly with 2016 International
  Conference on Photonics in Switching (PS)}.\hskip 1em plus 0.5em minus
  0.4em\relax IEEE, 2016, pp. 1--3.

\bibitem{ravindran20175g}
R.~Ravindran, A.~Chakraborti, S.~O. Amin, A.~Azgin, and G.~Wang, ``{5G-ICN:
  Delivering ICN Services over 5G Using Network Slicing},''
  \emph{Communications Magazine}, vol.~55, no.~5, pp. 101--107, 2017.

\bibitem{SESAME}
\BIBentryALTinterwordspacing
``{SESAME Project},'' {H2020 EU project}, accessed on 25.03.2018. [Online].
  Available: \url{http://www.sesame-h2020-5g-ppp.eu/Home.aspx}
\BIBentrySTDinterwordspacing

\bibitem{Anastacia}
\BIBentryALTinterwordspacing
``{ANASTACIA Project},'' {H2020 EU project}, accessed on 11.02.2018. [Online].
  Available: \url{http://www.anastacia-h2020.eu/}
\BIBentrySTDinterwordspacing

\bibitem{miEdge}
\BIBentryALTinterwordspacing
``{5G-MiEdge project: Millimeter-wave Edge Cloud as an Enabler for 5G
  Ecosystem},'' {H2020 EU\&Japan Project}, 2017, accessed on 15.02.2018.
  [Online]. Available: \url{https://5g-miedge.eu/}
\BIBentrySTDinterwordspacing

\bibitem{5GPagoda}
\BIBentryALTinterwordspacing
``{5G!Pagoda},'' {EU Japan collaboration project}, accessed on 19.02.2018.
  [Online]. Available: \url{https://5g-pagoda.aalto.fi/}
\BIBentrySTDinterwordspacing

\bibitem{InterIoT}
\BIBentryALTinterwordspacing
``{Inter-IoT Project},'' {H2020 EU project}, accessed on 15.02.2018. [Online].
  Available: \url{http://www.inter-iot-project.eu/}
\BIBentrySTDinterwordspacing

\bibitem{5g-monarch}
\BIBentryALTinterwordspacing
``{5G MoNArch Project},'' {H2020 EU project}, accessed on 17.02.2018. [Online].
  Available: \url{https://5g-monarch.eu/}
\BIBentrySTDinterwordspacing

\bibitem{5gEssence}
\BIBentryALTinterwordspacing
``{5G ESSENCE Project},'' {H2020 EU project}, accessed on 15.04.2018. [Online].
  Available: \url{https://5g-ppp.eu/5g-essence/}
\BIBentrySTDinterwordspacing

\bibitem{Matilda}
\BIBentryALTinterwordspacing
``{MATILDA Project},'' {H2020 EU project}, accessed on 15.04.2018. [Online].
  Available: \url{https://5g-ppp.eu/matilda/}
\BIBentrySTDinterwordspacing

\bibitem{5GCity}
\BIBentryALTinterwordspacing
``{5GCity Project},'' {H2020 EU project}, accessed on 22.02.2018. [Online].
  Available: \url{http://www.5gcity.eu/}
\BIBentrySTDinterwordspacing

\bibitem{MONICA}
\BIBentryALTinterwordspacing
``{MONICA Project},'' {H2020 EU project}, accessed on 11.03.2018. [Online].
  Available: \url{http://www.monica-project.eu/}
\BIBentrySTDinterwordspacing

\bibitem{Autopilot}
\BIBentryALTinterwordspacing
``{AUTOPILOT Project},'' {H2020 EU project}, accessed on 15.02.2018. [Online].
  Available: \url{http://autopilot-project.eu/}
\BIBentrySTDinterwordspacing

\bibitem{5Gcoral}
\BIBentryALTinterwordspacing
``{5G-CORAL Project},'' {H2020 EU project}, accessed on 15.03.2018. [Online].
  Available: \url{http://5g-coral.eu/}
\BIBentrySTDinterwordspacing

\bibitem{ETSIVRARA}
\BIBentryALTinterwordspacing
``{ETSI and VRARA cooperate on Virtual and Augmented Reality},'' {ETSI news
  event}, accessed on 04.05.2018. [Online]. Available:
  \url{http://www.etsi.org/news-events/}
\BIBentrySTDinterwordspacing

\bibitem{liu2014concert}
J.~Liu, T.~Zhao, S.~Zhou, Y.~Cheng, and Z.~Niu, ``{CONCERT: a cloud-based
  architecture for next-generation cellular systems},'' \emph{Wireless
  Communications}, vol.~21, no.~6, pp. 14--22, 2014.

\bibitem{mestres2017knowledge}
A.~Mestres, A.~Rodriguez-Natal, J.~Carner, P.~Barlet-Ros, E.~Alarc{\'o}n,
  M.~Sol{\'e}, V.~Munt{\'e}s-Mulero, D.~Meyer, S.~Barkai, M.~J. Hibbett
  \emph{et~al.}, ``{Knowledge-defined Networking},'' \emph{SIGCOMM Computer
  Communication Review}, vol.~47, no.~3, pp. 2--10, 2017.

\bibitem{crutcher2017hyperprofile}
A.~Crutcher, C.~Koch, K.~Coleman, J.~Patman, F.~Esposito, and P.~Calyam,
  ``{Hyperprofile-based Computation Offloading for Mobile Edge Networks},''
  \emph{arXiv preprint arXiv:1707.09422}, 2017.

\bibitem{ahmed2017bringing}
E.~Ahmed, A.~Ahmed, I.~Yaqoob, J.~Shuja, A.~Gani, M.~Imran, and M.~Shoaib,
  ``{Bringing Computation Closer toward the User Network: Is Edge Computing the
  Solution?}'' \emph{IEEE Communications Magazine}, vol.~55, no.~11, pp.
  138--144, 2017.

\bibitem{sorensen20155g}
L.~T. Sorensen, S.~Khajuria, and K.~E. Skouby, ``{5G Visions of User
  Privacy},'' in \emph{81st Vehicular Technology Conference (VTC
  Spring)}.\hskip 1em plus 0.5em minus 0.4em\relax IEEE, 2015, pp. 1--4.

\bibitem{cavoukian2014regulartor}
A.~Cavoukian and M.~Chibba, ``{A Regulartor's Perspective: Leading the way with
  Privacy by Design},'' \emph{Cyber security in future Internet, security and
  privacy by design. OUTLOOK, Visions and research for the wireless world},
  no.~11, 2014.

\bibitem{pitt2013trust}
D.~Pitt, ``{Trust in the Cloud: The Role of SDN},'' \emph{Network Security},
  vol. 2013, no.~3, pp. 5--6, 2013.

\bibitem{aikat2017rethinking}
J.~Aikat, A.~Akella, J.~S. Chase, A.~Juels, M.~K. Reiter, T.~Ristenpart,
  V.~Sekar, and M.~Swift, ``{Rethinking Security in the Era of Cloud
  Computing},'' \emph{Security \& Privacy}, vol.~15, no.~3, pp. 60--69, 2017.

\bibitem{li2016mobile}
H.~Li, G.~Shou, Y.~Hu, and Z.~Guo, ``{Mobile Edge Computing: Progress and
  Challenges},'' in \emph{International Conference on Mobile Cloud Computing,
  Services, and Engineering (MobileCloud)}.\hskip 1em plus 0.5em minus
  0.4em\relax IEEE, 2016, pp. 83--84.

\end{thebibliography}
\vfill

%

\begin{IEEEbiography}[{\includegraphics[width=1in,height=1.25in,clip,keepaspectratio]{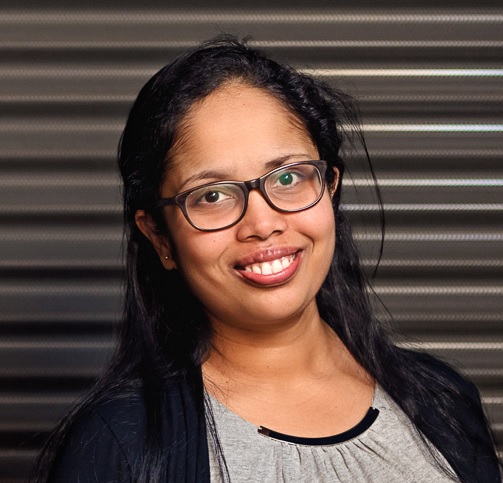}}]{Pawani Porambage}
Pawani Porambage is a Ph.D. student the Centre for Wireless Communications, University of Oulu, Finland. She obtained her Bachelor Degree in Electronics and Telecommunication Engineering in 2010 from University of Moratuwa, Sri Lanka and her Master’s Degree in Ubiquitous Networking and Computer Networking in 2012 from University of Nice Sophia-Anipolis, France. Her main research interests include lightweight security protocols, security and privacy on IoT and MEC, and Wireless Sensor Networks.
\end{IEEEbiography}

\begin{IEEEbiography}
[{\includegraphics[width=1in,height=1.25in,clip,keepaspectratio]{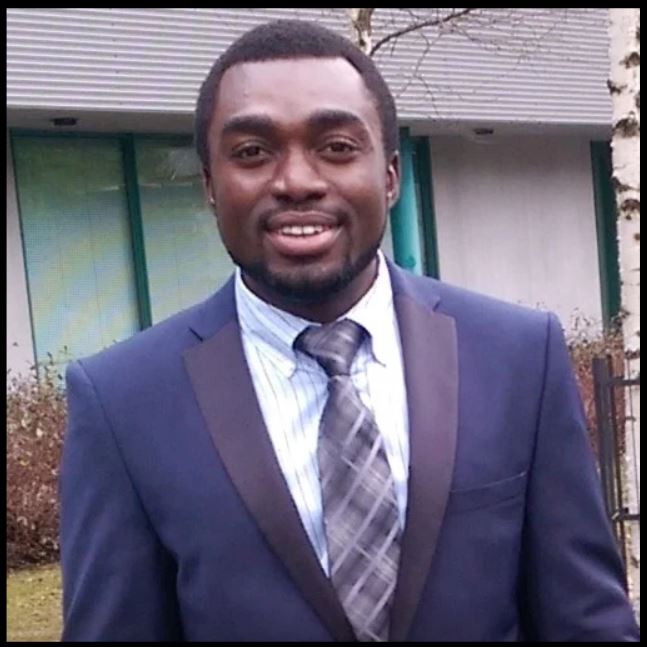}}]{Jude Okwuibe}
Jude Okwuibe received his BSc in Telecommunications and Wireless Technologies from the American University of Nigeria, Yola, in 2011. In 2015, Okwuibe received a master degree in Wireless Communications Engineering from the University of Oulu, Finland. Jude is currently doing a doctoral programme in Communications Engineering at the University of Oulu Graduate School (UniOGS), Finland. His research interests are 5G and future networks, IoT, SDN, Network security, and biometric verifications.
\end{IEEEbiography}

\begin{IEEEbiography}[{\includegraphics[width=1in,height=1.25in,clip,keepaspectratio]{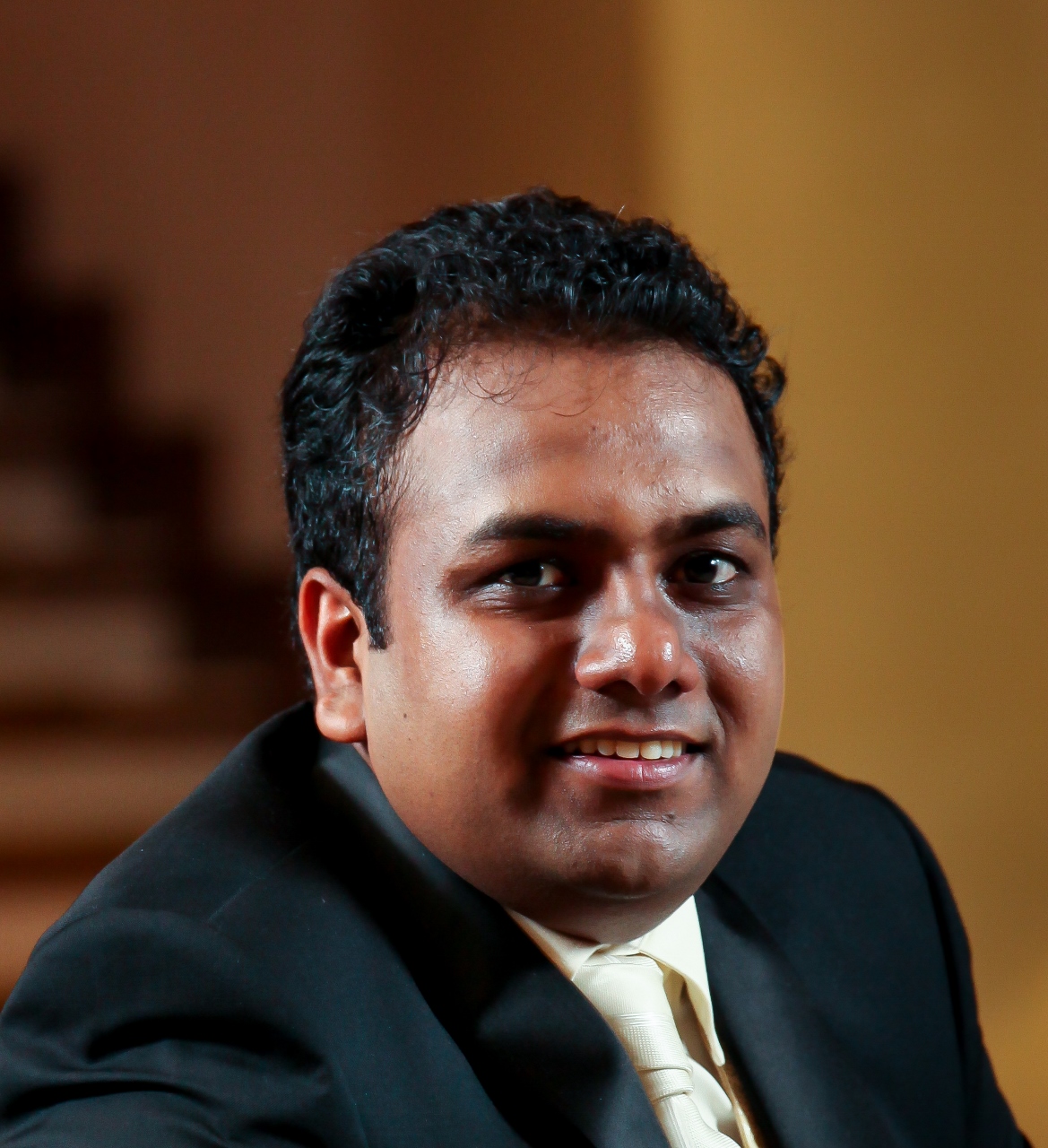}}]{Madhusanka Liyanage}
received his B.Sc. degree (First Class Honours) in electronics and telecommunication engineering from the University of Moratuwa, Moratuwa, Sri Lanka, in 2009, the M.Eng. degree from the Asian Institute of Technology, Bangkok, Thailand, in 2011, the M.Sc. degree from the University of Nice Sophia Antipolis, Nice, France, in 2011, and the Ph.D. degree in communication engineering from the University of Oulu, Oulu, Finland, in 2016. From 2011 to 2012, he worked a Research Scientist at the I3S Laboratory and Inria, Shopia Antipolis, France. He is currently a Post-Doctoral Researcher and a Project Manager at the Center for Wireless Communications, University of Oulu. He has been a Visiting Research Fellow at the Department of Computer Science, University of Oxford,  Data61, CSIRO, Sydney, Australia, the Infolabs21, Lancaster University, U.K., and Computer Science and Engineering, The University of New South Wales during 2015-2018. 

He has co-authored over 40 publications including two edited books with Wiley and one patent. He served as a Technical program Committee Members at EAI M3Apps 2016, 5GU 2017, EUCNC 2017, EUCNC 2018, MASS 2018, 5G-WF 2018, MCWN 2018  conferences and Technical program co-chair in SecureEdge workshop at IEEE CIT2017, MEC-IoT Workshop at 5GWF 2018 and BlockchainIoT workshop at Globecom 2018 conferences. He has also served as the session chair in a number of other conferences including IEEE WCNC 2013, CROWNCOM 2014, 5GU 2014, IEEE CIT 2017, IEEE PIMRC 2017. Moreover, He has received two best Paper Awards in the areas of SDMN security (at NGMAST 2015) and 5G Security (at IEEE CSCN 2017). Additionally, he has been awarded two research grants (IRC Postdoctoral Grant and Marie-Curie Fellowship) and 21 other prestigious awards/scholarships during his research career.

Dr. Liyanage has worked for more than twelve EU, international and national projects in ICT domain. He held responsibilities as a leader of work packages in several national and EU projects. Currently, he is the Finnish national coordinator for EU COST Action CA15127 on resilient communication services. In addition, he is/was serving as a management committee member for four other EU COST action projects namely EU COST Action IC1301, IC1303, CA15107 and CA16226. Liyanage has over three years’ experience in research project management, research group leadership, research project proposal preparation, project progress documentation and graduate student co-supervision/mentoring, skills. In  2015, 2016 and 2017, he won the Best Researcher Award at the Centre for Wireless Communications, University of Oulu for his excellent contribution in project management and dissemination activities. Additionally, two of the research projects (MEVICO and SIGMONA projects) received the CELTIC Excellence Award in 2013 and 2017 respectively.

Dr. Liyanage's research interests are SDN, IoT, Blockchain, MEC, mobile and virtual network security. Contact him at madhusanka.liyanage@oulu.fi 
\end{IEEEbiography}

\begin{IEEEbiography}[{\includegraphics[width=1in,height=1.25in,clip,keepaspectratio]{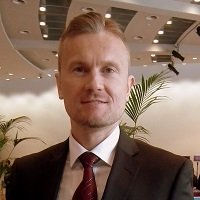}}]{Mika Ylianttila}
Prof. Mika Ylianttila is  a  full-time  professor  at  the  Centre  for  Wireless Communications  (CWC),  at  the  Faculty  of  Information  Technology and Electrical Engineering (ITEE), University of Oulu, Finland. Previously  he  was  the  director  of  the  Center  for  Internet  Excellence (2012–2015) and associate director of the MediaTeam research group (2009–2011),   and   professor   (pro   tem)   in   Information   networks (2005–2010).  He  is  also  adjunct  professor  in  Computer  Science  and Engineering  (since  2007).  He  received  his  doctoral  degree  on  Communications Engineering at the University of Oulu in 2005. He has coauthored   more   than   100   international   peer-reviewed   articles   on broadband  communications  networks  and  systems,  including  aspects on  network  security,  mobility  management,  distributed  systems  and novel applications. Research Interests include also 5G applications and services,  SDN  and  edge  computing.  He  is  a Senior Member of IEEE, and Editor in Wireless Networks journal.
\end{IEEEbiography}

\begin{IEEEbiography}[{\includegraphics[width=1in,height=1.25in,clip,keepaspectratio]{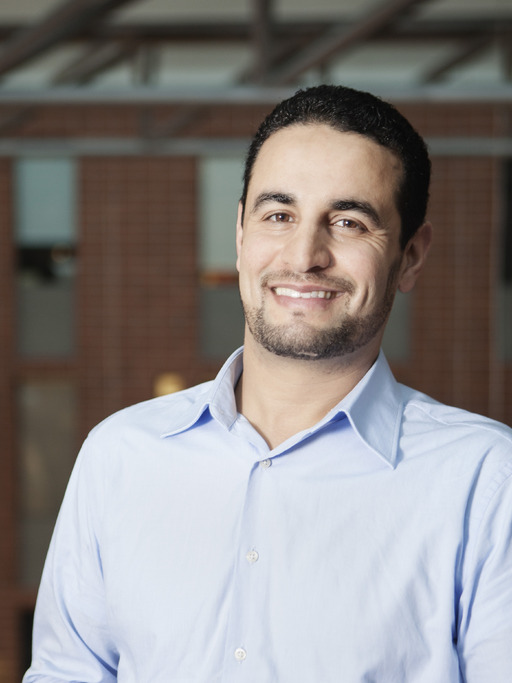}}]{Tarik Taleb}
Prof. Tarik Taleb is an IEEE Communications Society (ComSoc) Distinguished Lecturer and a senior member of IEEE. He is currently Professor at the School of Electrical Engineering, Aalto University, Finland. Prior to his current academic position, he was working as Senior Researcher and 3GPP Standards Expert at NEC Europe Ltd, Heidelberg, Germany. He was then leading the NEC Europe Labs Team working on R\&D projects on carrier cloud platforms, an important vision of 5G systems. Before joining NEC and till Mar. 2009, he worked as assistant professor at the Graduate School of Information Sciences, Tohoku University, Japan, in a lab fully funded by KDDI. From Oct. 2005 till Mar. 2006, he worked as research fellow at the Intelligent Cosmos Research Institute, Sendai, Japan. He received his B. E degree in Information Engineering with distinction, M.Sc. and Ph.D. degrees in Information Sciences from Tohoku Univ., in 2001, 2003, and 2005, respectively.

Prof. Taleb’s research interests lie in the field of architectural enhancements to mobile core networks (particularly 3GPP’s), mobile cloud networking, network function virtualization, software defined networking, mobile multimedia streaming, inter-vehicular communications, and social media networking. Prof. Taleb has been also directly engaged in the development and standardization of the Evolved Packet System as a member of 3GPP’s System Architecture working group. Prof. Taleb is a member of the IEEE Communications Society Standardization Program Development Board. As an attempt to bridge the gap between academia and industry, Prof. Taleb founded the “IEEE Workshop on Telecommunications Standards: from Research to Standards”, a successful event that got awarded “best workshop award” by IEEE Communication Society (ComSoC). Based on the success of this workshop, Prof. Taleb has also founded and has been the steering committee chair of the IEEE Conf. on Standards for Communications and Networking.

Prof. Taleb is the general chair of the 2019 edition of the IEEE Wireless Communications and Networking Conference (WCNC’19) to be held in Marrakech, Morocco. He is/was on the editorial board of the IEEE Transactions on Wireless Communications, IEEE Wireless Communications Magazine, IEEE Journal on Internet of Things, IEEE Transactions on Vehicular Technology, IEEE Communications Surveys \& Tutorials, and a number of Wiley journals. Till Dec. 2016, he served as chair of the Wireless Communications Technical Committee, the largest in IEEE ComSoC. He also served as Vice Chair of the Satellite and Space Communications Technical Committee of IEEE ComSoc (2006 - 2010). He has been on the technical program committee of different IEEE conferences, including Globecom, ICC, and WCNC, and chaired some of their symposia. 

Prof. Taleb is the (co)recipient of the 2017 IEEE Communications Society Fred W. Ellersick Prize (May 2017), the 2009 IEEE ComSoc Asia-Pacific Best Young Researcher award (Jun. 2009), the 2008 TELECOM System Technology Award from the Telecommunications Advancement Foundation (Mar. 2008), the 2007 Funai Foundation Science Promotion Award (Apr. 2007), the 2006 IEEE Computer Society Japan Chapter Young Author Award (Dec. 2006), the Niwa Yasujirou Memorial Award (Feb. 2005), and the Young Researcher's Encouragement Award from the Japan chapter of the IEEE Vehicular Technology Society (VTS) (Oct. 2003). Some of Prof. Taleb’s research work have been also awarded best paper awards at prestigious conferences. 

\end{IEEEbiography}

\vfill






\end{document}